\documentclass[12pt,a4paper]{article}

\usepackage[utf8]{inputenc}
\usepackage{amsmath}
\usepackage{amsfonts} 
\usepackage{mathtools} 
\usepackage{amssymb} 
\usepackage{subfig} 
\usepackage{graphicx}
\usepackage{xcolor}
\usepackage{cite} 
\usepackage{mathrsfs}
\usepackage{hyperref}
\usepackage{multirow}
\usepackage[normalem]{ulem}

\renewcommand{\refeq}[1]{eq.\,\eqref{#1}} 
\newcommand{\refeqs}[1]{eqs.\,\eqref{#1}} 
\newcommand{\abs}[1]{|#1|} 
\newcommand{\Abs}[1]{\left|#1\right|} 
\newcommand{\re}[1]{\text{Re}\left(#1\right)}
\newcommand{\im}[1]{\text{Im}\left(#1\right)}
\newcommand{\Hc}{\text{H.c.}}
\newcommand{\id}{\mathbf{1}}

\newcommand{\BR}[1]{\text{Br}\left(#1\right)}

\newcommand{\Zn}[1]{\mathbb{Z}_{#1}}
\newcommand{\ZZ}{\Zn{2}}
\newcommand{\cb}{c_\beta}
\renewcommand{\sb}{s_\beta}

\newcommand{\cba}{c_{\alpha\beta}}
\newcommand{\sba}{s_{\alpha\beta}}
\newcommand{\tb}{t_\beta}

\newcommand{\tbinv}{\tb^{-1}}

\newcommand{\vev}[1]{v_{#1}}
\newcommand{\ROTmat}{\mathcal R}
\newcommand{\ROTmatT}{\ROTmat^T}
\newcommand{\ROTmatinv}{\ROTmat^{-1}}
\newcommand{\ROT}[1]{\ROTmat_{#1}}
\newcommand{\HbROT}{\mathcal R_{\beta}^{\phantom{T}}}
\newcommand{\HbROTt}{\mathcal R_{\beta}^T}
\newcommand{\HbROTinv}{\mathcal R_{\beta}^{-1}}

\newcommand{\nHH}{{H}^0}
\newcommand{\nHR}{{R}^0}
\newcommand{\nHI}{{I}^0}
\newcommand{\nh}{\mathrm{h}}

\newcommand{\nH}{\mathrm{H}}
\newcommand{\nA}{\mathrm{A}}
\newcommand{\nS}{\mathrm{S}}
\newcommand{\cH}{\mathrm{H}^\pm}
\newcommand{\cHm}{\mathrm{H}^-}
\newcommand{\cHp}{\mathrm{H}^+}
\newcommand{\mNSc}{\mathcal M_0^2}

\newcommand{\mh}{m_{\nh}}

\newcommand{\mH}{m_{\nH}}
\newcommand{\mA}{m_{\nA}}
\newcommand{\mcH}{m_{\cH}}
\newcommand{\mS}{m_{\mathrm{S}}}
\newcommand{\nl}[1]{n_{#1}}
\newcommand{\nrl}[1]{\re{n_{#1}}}
\newcommand{\nrle}{\nrl{e}}\newcommand{\nrlm}{\nrl{\mu}}\newcommand{\nrlt}{\nrl{\tau}}
\newcommand{\noml}[1]{\frac{\nrl{#1}}{m_{#1}}}

 \newcommand{\FQ}{Q}\newcommand{\FL}{L}
\newcommand{\Fu}{u}\newcommand{\Fd}{d}
\newcommand{\Fl}{\ell}\newcommand{\Fn}{\nu}
\newcommand{\ferX}[3]{{#1}_{#2#3}}\newcommand{\ferXb}[3]{\bar #1_{#2#3}}

\newcommand{\dR}[1]{\ferX{\Fd}{R}{#1}}

\newcommand{\uR}[1]{\ferX{\Fu}{R}{#1}}

\newcommand{\lR}[1]{\ferX{\Fl}{R}{#1}}\newcommand{\lRb}[1]{\ferXb{\Fl}{R}{#1}}
\newcommand{\nL}[1]{\ferX{\Fn}{L}{#1}}\newcommand{\nLb}[1]{\ferXb{\Fn}{L}{#1}}

\newcommand{\SD}[1]{\Phi_{#1}^{\phantom{\dagger}}}
\newcommand{\SDd}[1]{\Phi_{#1}^\dagger}
\newcommand{\SDc}[1]{\Phi_{#1}^\ast}
\newcommand{\SDti}[1]{\tilde\Phi_{#1}^{\phantom{\dagger}}}

\newcommand{\Hv}{H_1}\newcommand{\Hvti}{\tilde H_1}
\newcommand{\Ho}{H_2}\newcommand{\Hoti}{\tilde H_2}

\newcommand{\weakferX}[3]{{#1}_{#2#3}^0}\newcommand{\weakferXb}[3]{\bar #1_{#2#3}^0}

\newcommand{\wQLb}[1]{\weakferXb{\FQ}{L}{#1}}

\newcommand{\wdR}[1]{\weakferX{\Fd}{R}{#1}}

\newcommand{\wuR}[1]{\weakferX{\Fu}{R}{#1}}
\newcommand{\wLLb}[1]{\weakferXb{\FL}{L}{#1}}

\newcommand{\wlR}[1]{\weakferX{\Fl}{R}{#1}}

\newcommand{\QLb}[1]{\ferXb{\FQ}{L}{#1}}
\newcommand{\LLb}[1]{\ferXb{\FL}{L}{#1}}

\newcommand{\CKM}{V}\newcommand{\CKMdag}{V^\dagger}

\newcommand{\Vc}[1]{\CKM^{\ast}_{#1}}

\newcommand{\PMNS}{U}\newcommand{\PMNSdag}{U^\dagger}

\newcommand{\matXF}[2]{{\rm #1}_{#2}}\newcommand{\matXFd}[2]{{\rm #1}_{#2}^\dagger}
\newcommand{\wmatXF}[2]{{\rm #1}_{#2}^{0}}
\newcommand{\basematM}{M}\newcommand{\basematN}{N}
\newcommand{\matNf}[1]{\matXF{\basematN}{#1}}\newcommand{\matNfd}[1]{\matXFd{\basematN}{#1}}
\newcommand{\matMf}[1]{\matXF{\basematM}{#1}}
\newcommand{\matND}{\matXF{\basematN}{d}}\newcommand{\matNDd}{\matXFd{\basematN}{d}}
\newcommand{\matNU}{\matXF{\basematN}{u}}\newcommand{\matNUd}{\matXFd{\basematN}{u}}
\newcommand{\matNL}{\matXF{\basematN}{\ell}}\newcommand{\matNLd}{\matXFd{\basematN}{\ell}}

\newcommand{\matMD}{\matXF{\basematM}{d}}
\newcommand{\matMU}{\matXF{\basematM}{u}}
\newcommand{\matML}{\matXF{\basematM}{\ell}}

\newcommand{\wmatMf}{\wmatXF{\basematM}{f}}
\newcommand{\wmatND}{\wmatXF{\basematN}{d}}
\newcommand{\wmatNU}{\wmatXF{\basematN}{u}}
\newcommand{\wmatNL}{\wmatXF{\basematN}{\ell}}

\newcommand{\wmatMD}{\wmatXF{\basematM}{d}}
\newcommand{\wmatMU}{\wmatXF{\basematM}{u}}
\newcommand{\wmatML}{\wmatXF{\basematM}{\ell}}

\newcommand{\matYukF}[2]{Y_{#1#2}}

\newcommand{\matYukU}[1]{\matYukF{u}{#1}}

\newcommand{\matYukD}[1]{\matYukF{d}{#1}}

\newcommand{\matYukL}[1]{\matYukF{\ell}{#1}}

\newcommand{\glFC}[1]{#1-g$\ell$FC}
\newcommand{\solA}{[A]}
\newcommand{\solB}{[B]}
\newcommand{\solBpm}{[B$_\pm$]}
\newcommand{\solBp}{[B$_+$]}
\newcommand{\solBm}{[B$_-$]}

\graphicspath{{Figs/}}

\begin{document}

 \hfill\begin{minipage}[r]{0.3\textwidth}\begin{flushright}  IFIC/20-24\\ CFTP/20-007 \end{flushright} \end{minipage}

\begin{center}

\vspace{0.50cm}
{\large\bf {Electron and muon $g-2$ anomalies in general flavour conserving two Higgs doublets models}}

\vspace{0.50cm}
Francisco J. Botella  $^{a,}$\footnote{\texttt{Francisco.J.Botella@uv.es}}, 
Fernando Cornet-Gomez  $^{a,}$\footnote{\texttt{Fernando.Cornet@ific.uv.es}}, 
Miguel Nebot $^{b}$\footnote{\texttt{miguel.r.nebot.gomez@tecnico.ulisboa.pt}}
\end{center}
\vspace{0.50cm}
\begin{flushleft}
\emph{$^a$ Departament de F\' \i sica Te\`orica and IFIC, Universitat de Val\`encia-CSIC,\\ \quad E-46100, Burjassot, Spain.} \\
\emph{$^b$ Departamento de F\'\i sica and Centro de F\' \i sica Te\' orica de Part\' \i culas (CFTP),\\
\quad Instituto Superior T\' ecnico (IST), U. de Lisboa (UL),\\ 
\quad Av. Rovisco Pais 1, P-1049-001 Lisboa, Portugal.} 
\end{flushleft}
\vspace{0.5cm}
\date{\today}
\begin{abstract}
In general two Higgs doublet models (2HDMs) without scalar flavour changing neutral couplings (SFCNC) in the lepton sector, the electron, muon and tau interactions can be decoupled in a robust framework, stable under renormalization group evolution. In this framework, the breaking of lepton flavour universality (LFU) goes beyond the mass proportionality, opening the possibility to accommodate in a simple manner a different behaviour among charged leptons. We analyze simultaneously the electron and muon $(g-2)$ anomalies in the context of these general flavour conserving models in the leptonic sector (g$\ell$FC). We consider two different models, I-g$\ell$FC and II-g$\ell$FC, in which the quark Yukawa couplings coincide, respectively, with the ones in type I and in type II 2HDMs.
We find two types of solutions that fully reproduce both $(g-2)$ anomalies, and which are compatible with experimental constraints from LEP and LHC, from LFU, from flavour and electroweak physics, and with theoretical constraints in the scalar sector. 
In the first type of solution, all the new scalars have masses in the 1--2.5 TeV range, the vacuum expectation values (vevs) of both doublets are quite similar in magnitude, and both anomalies are dominated by two loop Barr-Zee contributions. This solution appears in both models.
There is a second type of solution, where one loop contributions are dominant in the muon anomaly, all new scalars have masses below 1 TeV, and the ratio of vevs is in the range 10--100. The second neutral scalar $\mathrm{H}$ is the lighter among the new scalars, with a mass in the 210--390 GeV range while the pseudoscalar $\mathrm{A}$ is the heavier, with a mass in the range 400--900 GeV. The new charged scalar $\mathrm{H}^\pm$ is almost degenerate either with the scalar or with the pseudoscalar. This second type of solution only appears in the I-g$\ell$FC model. Both solutions require the soft breaking of the $\mathbb{Z}_{2}$ symmetry of the Higgs potential.
\end{abstract}


\section{Introduction\label{SEC:Introduction}}
%
After an improved determination of the fine structure constant \cite{Parker:2018vye}, a new anomaly has emerged \cite{Davoudiasl:2018fbb} concerning the anomalous magnetic moment of the electron $a_e=(g_e-2)/2$: there is a discrepancy among the experimental determination and the Standard Model (SM) prediction \cite{Aoyama:2012wj,Aoyama:2012wk,Laporta:2017okg,Aoyama:2017uqe,Volkov:2019phy,Terazawa:2018pdc},
\begin{equation}\label{eq:ae}
\delta a_e\equiv a_e^{\rm Exp}-a_e^{\rm SM}=-(8.7\pm 3.6)\times 10^{-13}\,.
\end{equation}
Another well known and long standing anomaly concerns the anomalous magnetic moment of the muon \cite{Bennett:2006fi,Jegerlehner:2009ry,Davier:2010nc,Davier:2019can,Blum:2018mom,Aoyama:2020ynm,Roig:2019reh},
\begin{equation}\label{eq:am}
\delta a_\mu\equiv a_\mu^{\rm Exp}-a_\mu^{\rm SM}=(2.7\pm 0.9)\times 10^{-9}\,.
\end{equation}
It is to be noticed that the anomalies in \refeqs{eq:ae} and \eqref{eq:am} have opposite sign.\\
Because of this difference of sign, several New Physics solutions addressing \refeq{eq:am} tend to be eliminated as solutions to both \refeqs{eq:am} and \eqref{eq:ae}. In particular, many popular models in which the anomaly scales with the square of the lepton mass \cite{Giudice:2012ms} tend to generate too large $\delta a_e$ with the wrong sign. Some authors \cite{Crivellin:2018qmi} argue that if the origin of both anomalies is beyond the SM, the corresponding model must incorporate some sort of effective decoupling between $\mu$ and $e$. Recent beyond-SM explanations of both anomalies can be found in  \cite{Liu:2018xkx,Han:2018znu,Endo:2019bcj,Bauer:2019gfk,Badziak:2019gaf,Hiller:2019mou,CarcamoHernandez:2020pxw,Haba:2020gkr,Bigaran:2020jil,Calibbi:2020emz,Chen:2020jvl,Jana:2020pxx,Hati:2020fzp,Dutta:2020scq,Sabatta:2019nfg}. 
A minimal extension of the SM is the two Higgs doublets model (2HDM) \cite{Lee:1973iz} which introduces, in general, a new set of flavour structures in the Yukawa sector. Those structures could implement the decoupling between $\mu$ and $e$ required to explain $\delta a_\mu$ and $\delta a_e$. Of course, the most popular 2HDMs shaped by symmetries \cite{Branco:2011iw,Ivanov:2017dad}, the so-called 2HDMs of types I, II, X and Y \cite{Glashow:1976nt,Haber:1978jt,Barger:1989fj}, do not implement in a straightforward way this decoupling between $\mu$ and $e$, since the new Yukawa couplings in the lepton sector are proportional to the charged lepton mass matrix.\\ 
Going one step further in generality, the so-called ``Aligned'' 2HDM (A2HDM) \cite{Pich:2009sp} gives up stability of the model under the renormalization group evolution (RGE) \cite{Ferreira:2010xe} (the model is not shaped by a symmetry). The A2HDM cannot, however, incorporate some effective decoupling between $\mu$ and $e$ since the new Yukawa structures are still proportional to the fermion mass matrices. It is nevertheless interesting to note that the lepton sector of the A2HDM is stable under one loop RGE\footnote{As in the SM, one is assuming massless neutrinos.} \cite{Botella:2015yfa,Braeuninger:2010td,Jung:2010ik}: scalar flavour changing neutral couplings (SFCNC), absent at tree level, do not appear at one loop.\\
A generalization of the A2HDM is the general flavour conserving (gFC) 2HDM where, at tree level, all Yukawa couplings are diagonal in the fermion mass basis \cite{Penuelas:2017ikk,Botella:2018gzy,Rodejohann:2019izm}. As in the A2HDM, it has been shown that the charged lepton sector of the gFC-2HDM is one loop stable under RGE, in the sense that SFCNC, absent at tree level, are not generated at one loop \cite{Botella:2018gzy}. This implies that a well behaved and minimal 2HDM that can implement the effective decoupling among $\mu$ and $e$ is a gFC-2HDM in the leptonic sector. Since this is all what is required to address the two anomalies in \refeqs{eq:ae}--\eqref{eq:am}, we consider two minimal models in which the quark sector is a 2HDM of either type I or type II, while the lepton sector corresponds to a gFC-2HDM. We refer to them as models \glFC{I} and \glFC{II} respectively. Note that these models do not have SFCNC at tree level neither in the quark nor in the lepton sectors. Additionally, the new Yukawa couplings in the lepton sector are independent of the charged lepton mass matrix. In the appropriate limits, model \glFC{I} can reproduce 2HDMs of types I and X while, similarly, model \glFC{II} can reproduce 2HDMs of types II and Y. In this sense model \glFC{I} is a generalization of 2HDMs of types I and X, while model \glFC{II} is instead a generalization of 2HDMs of types II and Y. The convenience of adopting this kind of generalized flavour conserving 2HDMs for phenomenological analyses was advocated in \cite{Botella:2018gzy}. 

The paper is organised as follows. In section \ref{SEC:Models} the models are presented in detail. In section \ref{SEC:g-2}, the one and two loop contributions to $a_\ell$ are revisited. In a simplified analysis it is shown that, with dominating two loop contributions, a new simple scaling law follows:
\begin{equation}\label{eq:ae:am:scaling}
\frac{\delta a_{e}}{\delta a_{\mu}}=\frac{m_{e}\,\nrle}{m_{\mu}\,\nrlm}\,,
\end{equation}
with $\nl{e}$, $\nl{\mu}$, the new Yukawa couplings of the charged leptons, in the lepton mass basis. In order to solve the discrepancies in \refeqs{eq:ae}--\eqref{eq:am} through the two loop contributions, the scaling in \refeq{eq:ae:am:scaling} requires
\begin{equation}\label{eq:ae:am:scaling:num}
\nrl{\mu}=-\left(15.11_{-7.56}^{+15.11}\right) \nrl{e}
\end{equation}
in the framework of models \glFC{I} and \glFC{II}. Besides solutions with dominating two loop contributions, an additional possibility with relevant one loop contributions is also analysed (similarly to \cite{Davoudiasl:2018fbb}). In section \ref{SEC:Constraints}, a number of constraints, relevant for a full analysis, is addressed in detail. In section \ref{SEC:Results}, the main results of such a full analysis are presented and discussed. Details concerning some aspects of the different sections are relegated to the appendices.

\section{The \glFC{I} and \glFC{II} models \label{SEC:Models}}
%
In 2HDMs, the Yukawa sector of the SM is extended to
\begin{multline}\label{eq:LagrYuk}
\mathscr L_{\rm Y}=
 -\wQLb{}\left(\SD{1}\matYukD{1}+\SD{2}\matYukD{2}\right)\wdR{}
 -\wQLb{}\left(\SDti{1}\matYukU{1}+\SDti{2}\matYukU{2}\right)\wuR{}\\
 -\wLLb{}\left(\SD{1}\matYukL{1}+\SD{2}\matYukL{2}\right)\wlR{}+\Hc
\end{multline}
where $\SDti{j}=i\sigma_2\SDc{j}$, and, as in the SM, neutrinos are massless (in the leptonic sector only two flavour structures are present). The vacuum expectation values $\vev{j}$ of the scalar fields $\SD{j}$ are in general non-vanishing; expanding around the vacuum appropriate for electroweak symmetry breaking,
\begin{equation}\label{eq:SDgen}
\SD{j}=e^{i\theta_j}\begin{pmatrix}\varphi^+_j\\ \frac{\vev{j}+\rho_j+i\eta_j}{\sqrt 2}\end{pmatrix}\,.
\end{equation}
The so-called Higgs basis \cite{Georgi:1978ri,Donoghue:1978cj,Botella:1994cs} is defined by 
\begin{equation}\label{eq:HiggsBasis:01}
\begin{pmatrix}\Hv\\ \Ho\end{pmatrix}=\HbROT\,
\begin{pmatrix}e^{-i\theta_1}\SD{1}\\ e^{-i\theta_2}\SD{2}\end{pmatrix},\quad \text{with}\quad 
\HbROT=\begin{pmatrix}\phantom{-}\cb & \sb\\ -\sb & \cb \end{pmatrix},\ \HbROTt=\HbROTinv,
\end{equation}
in such a way that only one of the scalar doublets has a non-vanishing vacuum expectation value: $\langle \Hv\rangle=\frac{\vev{}}{\sqrt 2}\left(\begin{smallmatrix}0\\ 1\end{smallmatrix}\right)$, $\langle \Ho\rangle=\left(\begin{smallmatrix}0\\ 0\end{smallmatrix}\right)$. In \refeq{eq:HiggsBasis:01}, we have used  $\cb\equiv\cos\beta=\vev{1}/\vev{}$, $\sb\equiv \sin \beta =\vev{2}/\vev{}$, with $\vev{}^2=\vev{1}^2+\vev{2}^2=\frac{1}{\sqrt{2}G_F}$.
Expanding around the vacuum
\begin{equation}
H_{1}=\begin{pmatrix} G^+\\ \frac{v+\nHH+iG^0}{\sqrt{2}}\end{pmatrix},\quad 
H_{2}=\begin{pmatrix} \cHp\\ \frac{\nHR+i\nHI}{\sqrt{2}}\end{pmatrix},
\end{equation}
the would-be Goldstone bosons $G^{0}$, $G^{\pm}$ and the physical charged scalar $H^{\pm}$ are already identified. The neutral scalars $\{\nHH,\nHR,\nHI\}$ are not, in general, the mass eigenstates.\\
It is in the Higgs basis where the Yukawa couplings have the simplest interpretation:
\begin{multline}
\mathscr L_{\rm Y}=
 -\frac{\sqrt{2}}{v}\wQLb{}\left(\Hv\wmatMD+\Ho\wmatND\right)\wdR{}
 -\frac{\sqrt{2}}{v}\wQLb{}\left(\Hvti\wmatMU+\Hoti\wmatNU\right)\wuR{}\\
 -\frac{\sqrt{2}}{v}\wLLb{}\left(\Hv\wmatML+\Ho\wmatNL\right)\wlR{}+\Hc \ .
\end{multline}
Since only the neutral component of $\Hv$ has a non-vanishing vacuum expectation value, the Yukawa couplings $\wmatMf$, for all the fermions $f=u,d,\ell$, will be the corresponding mass matrices. Going directly to the fermion mass bases, we obtain the relevant new Yukawa structures
\begin{multline}
\mathscr L_{\rm Y}= 
 -\frac{\sqrt{2}}{v}\QLb{}\left(\Hv\matMD+\Ho\matND\right)\dR{}
 -\frac{\sqrt{2}}{v}\QLb{}\left(\Hvti\matMU+\Hoti\matNU\right)\uR{}\\
 -\frac{\sqrt{2}}{v}\LLb{}\left(\Hv\matML+\Ho\matNL\right)\lR{}+\Hc\,,
\end{multline}
where $\matMf{f}$ are the diagonal fermion mass matrices for $f=u,d,\ell$ and $\matNf{f}$ are the new flavour structures that may be able to explain the electron and muon anomalies in \refeqs{eq:ae}--\eqref{eq:am}. As motivated previously, we consider two models.
\begin{itemize}
\item Model \glFC{I} is defined by\footnote{Here and in the following, $\tb\equiv \tan\beta$ and $\tbinv\equiv\cot\beta$.}
\begin{equation}
\matNU=\tbinv\matMU,\quad \matND=\tbinv\matMD,\quad \matNL=\text{diag}(\nl{e},\nl{\mu},\nl{\tau})\,.
\end{equation}
The couplings $\matNU,\matND$ are the same as in 2HDMs of types I or X.
\item Model \glFC{II} is defined by
\begin{equation}
\matNU=\tbinv\matMU,\quad \matND=-\tb\matMD,\quad \matNL=\text{diag}(\nl{e},\nl{\mu},\nl{\tau})\,.
\end{equation}
The couplings $\matNU,\matND$ are the same as in 2HDMs of types II or Y.
\end{itemize}
In both models $\matNL$ is diagonal, arbitrary and stable at one loop level under RGE, in the sense that it remains diagonal. Note that the effective decoupling among the new couplings of $e$ and $\mu$ that is required in order to explain the $g-2$ anomalies is simply obtained from the independence of $\nl{e}$ and $\nl{\mu}$.\\ 
To complete the definition of the model, in accordance with the fact that the quark sector is a type I or type II 2HDM, we adopt a $\ZZ$ symmetric scalar potential
\begin{multline}\label{eq:ScalarPotential:general:01}
\mathcal V(\SD{1},\SD{2})= 
\mu_{11}^2\SDd{1}\SD{1}+\mu_{22}^2\SDd{2}\SD{2}+\left(\mu_{12}^2\SDd{1}\SD{2}+\Hc\right)
+\lambda_1(\SDd{1}\SD{1})^2+\lambda_2(\SDd{2}\SD{2})^2\\
+2\lambda_3(\SDd{1}\SD{1})(\SDd{2}\SD{2})+2\lambda_4(\SDd{1}\SD{2})(\SDd{2}\SD{1})+\left(\lambda_5(\SDd{1}\SD{2})^2+\Hc\right)\,.
\end{multline}%
For $\mu_{12}^2\neq 0$, the $\ZZ$ symmetry is softly broken. 
This potential generates the mass matrix of the neutral scalars $\mNSc$, which is diagonalised by a $3\times 3$ real orthogonal matrix $\ROTmat$
\begin{equation}\label{eq:NeutralMass:Diag:01}
\ROTmatT\,\mNSc\,\ROTmat=\text{diag}(\mh^2,\mH^2,\mA^2)\,,\quad \ROTmatinv=\ROTmatT\,.
\end{equation}
The physical neutral scalars $\{\nh,\nH,\nA\}$ are:
\begin{equation}\label{eq:ScalarROT:00}
\begin{pmatrix}\nh\\ \nH\\ \nA\end{pmatrix}=\ROTmatT\begin{pmatrix}\nHH\\ \nHR\\ \nHI\end{pmatrix}\,.
\end{equation}
The Yukawa couplings of the neutral scalars are flavour conserving\footnote{The general form of the Yukawa couplings is given, for completeness, in appendix \ref{APP:Yukawa}.}:
\begin{equation}\label{eq:YukawaNeutral:gen}
\mathscr L_{N}=-\sum_{S=\nh,\nH,\nA}\,\sum_{f=u,d,\ell}\sum_{j=1}^3\frac{m_{f_j}}{\vev{}}\,S\,\bar f_j(a_{f_j}^S+ib_{f_j}^S\gamma_5)f_j\,.
\end{equation}
In the following we focus on a simplified case: we assume that (i) there is no CP violation in the scalar sector and (ii) the new Yukawa couplings 
are real%
, $\im{\nl{\ell}}=0$. 
In the scalar sector, this corresponds to 
\begin{equation}
\ROTmat=\begin{pmatrix}\sba & -\cba & 0\\ \cba & \sba & 0\\ 0 & 0 & 1\end{pmatrix}\,,
\end{equation}
with $\sba\equiv \sin(\alpha-\beta)$ and $\cba\equiv \cos(\alpha-\beta)$, where $\alpha-\frac{\pi}{2}$ is the mixing angle parametrizing the change of basis from the fields in \refeq{eq:SDgen} to the mass eigenstates in \refeq{eq:ScalarROT:00}. The alignment limit, in which $\nh$ has the same couplings of the SM Higgs, corresponds to $\sba\to 1$. 
Table \ref{TAB:neutralcouplings} collects the Yukawa couplings, as expressed in \refeq{eq:YukawaNeutral:gen}, in both models \glFC{I} and \glFC{II}.
\begin{table}[h!tb]
\begin{center}
\begin{tabular}{c c|c|c|c|c|c|c|}
\cline{3-8}
                                        &   & $a_u^S$ & $b_u^S$ & $a_d^S$ & $b_d^S$ & $a_{\ell}^S$ & $b_{\ell}^S$ \\ \hline
\multicolumn{1}{|c|}{\multirow{3}{*}{\glFC{I}}} & $\nh$ & $\sba+\cba\tbinv$ & 0       & $\sba+\cba\tbinv$ & 0       & $\sba+\cba\noml{\ell}$ & 0       \\ \cline{2-8} 
\multicolumn{1}{|c|}{}                  & $\nH$ & $-\cba+\sba\tbinv$ & 0       & $-\cba+\sba\tbinv$ & 0       & $-\cba+\sba\noml{\ell}$ & 0       \\ \cline{2-8} 
\multicolumn{1}{|c|}{}                  & $\nA$ & 0       & $-\tbinv$ & 0       &  $+\tbinv$  & 0       &   $\noml{\ell}$ \\ \hline \hline
\multicolumn{1}{|c|}{\multirow{3}{*}{\glFC{II}}} & $\nh$ & $\sba+\cba\tbinv$ & 0       & $\sba-\cba\tb$ & 0       & $\sba+\cba\noml{\ell}$ & 0       \\ \cline{2-8} 
\multicolumn{1}{|c|}{}                  & $\nH$ & $-\cba+\sba\tbinv$ & 0       & $-\cba-\sba\tb$ & 0       & $-\cba+\sba\noml{\ell}$ & 0       \\ \cline{2-8} 
\multicolumn{1}{|c|}{}                  & $\nA$ & 0       &   $-\tbinv$  & 0       &    $-\tb$     & 0       &   $\noml{\ell}$ \\ \hline
\end{tabular}
\caption{Fermion couplings to neutral scalars.\label{TAB:neutralcouplings}}
\end{center}
\end{table}
The absence of CP violation is clear from the exact relation $a_{f}^{S}b_{f}^{S}=0$ \cite{Nebot:2015wsa}; one important consequence of this simplification is the absence of new contributions generating electric dipole moments (EDMs), in particular contributions to the electron EDM $d_e$, which is quite constrained: $\abs{d_e}<1.1\times 10^{-29}$ e$\cdot$cm \cite{Andreev:2018ayy,Han:2015yys}. 
The Yukawa couplings of $\cH$ are of the form
\begin{equation}
\mathscr L_{Ch}=-\frac{1}{\sqrt{2}\vev{}}\sum_{f=q,l}\sum_{j,k=1}^3\left\{\cHm\ferXb{f}{-\frac{1}{2}}{,j}(\alpha_{jk}^f+i\beta_{jk}^f\gamma_5)\ferX{f}{\frac{1}{2}}{,k}+\cHp\ferXb{f}{\frac{1}{2}}{,k} (\alpha_{jk}^{f*}+i\beta_{jk}^{f*}\gamma_5)\ferX{f}{-\frac{1}{2}}{,j}\right\}
\end{equation} 
where $q_{+\frac{1}{2},j}=u_{j}$, $q_{-\frac{1}{2},j}=d_{j}$, $l_{+\frac{1}{2},j}=\nu_{j}$, $l_{-\frac{1}{2},j}=\ell_{j}$, and the corresponding couplings are given in Table \ref{TAB:chargedcouplings}. 
\begin{table}[h!tb]
\begin{center}
\begin{tabular}{c|c|c|c|c|}
\cline{2-5}
                                  & $\alpha_{ij}^q$                         & $\beta_{ij}^q$                         & $\alpha_{ij}^l$         & $\beta_{ij}^l$         \\ \hline
\multicolumn{1}{|c|}{\glFC{I}}  & $\Vc{ji}\,\tbinv (m_{u_j}-m_{d_i})$      & $\Vc{ji}\, \tbinv (m_{u_j}+m_{d_i})$     & $-\nrl{\ell_i}\delta_{ij}$ & $\nrl{\ell_i}\delta_{ij}$ \\ \hline
\multicolumn{1}{|c|}{\glFC{II}} & $ \Vc{ji}\,(\tbinv m_{u_j}+\tb m_{d_i})$ & $\Vc{ji}\,(\tbinv m_{u_j}-\tb m_{d_i})$ & $-\nrl{\ell_i}\delta_{ij}$ & $\nrl{\ell_i}\delta_{ij}$ \\ \hline
\end{tabular}
\caption{Fermion couplings to $\cH$.\label{TAB:chargedcouplings}}
\end{center}
\end{table}
Note that the Yukawa couplings of the charged leptons in Tables \ref{TAB:neutralcouplings} and \ref{TAB:chargedcouplings} are the same in both models \glFC{I} and \glFC{II}.


\section{The new contributions to $\delta a_\ell$\label{SEC:g-2}}
\begin{center}
\begin{figure}[!htb]
\includegraphics[width=0.25\textwidth]{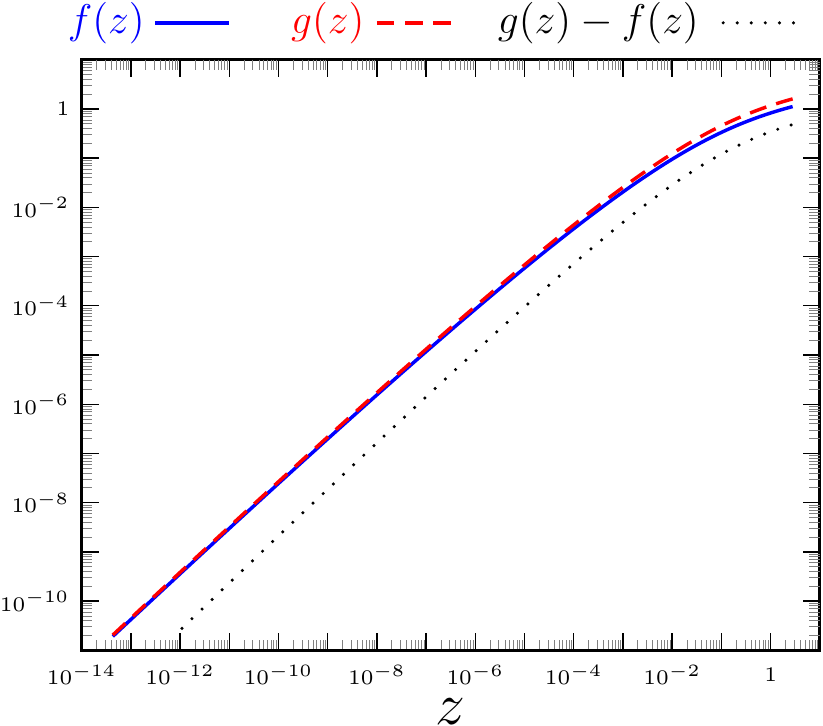}\qquad
\includegraphics[width=0.3\textwidth]{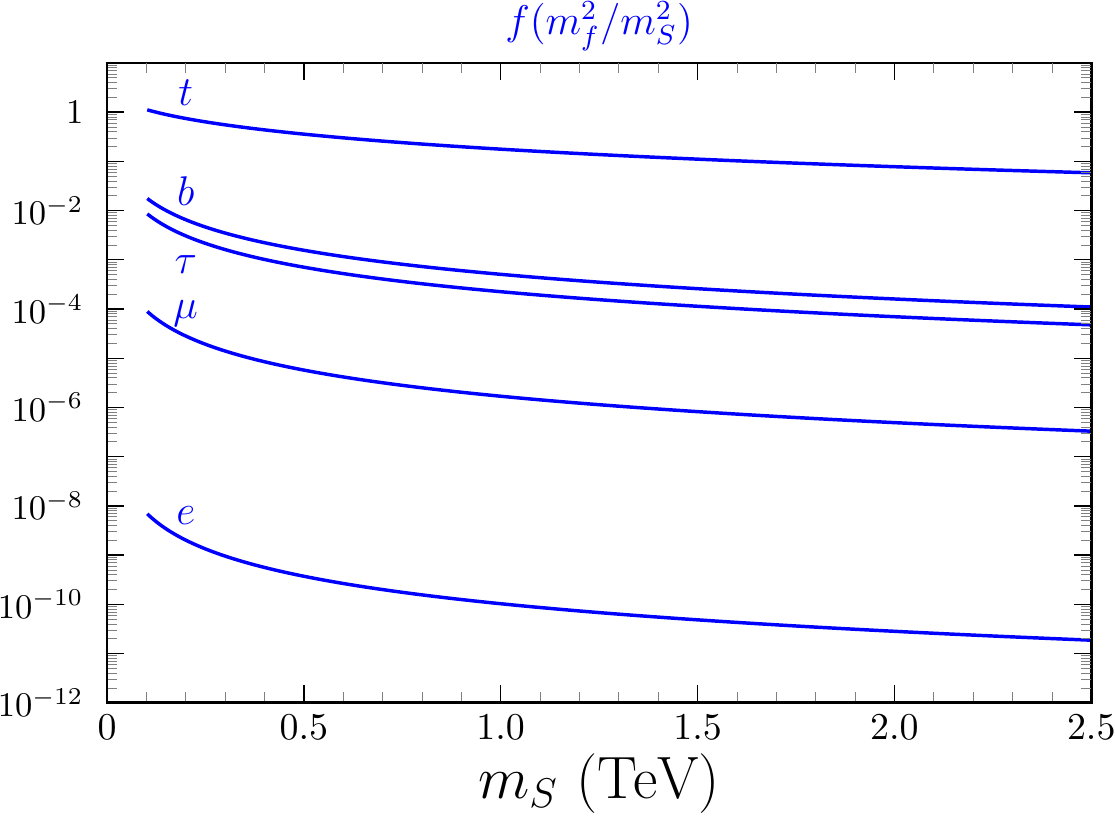}\qquad
\includegraphics[width=0.3\textwidth]{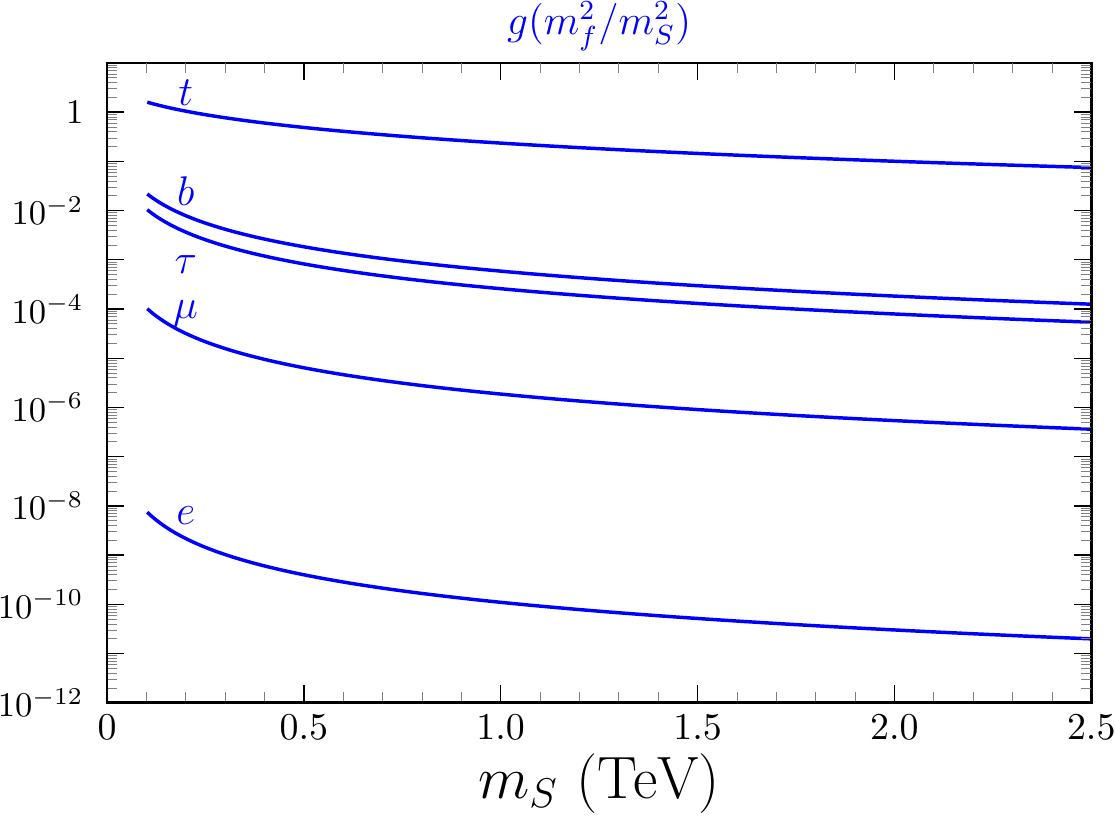}
\caption{Loop functions.\label{fig:2loopfunctions}}
\end{figure}
\end{center}

%
The full prediction $a_\ell^{\rm Th}$ of the anomalous magnetic moments of $\ell=e,\mu$ has the form
\begin{equation}
a_\ell^{\rm Th}=a_\ell^{\rm SM}+\delta a_\ell\,,
\end{equation}
with $a_\ell^{\rm SM}$ the SM contribution and $\delta a_\ell$ the corrections due to the model. 
To solve the discrepancies in \refeqs{eq:ae}--\eqref{eq:am}, the aim is to obtain $\delta a_e\simeq \delta a_e^{\rm Exp}$ and $\delta a_\mu\simeq \delta a_\mu^{\rm Exp}$ within models \glFC{I} and \glFC{II}. It is convenient to introduce $\Delta_\ell$ following
\begin{equation}\label{eq:alKl}
\delta a_\ell=K_\ell\,\Delta_\ell,\qquad K_\ell=\frac{1}{8\pi^2}\left(\frac{m_\ell}{\vev{}}\right)^2=\frac{1}{8\pi^2}\left(\frac{gm_\ell}{2M_W}\right)^2\,.
\end{equation}
The quantities $K_\ell$ collect the typical factors arising in one loop contributions; since $K_e\simeq 5.5\times 10^{-14}$ and $K_\mu\simeq 2.3\times 10^{-9}$, in order to reproduce the anomalies we roughly need
\begin{equation}
\Delta_e\simeq -16\,,\qquad \Delta_\mu\simeq 1\,.
\end{equation}
It is well known that in the type of models considered here, both one loop \cite{Leveille:1977rc} or two loop Barr-Zee contributions \cite{Barr:1990vd,Chang:1990sf,Cheung:2001hz,Cheung:2009fc,Ilisie:2015tra,Cherchiglia:2016eui} can be dominant. Complete expressions used in the full analyses of section \ref{SEC:Results}, can be found in appendix \ref{APP:Dipoles}. For the moment, we consider in this section two approximations: we only keep leading terms in a $(m_\ell/\mS)^2$ expansion (for the different scalars $\mathrm{S}=\nh,\nH,\nA)$, and the alignment limit $\sba\to 1$. With these approximations, the one loop contribution to $\Delta_\ell$ in \refeq{eq:alKl} is
\begin{equation}\label{delta_l_1_loop}%
\Delta_{\ell}^{(1)}\simeq \nl{\ell}^{2}\left(\frac{I_{\ell\nH}}{\mH^{2}}-\frac{I_{\ell\nA}-2/3}{\mA^{2}}-\frac{1}{6{\mcH}^{2}}\right)
\end{equation}%
where%
\begin{equation}%
I_{\ell\mathrm{S}}=-\frac{7}{6}-2\ln \left(\frac{m_{\ell}}{\mS}\right)\,.
\label{1_loop_function}
\end{equation}%
Equation \eqref{delta_l_1_loop} applies to both model \glFC{I} and \glFC{II}. We do not consider light scalars or pseudoscalars (see reference \cite{Jana:2020pxx}): in the different analyses it is assumed that $\nh$ is the lightest scalar, i.e. $\mh<\mH,\mA$. For a typical range $\mS\in[0.2;\, 2.0]$ TeV, the loop functions $I_{\ell\mathrm{S}}$ obey
\begin{equation}%
I_{e\mathrm{S}}\in [24.6;29.2]\,,\qquad I_{\mu\mathrm{S}}\in [13.9;18.5]\,,
\label{Values_1_loop_function}
\end{equation}%
and thus the dominant contributions to $\Delta_\ell^{(1)}$ in \refeq{delta_l_1_loop} are the logarithmically enhanced contributions from $\nH$ and $\nA$. Then, $\Delta_e\simeq -16$ can only arise from the negative sign of the $\nA$ pseudoscalar contribution: $\Delta_e\simeq -[\nrle]^2I_{e\nA}/\mA^2$. Taking into account the $I_{e\nA}$ value in \refeq{Values_1_loop_function}, it would require $[\nrle]^2\sim \mA^2$, which can easily violate perturbativity requirements in the Yukawa sector or contraints from resonant dilepton searches. Consequently, we do not expect an explanation of $\delta a_e$ in terms of one loop contributions. For $\delta a_\mu$, any relevant one loop contribution in \refeq{delta_l_1_loop} should arise from the $\nH$ contribution attending, again, to the required sign and the logarithmically enhanced value of $I_{\mu\nH}$ in \refeq{Values_1_loop_function}: $\Delta_\mu\simeq [\nrlm]^2I_{\mu\nH}/\mH^2$. For $I_{\mu\nH}\simeq 16$ such a contribution needs $[\nrlm]^2\sim [\mH/4]^2$, that is a not too heavy $\nH$ (in order to have reasonably perturbative $\nl{\mu}$) and $\mA>\mH$ in order to avoid cancellations with wrong sign contributions. In the same approximation (leading $m_\ell/\mS$ terms and $\sba\to 1$), the two loop contributions are dominated by Barr-Zee diagrams in which the internal fermion loop is connected with the external lepton via one virtual photon and one virtual neutral scalar $\nH$ or $\nA$. The leading contribution to $\Delta_\ell$ in \refeq{eq:alKl} is (for detailed expressions, see appendix \ref{APP:Dipoles})
\begin{equation}\label{delta_l_2_loops}%
\Delta_{\ell}^{(2)}=-\left( \frac{2\alpha}{\pi}\right) \left(
\frac{\nl{\ell}}{m_{\ell}}\right) F\,.
\end{equation}%
The factor $F$ depends on the masses of the fermions in the closed loop, on the couplings of those fermions to $\nH$ and $\nA$, and, of course, on $\mH$ and $\mA$; it is consequently different in models \glFC{I} and \glFC{II}:
\begin{equation}\label{Barr_II_function}
\begin{aligned}
F_{\rm I} &=\frac{\cot\beta}{3}\left[4(f_{t\nH}+g_{t\nA}) + (f_{b\nH}-g_{b\nA})\right] + \frac{\nrlt}{m_{\tau}}(f_{\tau\nH}-g_{\tau\nA})\,,    \\
F_{\rm II} &=\frac{\cot\beta }{3}\left[4(f_{t\nH}+g_{t\nA}) -\tan^{2}\beta (f_{b\nH}-g_{b\nA})\right] + \frac{\nrlt}{m_{\tau}}(f_{\tau\nH}-g_{\tau\nA})\,, 
\end{aligned}
\end{equation}
where
\begin{equation}
f_{f\mathrm{S}}\equiv f\left(\frac{m_{f}^{2}}{\mS^{2}}\right)\,,\quad g_{f\mathrm{S}}\equiv g\left(\frac{m_{f}^{2}}{\mS^{2}}\right) \,.
\end{equation}
The functions $f(z)$ and $g(z)$ are defined in appendix \ref{APP:Dipoles}; they are represented in Figure \ref{fig:2loopfunctions}.
Their main features are: (i) $f(z)\simeq g(z)$ in the whole range of interest, (ii) the largest values correspond to the heavier fermion (the top quark), (iii) the values of $f$ and $g$ for the top quark contributions vary between $0.1$ and $1$ in the relevant range of scalar masses. Considering the dominant top quark terms, for $\tb\simeq 1$ and $\mH\simeq \mA$, it is easy to realize that for $\mH\sim 1-2$ TeV, $\delta a_e$ can be explained with Yukawa couplings $\nrle\sim 3-7$ GeV ($\nrle>0$ gives the right sign of $\delta a_e$). If we assume that $\delta a_\mu$ must also be explained by the same kind of dominant Barr-Zee two loop contributions, which are independent of the specific charged lepton, it is straightforward that
\begin{equation}\label{scaling_of_the_anomaly_1}%
\delta a_{\mu}=\frac{m_{\mu}\,\nrlm}{m_{e}\,\nrle}\delta a_{e}\,.
\end{equation}%
With this relation, the origin of the different signs of $\delta a_e$ and $\delta a_\mu$ relies on the freedom to have $\nrle$ and $\nrlm$ with opposite signs, $\nrlm\simeq -15\nrle$, as anticipated in \refeq{eq:ae:am:scaling}. In terms of $\nrlm$, with the same assumptions ($\tb\sim 1$, $\mA\sim\mH\sim 1-2$ TeV), $\nrlm\in-[45;105]$ GeV. The previous arguments apply to both models, \glFC{I} and \glFC{II}, since $4(f_{t\nH}+g_{t\nA})$ is the dominant term in both $F_{\rm I}$ and $F_{\rm II}$.\\ 
Attending to the flavour constraints discussed in section \ref{SEC:Constraints} ($B_d$ and $B_s$ meson mixings, $b\to s\gamma$ radiative decays), $\tb\ll 1$ are excluded in 2HDMs of types I and II, and thus also in \glFC{I} and \glFC{II} models, there is no need to discuss the $\tb\ll 1$ regime.\\ 
Let us now analyse the two loop Barr-Zee contributions in \refeq{Barr_II_function} for large values of $\tb$. As a reference, consider the analysis above with $\tb\sim 1$ and $\mA\sim\mH\sim 1-2$ TeV; for definiteness we now take $\tb=50$. For large $\tb$, it is clear that these contributions in models \glFC{I} and \glFC{II} are quite different. Starting with model \glFC{I}, in order to maintain the right value of $\delta a_e$, the $\tb$ suppression in $\nrle\tbinv(f_{t\nH}+g_{t\nA})$ can be compensated with smaller $\mH$, $\mA$, and larger $\nrle$. For example, $\mA\sim\mH\sim 200$ GeV gives an increase of the loop functions by a factor of 10 with respect to $\mA\sim\mH\sim 1-2$ TeV; increasing then $\nrle$ by a factor of 5, the suppression $\tbinv=1/50$ is compensated. Therefore, the discrepancy in $\delta a_e$ can be explained in the \glFC{I} model through two loop contributions, for large values of $\tb$ and $\nrle\sim 15-35$ GeV. The question now is if one can explain, with the two loop contributions, the muon anomaly $\delta a_\mu$. Attending to \refeq{scaling_of_the_anomaly_1}, one would need $\nrlm\in -[225; 505]$ GeV, which would be in conflict with perturbativity requirements in the Yukawa sector. However, as the discussion on one loop contributions after \refeq{delta_l_1_loop} shows, for light $\mH$, e.g. $\mH\in[200;400]$ GeV, $\delta a_\mu$ can be obtained with $\nH$-mediated one loop contributions, and $\mA>\mH$ to avoid cancellations. One needs $\abs{\nrlm}\sim \mH/4$, in which case $\abs{\nrlm}\in [50;100]$ GeV is acceptable from the perturbativity point of view.\\
Summarizing the previous discussion, we envisage, at least, two kinds of solutions:
\begin{itemize}
\item The first is realized with scalars having masses in the 1--2 TeV range, $\tb\sim 1$, and both anomalies produced by two loop Barr-Zee contributions. The coupling of electrons to the new scalar and pseudoscalar, $\nrle$, should be in the few GeV range. Following \refeq{scaling_of_the_anomaly_1}, the corresponding muon coupling is larger. This first solution can appear, a priori, in both \glFC{I} and \glFC{II} models. In section \ref{SEC:Results} we refer to this first type of solution as ``solution \solA''.
\item The second solution corresponds to a lighter $\nH$, $\mH\in[200;400]$ GeV and a heavier $\nA$; the required values of $\tb$ are larger, $\tb\gg 1$. In this second solution, the electron anomaly is obtained with two loop contributions while the muon anomaly is one loop controlled; contrary to the first solution, there is no linear relation among $\nrlm$ and $\nrle$. This second kind of solution can clearly appear in the \glFC{I} model, but in this simplified analysis it cannot be elucidated if this possibility is also open in the \glFC{II} model. Anticipating the results of the complete numerical analyses of section \ref{SEC:Results}, this will not be the case: within the \glFC{II} model there is no solution with large $\tb$ and relatively light $\nH$. In section \ref{SEC:Results} we refer to this second type of solution as ``solution \solB''. Notice also that, a priori, this second kind of solution might be obtained with both signs of $\nrlm$.
\end{itemize}

\section{Constraints\label{SEC:Constraints}}
%
In this section we discuss the different constraints that can play a relevant role in the detailed analyses of section \ref{SEC:Results}. Each constraint is implemented as $\chi^2$ function; a global $\chi^2$ function, the sum of all separate contributions, is used to drive the analyses (for an efficient exploration of parameter space we employ Markov chain MonteCarlo techniques) and represent the relevant regions for different parameters and observables.
 The usual form $\chi^2_{\mathcal O}=\left(\frac{\mathcal O_{\rm Th}-\mathcal O_{\rm Exp}}{\sigma_{\rm Exp}}\right)^2$ is adopted for the constraint corresponding to an observable $\mathcal O$, where the experimental input is a measurement $\mathcal O_{\rm Exp}$ with uncertainty $\sigma_{\rm Exp}$ and the theoretical prediction is $\mathcal O_{\rm Th}$ (for correlated measurements or asymmetric uncertainties, appropriate modifications are incorporated). Not all constraints are implemented through the usual $\chi^2$ form: it is not adequate to incorporate the bounds for perturbativity requirements of the Yukawa couplings in subsection \ref{sSEC:Constraints:fermion} and the bounds obtained in LHC searches in subsection \ref{sSEC:Constraints:LHC}. In those cases, instead of sharp bounds or cuts, the bounds are implemented as described in the respective subsections. Furthermore, since $\delta a_e$ and $\delta a_\mu$ have a role much more important than the rest of constraints, they are also incorporated in a different manner (as described in detail below) to ensure that the analysis focuses on the ability of the model to reproduce values which are clearly non-SM.

\subsection{$\delta a_\ell$ constraints for the numerical analyses\label{sSEC:Constraints:g2}}
\begin{figure}[h!tb]
\begin{center}
\includegraphics[width=0.3\textwidth]{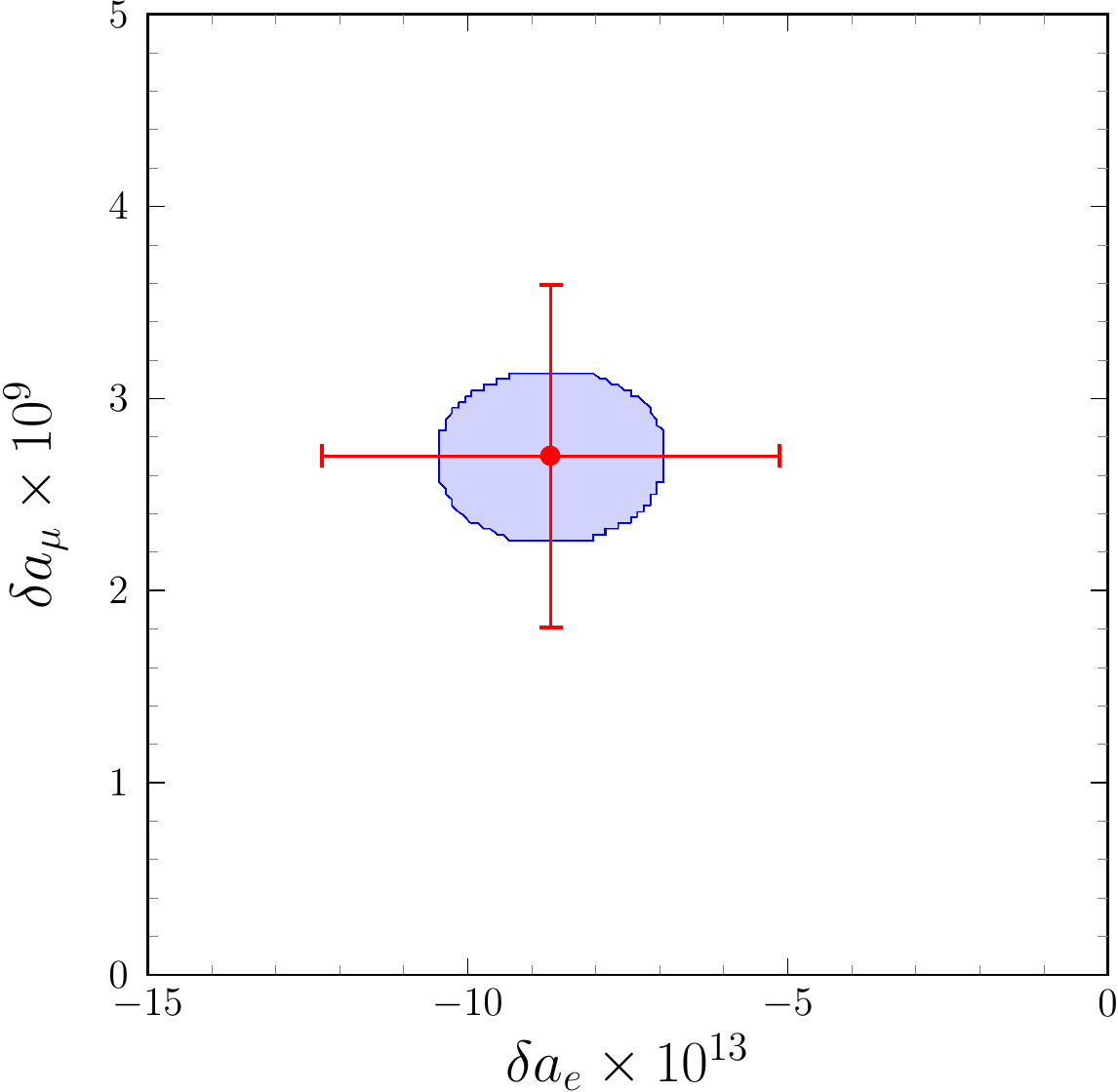}
\end{center}
\caption{Allowed $\delta a_\mu$ vs. $\delta a_e$ region.\label{fig:amu:ae}}
\end{figure}

The main motivation of this work is to accommodate the departures from SM expectations in the anomalous magnetic moments of both electron and muon. We now discuss how these departures are implemented as constraints in the analyses presented in section \ref{SEC:Results}. The $g-2$ anomalies $\delta a_\ell^{\rm Exp}=a_\ell^{\rm Exp}-a_\ell^{\rm SM}$ in \refeqs{eq:ae}--\eqref{eq:am} are
\begin{equation}\label{eq:g2l:anomalies}
\delta a_e^{\rm Exp}=-(8.7\pm 3.6)\times 10^{-13}\,,\quad \delta a_\mu^{\rm Exp}=(2.7\pm 0.9)\times 10^{-9}\,.
\end{equation}
The theoretical prediction in the present models is $a_\ell^{\rm Th}=\delta a_\ell+a_\ell^{\rm SM}$ and thus a simple and natural measure of their ability to accommodate the experimental results is a $\chi^2$ function
\begin{equation}\label{eq:chi2:g-2}
\chi^2_0(\delta a_e,\delta a_\mu)=\left(\frac{\delta a_e-c_e}{\sigma_e}\right)^2+\left(\frac{\delta a_\mu-c_\mu}{\sigma_\mu}\right)^2\,,
\end{equation}
where $\delta a_\ell^{\rm Exp}=c_\ell\pm \sigma_\ell$ in \refeq{eq:g2l:anomalies}.\\ 
The interest in explanations of the experimental results in terms of non-SM contributions is due to the $3-4\sigma$ deviation $\chi^2_0(0,0)\simeq 15$. For the numerical exploration of the regions in parameter space which could provide such an explanation, rather than including a contribution $\chi^2_0(\delta a_e,\delta a_\mu)$ in the global $\chi^2$, we impose a stronger requirement: instead of $\chi^2_0(\delta a_e,\delta a_\mu)$ we include 
\begin{equation}\label{eq:chi2:g-2:final}
\chi^2(\delta a_e,\delta a_\mu)= \left\{\begin{matrix} 0,& \text{if }\chi^2_0(\delta a_e,\delta a_\mu)\leq\frac{1}{4},\\ 10^6\times(\chi^2_0(\delta a_e,\delta a_\mu)-\frac{1}{4}),& \text{if } \chi^2_0(\delta a_e,\delta a_\mu)>\frac{1}{4},\end{matrix}\right.
\end{equation}
in order to guarantee that the models reproduce both anomalies simultaneously within less than $\frac{1}{2}\sigma_\ell$ of the central values (in the regions of interest, \refeq{eq:chi2:g-2:final} approximates a ``sharp'' box function). This approach is adopted in order to ensure that, when representing allowed regions at a given confidence level in the next section, they do not include regions where one or both anomalies are only partially reproduced. For illustration, Figure \ref{fig:amu:ae} shows the allowed region obtained in the complete numerical analyses (which is identical in both models); that is, in the results of section \ref{SEC:Results}, within all the represented allowed regions, the values of $\delta a_e$ and $\delta a_\mu$ belong to the allowed region of Figure \ref{fig:amu:ae}.
Notice, finally, that the SM prediction $a_\ell^{\rm SM}$ includes Higgs-mediated contributions: since these are just the $\nh$ mediated contributions for exact alignment $\sba=1$, they have to be subtracted from the New Physics contributions to $\delta a_\ell$ mediated by $\nh$ (quantitatively, however, this subtlety is rather irrelevant).

\subsection{Scalar sector\label{sSEC:Constraints:}}
\noindent For the scalar sector, we use the set of independent parameters $\{\vev{},\mh,\mH,\mA,\mcH,\tb,\alpha-\beta,\mu_{12}^2\}$ (from which the quartic parameters $\lambda_j$ are obtained) with $\vev{}=246$ GeV and $\mh=125$ GeV. We require the potential to be bounded from below following \cite{Ivanov:2015nea}, we also require the quartic parameters to respect perturbativity and perturbative unitarity in $2\to 2$ scattering \cite{Kanemura:1993hm,Akeroyd:2000wc,Ginzburg:2005dt,Horejsi:2005da} (see also \cite{Kanemura:2015ska,Grinstein:2015rtl,Nebot:2020niz}), and finally the corrections to the oblique parameters $S$ and $T$, which depend on the scalar masses and $\alpha-\beta$, have to be in agreement with electroweak precision data \cite{Grimus:2008nb}.

\subsection{Fermion sector\label{sSEC:Constraints:fermion}}
In the fermion sector, the new couplings $\nl{\ell}$ arise from the dimensionless Yukawa couplings $\matYukL{j}$ in \refeq{eq:LagrYuk}. If one required that Yukawa couplings remain perturbative, for example not exceeding $\mathcal O(1)$ values, this would translate into $\nl{\ell}$'s smaller than $\vev{}/\sqrt{2}\simeq 174$ GeV. We adopt a more conservative approach, and include a contribution of the following form to the global $\chi^2$ function driving the numerical analyses
\begin{equation}\label{eq:Pert:nl}
\chi^2_{\rm Pert}(\nl{\ell})=\left\{\begin{matrix}0,\text{ for }\abs{\nl{\ell}}\leq n_0,\\ \left(\frac{\abs{\nl{\ell}}-n_0}{\sigma_{n_0}}\right)^2,\text{ for }\abs{\nl{\ell}}> n_0.\end{matrix}\right.
\end{equation}
We choose $n_0=95$ GeV and $\sigma_{n_0}=1$ GeV. One could have adopted a crude requirement such as imposing for example $\abs{\nl{\ell}}\leq 100$ GeV with a sharp cut: \refeq{eq:Pert:nl} is simply a smooth version (more convenient for numerical purposes) of that kind of requirement. 
%
\subsection{Higgs signal strengths\label{sSEC:Constraints:Higgs}}
\noindent Concerning the 125 GeV Higgs-like scalar, agreement with the observed production $\times$ decay signal strengths of the usual channels is also imposed \cite{Khachatryan:2016vau,Aaboud:2017xsd,Sirunyan:2017elk,Sirunyan:2017khh}. The measured signal strengths, with uncertainties reaching the 10\% level, tend to favour the alignment limit in the scalar sector; it is to be noticed that since the models require $\abs{\nrle}\gg m_e$ and $\abs{\nrlm}\gg m_\mu$, the Higgs measurements in the $\mu^+\mu^-$ channel such as \cite{Khachatryan:2014aep} and \cite{Aaboud:2017ojs} are even more effective in forcing that alignment limit. Constraints on the total width $\Gamma_\nh$, arising from off-shell (ggF+VBF)$\to\nh^{(\ast)}\to WW^{(\ast)}$ \cite{Kauer:2012hd}, are also included \cite{Aad:2015xua,Khachatryan:2016ctc}, even if in the models considered here their effect is negligible in the alignment limit. For additional details, see \cite{Botella:2015hoa,Botella:2018gzy,Nebot:2018nqn}.
%
\subsection{$\cH$ mediated contributions\label{sSEC:Constraints:Charged}}
\noindent Flavour transitions mediated by $W^\pm$ can receive new contributions where $W^\pm\to \cH$. For tree level processes involving leptons, one refers to ``Lepton Flavour Universality'' constraints; we also consider constraints at the loop level in the quark sector.\\ 
One may also worry about too large $\cH$-mediated contributions to processes like $\ell_j\to\ell_k\gamma$: since in the present models we are considering massless neutrinos, lepton family numbers are conserved -- i.e. there is a $[U(1)]^3$ symmetry --, and such processes are absent.
%
\subsubsection{Lepton Flavour Universality\label{sSEC:Constraints:Charged:LFU}}
Contributions mediated by $\cH$ modify the leptonic decays $\ell_j\to\ell_k\nu\bar\nu$:
\begin{multline}
\Gamma(\ell_j\to\ell_k\nu\bar\nu)=\\ 
\frac{G_F^2}{192\pi^3}m_{\ell_j}^5\,\mathrm{f}(x_{kj})\left(1+\frac{1}{4}\Abs{g^{\rm S,RR}_{j\to k}}^2+2\re{g^{\rm S,RR}_{j\to k}}\frac{m_{\ell_k}}{m_{\ell_j}}\frac{\mathrm{g}(x_{kj})}{\mathrm{f}(x_{kj})}\right)\times (1+\Delta^{\ell_j\ell_k}_{\text{RC}}),
\end{multline}
where $\mathrm{f}(x)$ and $\mathrm{g}(x)$ are the usual phase space integrals\footnote{$\mathrm{f}(x)=1-8x+8x^3-x^4-12x^2\ln x$, and $\mathrm{g}(x)=1+9x-9x^2-x^3+6x(1+x)\ln x$.} \cite{Kuno:1999jp}, $x_{kj}\equiv (m_{\ell_k}/m_{\ell_j})^2$) and $\Delta^{\ell_j\ell_k}_{\text{RC}}$ correspond to QED radiative corrections and most importantly,
\begin{equation}
g^{\rm S,RR}_{j\to k}=-\frac{n_{\ell_j}^\ast n_{\ell_k}}{\mcH^2}\,.
\end{equation}
The notation $g^{\rm S,RR}_{j\to k}$ reflects the fact that in the present models the new contributions only affect, in an effective description, the operator  $\bar\nu_L\ell_{jR}\,\bar\ell_{kR}\nu_L$. The ratios
\begin{equation}
R{\scriptstyle\left[\frac{\ell_a\to\ell_b}{\ell_\alpha\to \ell_\beta}\right]}\equiv \frac{\Gamma(\ell_a\to\ell_b\nu\bar\nu)}{\Gamma(\ell_a\to\ell_b\nu\bar\nu)_{\rm SM}}\frac{\Gamma(\ell_\alpha\to\ell_\beta\nu\bar\nu)_{\rm SM}}{\Gamma(\ell_\alpha\to\ell_\beta\nu\bar\nu)}
\end{equation}
give  the following constraints \cite{Tanabashi:2018oca}:
\begin{equation}\label{eq:LFU:lep:01}
R{\scriptstyle\left[\frac{\tau\to\mu}{\tau\to e}\right]}=1+(3.8\pm 3.2)\times 10^{-3}\,,\qquad
R{\scriptstyle\left[\frac{\tau\to e}{\mu\to e}\right]}=1+(2.4\pm 2.3)\times 10^{-3}\,.
\end{equation}
In addition, measurements of decay spectra with polarized leptons impose
\begin{equation}\label{eq:LFU:lep:02}
\Abs{g^{\rm S,RR}_{\mu\to e}}<0.035\text{ at }90\%\text{ CL},\quad
\Abs{g^{\rm S,RR}_{\tau\to\mu}}<0.72\text{ at }95\%\text{ CL},\quad 
\Abs{g^{\rm S,RR}_{\tau\to e}}<0.7\text{ at }95\%\text{ CL}.
\end{equation}
Besides purely leptonic decays $\ell_j\to\ell_k\bar\nu\nu$, leptonic decay modes like $K,\pi\to e\nu,\mu\nu$ and $\tau\to K\nu,\pi\nu$, provide additional constraints on the different $n_\ell$ (together with the $\tb$ dependence of the quark couplings with $\cH$). In particular, we consider ratios
\begin{equation}
R^P_{\ell_1\ell_2}=\frac{\Gamma(P^+\to\ell_1^+\nu)}{\Gamma(P^+\to\ell_1^+\nu)_{\text{\sc SM}}}\frac{\Gamma(P^+\to\ell_2^+\nu)_{\text{\sc SM}}}{\Gamma(P^+\to\ell_2^+\nu)}=\frac{\Abs{1-\Delta^P_{\ell_1}}^2}{\Abs{1-\Delta^P_{\ell_2}}^2},
\end{equation}
where the quark content of $P^+$ is $u_i\bar d_j$ and
\begin{equation}
\Abs{1-\Delta^P_{\ell_a}}^2 \equiv \Abs{1-\frac{M_P^2}{M_{H^\pm}^2}\frac{k_{u}m_{u_i}+k_{d}^\ast m_{d_j}}{m_{u_i}+m_{d_j}}\frac{n_{\ell_a}}{m_{\ell_a}}}^2 ,
\end{equation}
with $k_u=k_d=\tbinv$ in model \glFC{I} and $k_u=-k_d^{-1}=\tbinv$ in model \glFC{II}. Notice the enhanced sensitivity of these observables due to the $\frac{n_{\ell_a}}{m_{\ell_a}}$ factor: unlike the SM amplitude, the new $\cH$-mediated amplitude is not helicity suppressed. For ratios involving $\tau^+\to P^+\nu$ decays, the expressions are unchanged. The actual constraints \cite{Tanabashi:2018oca,Cirigliano:2007xi,Pich:2013lsa} read
\begin{equation}
\begin{aligned}
&R^\pi_{\mu e}=1+(4.1\pm 3.3)\times 10^{-3},\quad 
&R^\pi_{\tau \mu}=1-(5.9\pm 5.9)\times 10^{-3},\\
&R^K_{\mu e}=1-(4.8\pm 4.7)\times 10^{-3},\quad 
&R^K_{\tau \mu}=1-(2.2\pm 1.4)\times 10^{-2}.
\end{aligned}
\end{equation}
All these LFU violating effects scale with $1/\mcH^2$ and therefore one expects that in both models, \glFC{I} and \glFC{II}, the effects for large $\mcH$ are much more suppressed, including in particular the solution \solA\ region introduced in section \ref{SEC:g-2}. This is quite clear in the pure leptonic decays, where the most relevant constraints, \refeq{eq:LFU:lep:01} and $\Abs{g^{\rm S,RR}_{\mu\to e}}$ in \refeq{eq:LFU:lep:02}, can be comfortably satisfied, giving a contribution to the corresponding $\chi^2$ at a level similar to the SM. Since solution \solA\ corresponds to $\tb\sim 1$, the effects in semileptonic processes are similar in both models, with the effects in kaons larger by a factor of 10 than the effects in pions. The leading contribution to $R^K_{\mu e}-1$ is of the order of the uncertainty: since in that channel there is essentially a change of sign between the contributions in models \glFC{I} and \glFC{II}, it turns out that in the \glFC{II} case the corresponding $\chi^2$ value can improve over the SM one, while in the \glFC{I} case it is the other way around. In any case, for solution \solA, these differences are small. For solution \solB, the situation is different since we have:
\begin{align}
&\text{solution \solB, model \glFC{I},}\qquad \Delta_\ell^K\sim \frac{M_K^2}{m_{\ell}}\frac{n_{\ell}}{M_{H^\pm}^2\,\tb},\\
&\text{solution \solB, model \glFC{II},}\qquad \Delta_\ell^K\sim -\frac{M_K^2}{m_{\ell}}\frac{n_{\ell}\,\tb}{M_{H^\pm}^2},
\end{align}
considering that it requires $\tb\gg 1$ and smaller $\mcH$. Clearly, lower values of $\mcH$ can be compensated by large values of $\tb$ in model \glFC{I} , and solution \solB\ is similar to \solA\ concerning this constraint. On the contrary, in model \glFC{II}, lower values of $\mcH$ and larger values of $\tb$ enhance the new contributions: this observable is highly relevant to eliminate solution \solB\ in model \glFC{II}.\\
The new scalars can also give one loop corrections to $Z\to\ell^+\ell^-$ decays. In the parameter space region corresponding to solution \solA, one can easily check that these new contributions are at least a factor of 30 smaller than the experimental uncertainties (in the limit $\mA=\mH=\mcH\gg M_Z$ they decouple, see \cite{Chun:2016hzs}); in the parameter space of solution \solB, the new contributions are larger, but still below uncertainties.
%
\subsubsection{$b\to s\gamma$ and $B_q^0$--$\bar B_q^0$\label{sSEC:Constraints:Charged:Quarks}}
As loop level transitions mediated by the charged scalar, we consider contributions to the mixing in $B_d$ and $B_s$ meson systems (in particular to the dispersive part of the mixing, which controls the mass differences) and contributions to the radiative decay $b\to s\gamma$. In both cases, concerning the dependence on CKM factors of the new contributions involving $\cH$, it is clear from Table \ref{TAB:chargedcouplings} that they are analog to their SM counterparts; this implies, for example, that there is no need to worry about new contributions to CP asymmetries in $B_d\to J/\Psi K_S$ or $B_s\to J/\Psi\Phi$. Contributions to the mentioned mass differences in $B_d$ and $B_s$ are required to not exceed the 2-3\% level (that is already below the current level of theoretical uncertainty in the relevant matrix elements obtained from lattice QCD computations). For $b\to s\gamma$, we impose that the correction to the usual $\Gamma(B\to X_s\gamma)_{E_\gamma>1.6\text{GeV}}$ is below the experimental uncertainty. Both observables are insensitive to scalar-lepton couplings, they can only constrain $m_{\cH}$ and $\tb$. For  $m_{\cH}$ the effect is straightforward: for large values of $m_{\cH}$, the new contributions are suppressed. Concerning $\tb$, dominant new contributions with virtual top quarks are further enhanced or suppressed by the $\tbinv$ dependence in Table \ref{TAB:neutralcouplings}: altogether, one expects that these two constraints tend to disfavour $\tb\ll 1$ and light $\cH$. We refer to \cite{Misiak:2006zs,Crivellin:2013wna,Botella:2014ska} for further details.
%
\subsection{$e^+e^-\to \mu^+\mu^-,\tau^+\tau^-$ at LEP \label{sSEC:Constraints:LEP}}
\noindent LEP measured $e^+e^-\to \mu^+\mu^-,\tau^+\tau^-$ with center-of-mass energies up to $\sqrt{s}=208$ GeV: although $s$-channel contributions with virtual $\nH$ and $\nA$ do not interfere with SM $\gamma$ and $Z$ mediated contributions, for light $\nH$, $\nA$, the resonant enhancement together with the large couplings to leptons might give predictions in conflict with data (e.g. \cite{Schael:2006wu}). The effect of these LEP constraints is, essentially, to forbid values of $m_\nH$, $m_\nA$ below $210-215$ GeV.
%
\subsection{LHC searches\label{sSEC:Constraints:LHC}}
\noindent We consider constraints from LHC searches of scalars, in particular
\begin{itemize}
\item searches of dilepton resonances \cite{Aaboud:2017buh,Aaboud:2019sgt,Sirunyan:2019tkw,Aaboud:2017sjh,Khachatryan:2016qkc,Sirunyan:2018zut} which give constraints on $\sigma(pp\to S)_{[\rm ggF]}\times\BR{S\to \ell^+\ell^-}$, $S=\nH,\nA$ and $\ell=\mu,\tau$, where the production cross section $\sigma(pp\to S)_{[{\rm ggF}]}$ corresponds to gluon-gluon fusion,
\item and searches of charged scalars \cite{Aaboud:2016dig,Aaboud:2018cwk,Sirunyan:2019hkq,Sirunyan:2019arl,Sirunyan:2020hwv} which give constraints on $\sigma(pp\to\cH tb)\times\BR{\cH\to f}$, $f=\tau\nu,tb$.
\end{itemize}
For production, the narrow width approximation (NWA) is considered; the widths of $\nH$, $\nA$ and $\cH$ can reach $\sim 10\%$ of their respective masses: if one incorporates finite width effects through the convolution of the cross section computed in the NWA with a (relativistic) Breit-Wigner distribution for the scalars, the computed signal would be partially ``diluted''. In this sense, using the NWA is conservative since it gives stronger pointwise bounds. The constraints are incorporated as contributions of the following form in the global $\chi^2$: for each ``production $\times$ decay'' channel with experimental bound $[\sigma\times \text{Br}]_{\rm Exp}$ and theoretical prediction $[\sigma\times \text{Br}]_{\rm Th}$, the contribution is given by
\begin{equation}\label{eq:chi2:LHC:final}
\chi^2([\sigma\times \text{Br}]_{\rm Th})= \left\{\begin{matrix} 0,& \text{if }[\sigma\times \text{Br}]_{\rm Th}\leq 0.9\times[\sigma\times \text{Br}]_{\rm Exp},\\ 10^3\times\left(\frac{[\sigma\times \text{Br}]_{\rm Th}}{[\sigma\times \text{Br}]_{\rm Exp}}-0.9\right),& \text{if } [\sigma\times \text{Br}]_{\rm Th}> 0.9\times[\sigma\times \text{Br}]_{\rm Exp}.\end{matrix}\right.
\end{equation}
Equation \eqref{eq:chi2:LHC:final} is a convenient smooth approximation of a ``sharp'' bound/cut.\\ 
Production cross sections incorporate corrections associated to the modified fermion-scalar vertices in the following manner. For generic interaction terms
\begin{equation}\label{eq:Sqq}
\mathscr L_{S\bar qq}=-\frac{m_t}{\vev{}}S\bar t(a_t^S+ib_t^S\gamma_5)t-\frac{m_b}{\vev{}}S\bar b(a_b^S+ib_b^S\gamma_5)b\,,
\end{equation}
the gluon-gluon fusion production cross section reads
\begin{multline}
\sigma[pp\to S]_{\rm ggF}=\\
\sigma[pp\to S]_{[\rm ggF]}^{\text{SM-like}}\times
\frac{\abs{a_t^Sm_tF(x_t)+a_b^Sm_bF(x_b)}^2+\abs{b_t^Sm_t\hat F(x_t)+b_b^Sm_b\hat F(x_b)}^2}{\abs{m_t\,F(x_t)+m_b\,F(x_b)}^2},
\end{multline}
with $x_q\equiv (m_q/m_S)^2$, and $F(x)$ and $\hat F(x)$ the loop functions corresponding to scalar or pseudoscalar couplings, respectively; $\sigma[pp\to S]_{[\rm ggF]}^{\text{SM-like}}$ can be found in \cite{deFlorian:2016spz,Harlander:2002wh,Ravindran:2003um,Pak:2011hs}. This simple recipe also gives sufficiently good agreement with results for a SM-Higgs-like neutral pseudoscalar, which can be found in \cite{Harlander:2002vv,Anastasiou:2002wq,Pak:2011hs,Ahmed:2016otz}. The couplings in \refeq{eq:Sqq} for $S=\nH,\nA$ in each model can be read in Table \ref{TAB:neutralcouplings}.\\
Similarly, for the production cross sections $pp\to H^\pm tb$ (i.e. $\cH$ in association with $tb$), we refer to \cite{Flechl:2014wfa,Degrande:2015vpa}, which provide results, labeled here $\sigma_{[\text{Ref}]}$, for a type II 2HDM with $\tb=1$. For arbitrary values of $\tb$, we use
\begin{equation}
\begin{aligned}
&\text{Model \glFC{I}: }\sigma_{\text{I}}(t_\beta)=\frac{(m_t/t_\beta)^2+(m_b/t_\beta)^2}{m_t^2+m_b^2}\times \sigma_{[\text{Ref}]}=\frac{1}{t_\beta^2}\sigma_{[\text{Ref}]},\\
&\text{Model \glFC{II}: }\sigma_{\text{II}}(t_\beta)=\frac{(m_t/t_\beta)^2+(m_b t_\beta)^2}{m_t^2+m_b^2}\times\sigma_{[\text{Ref}]}=\frac{1}{t_\beta^2}\frac{1+t_\beta^4 m_b^2/m_t^2}{1+m_b^2/m_t^2}\sigma_{[\text{Ref}]}.
\end{aligned}
\end{equation}
As an additional check, (i) the previous cross sections and (ii) the computations of the decay branching ratios of the scalars, have been compared with the results of {\sc\small MadGraph5\_aMC@NLO}\cite{Alwall:2014hca} at leading order. With {\sc\small FeynRules} \cite{Alloul:2013bka} and NLOCT \cite{Degrande:2014vpa,Degrande:2014qga}, the needed universal Feynrules Output at NLO of the \glFC{I} and \glFC{II} models is produced. A good agreement in the gluon-gluon fusion production cross section is found, given the fact that the {\sc\small MadGraph5\_aMC@NLO} calculation is at leading order (one loop in this case). For the branching ratios, there is complete agreement.
%

\section{Results\label{SEC:Results}}
\begin{figure}[!htb]
\begin{center}
\includegraphics[width=0.35\textwidth]{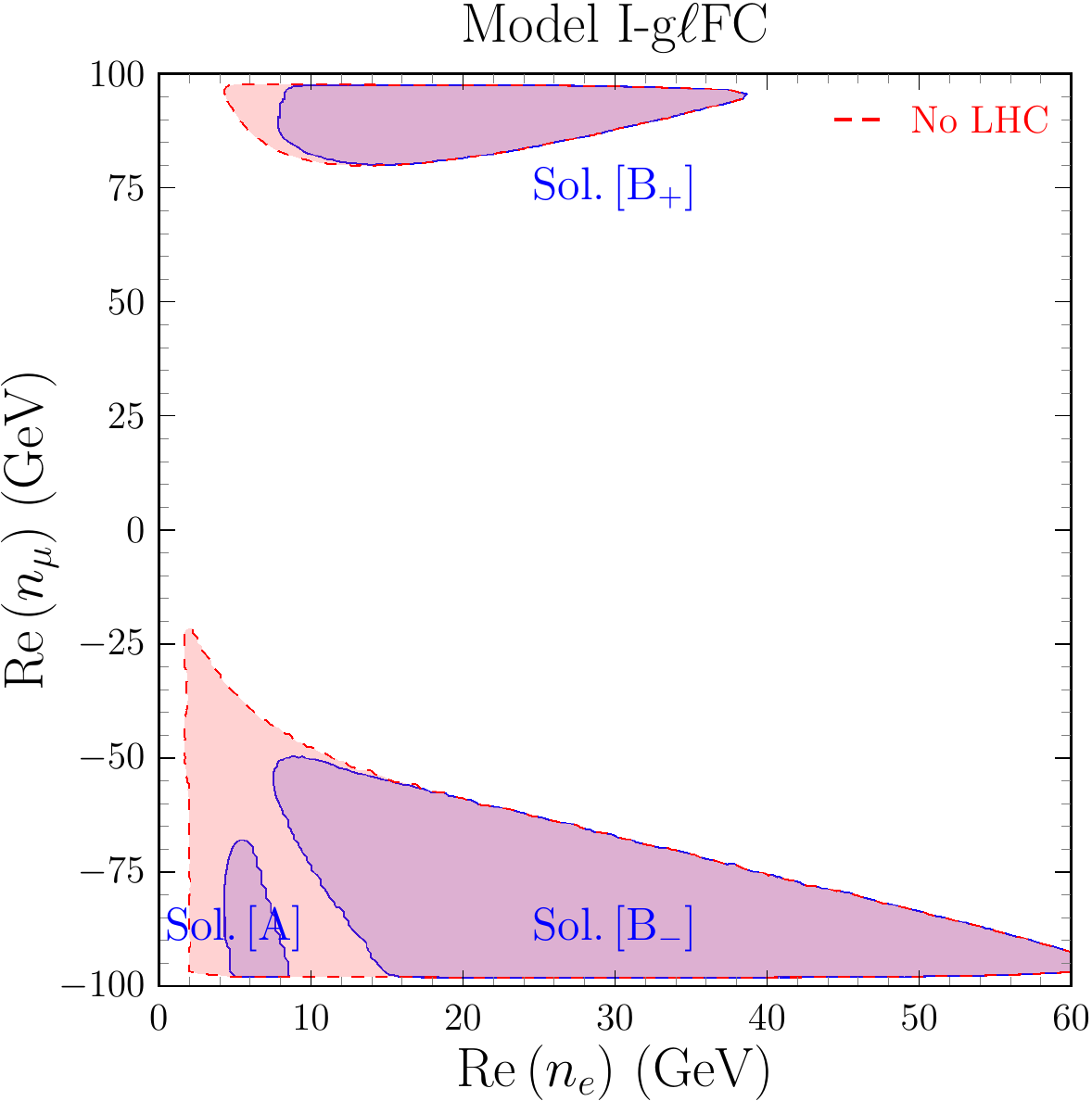}
\end{center}
\caption{$\nrlm$ vs. $\nrle$.\label{fig:typeI:nmu_vs_ne}}
\end{figure}

\begin{figure}[!htb]
\begin{center}
\subfloat[\label{fig:couplings:typeI:SBS:ne:tb}]{\includegraphics[width=0.25\textwidth]{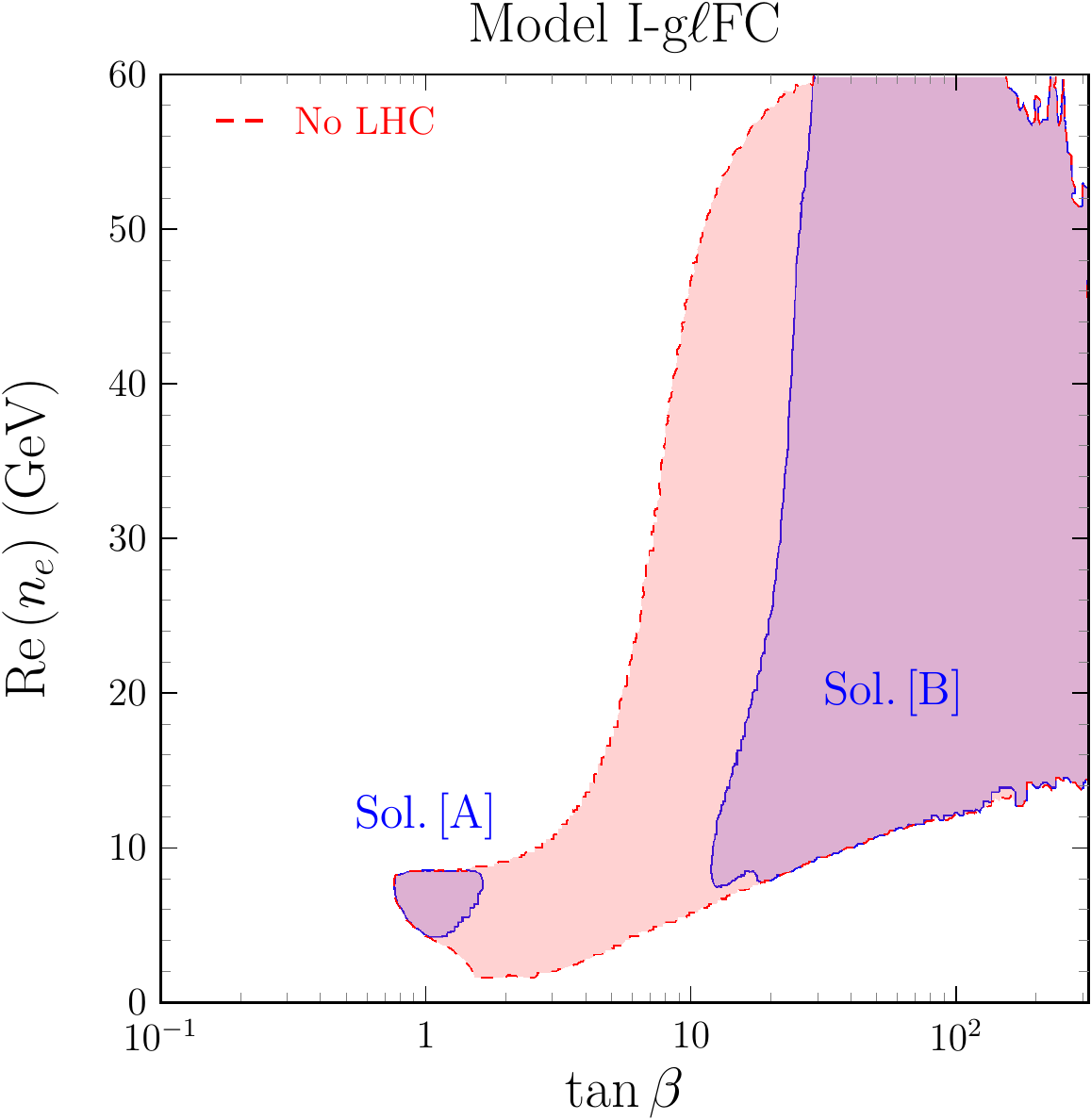}}\quad
\subfloat[\label{fig:couplings:typeI:SBS:nm:tb}]{\includegraphics[width=0.25\textwidth]{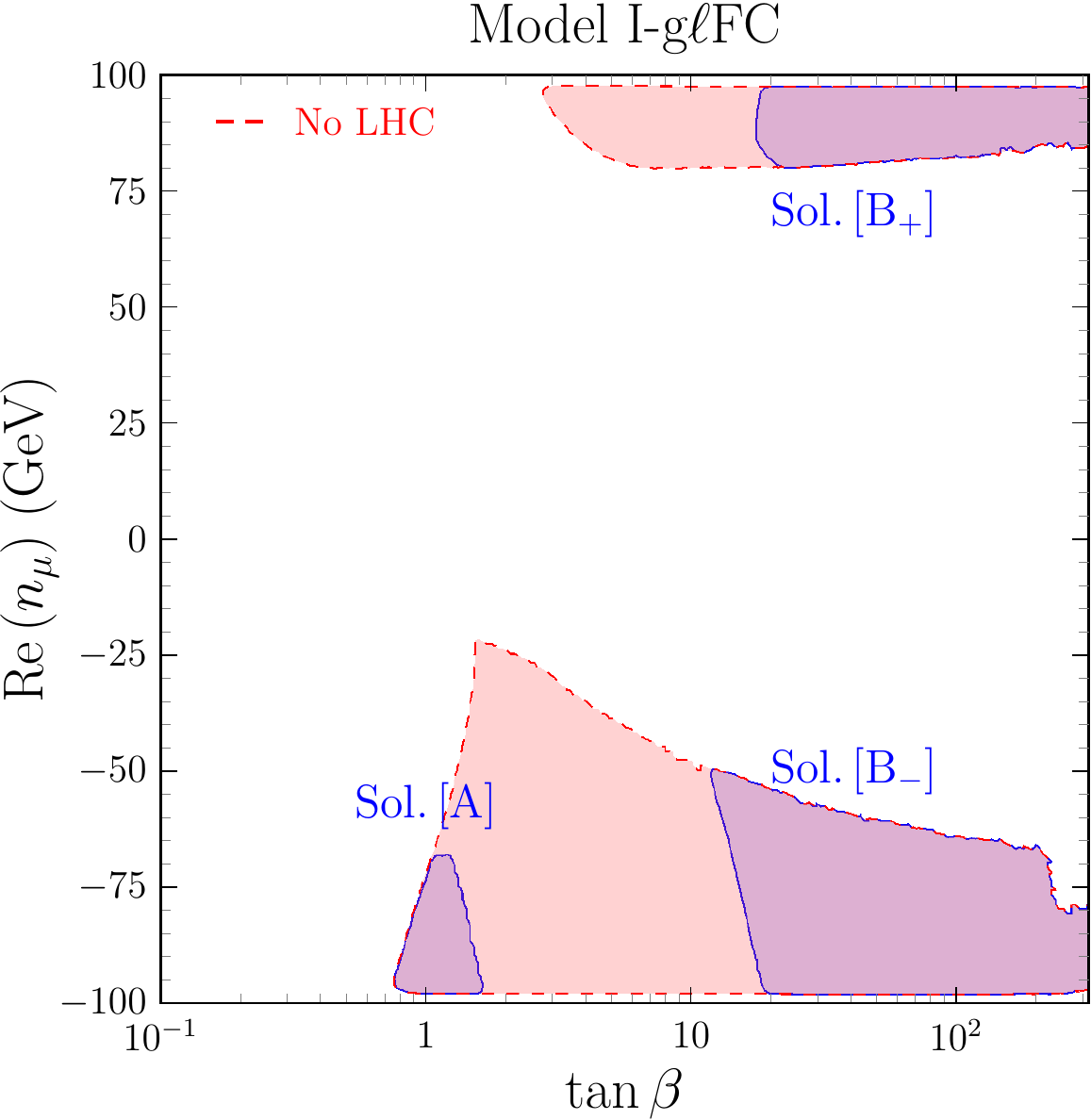}}\quad
\subfloat[\label{fig:couplings:typeI:SBS:nt:tb}]{\includegraphics[width=0.25\textwidth]{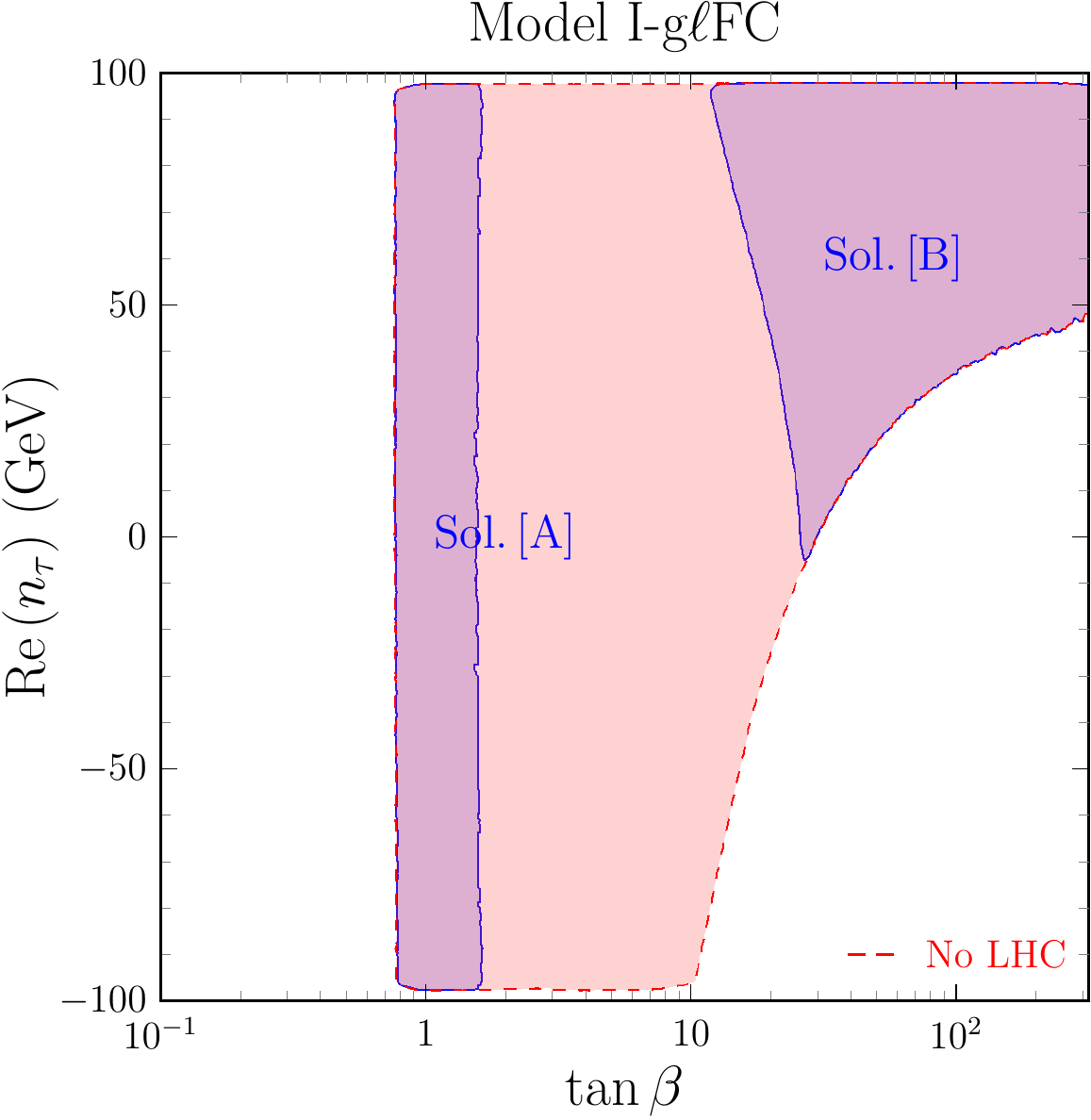}}\\
\subfloat[\label{fig:couplings:typeI:SBS:ne:mH}]{\includegraphics[width=0.25\textwidth]{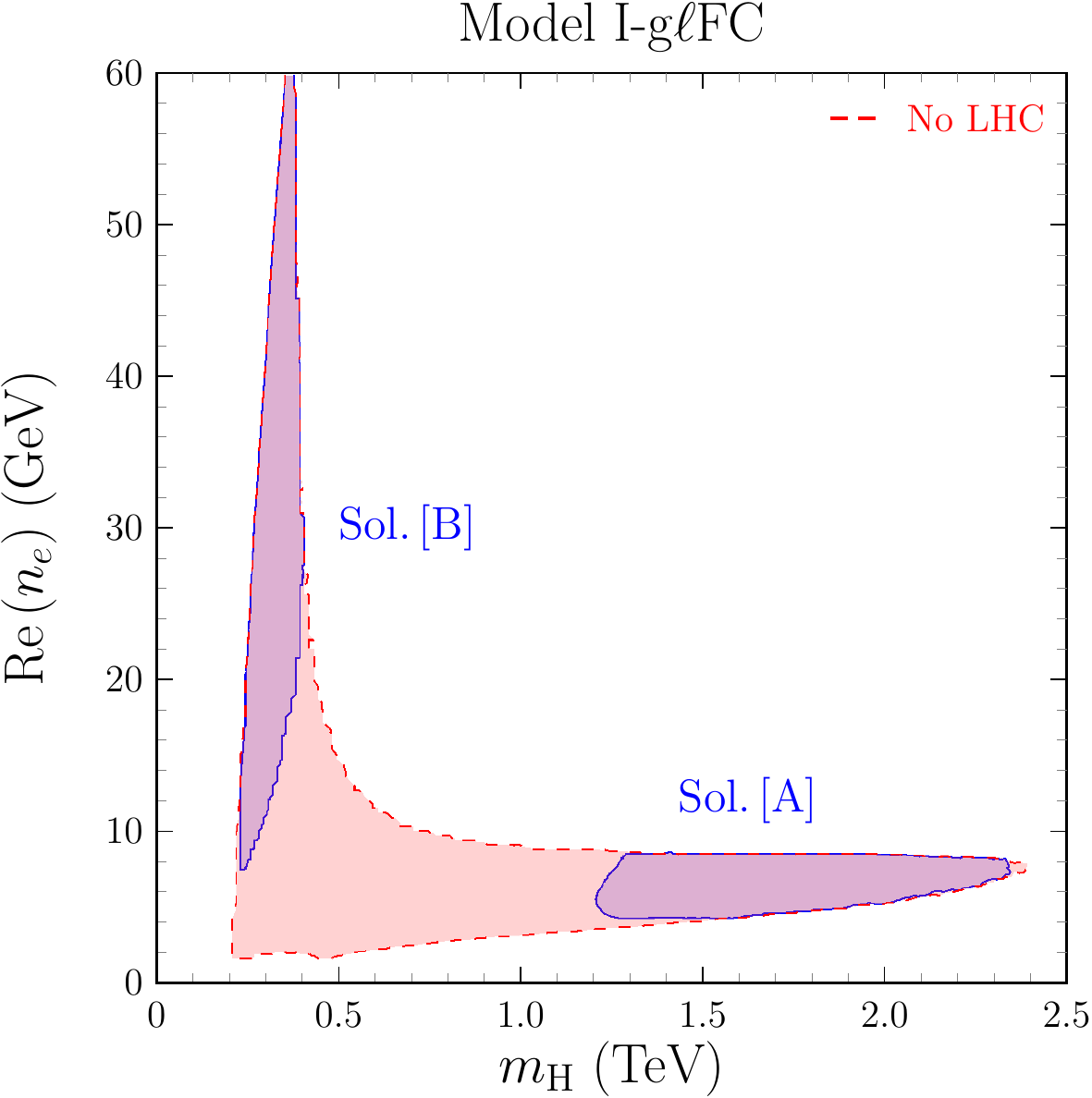}}\quad
\subfloat[\label{fig:couplings:typeI:SBS:nm:mH}]{\includegraphics[width=0.25\textwidth]{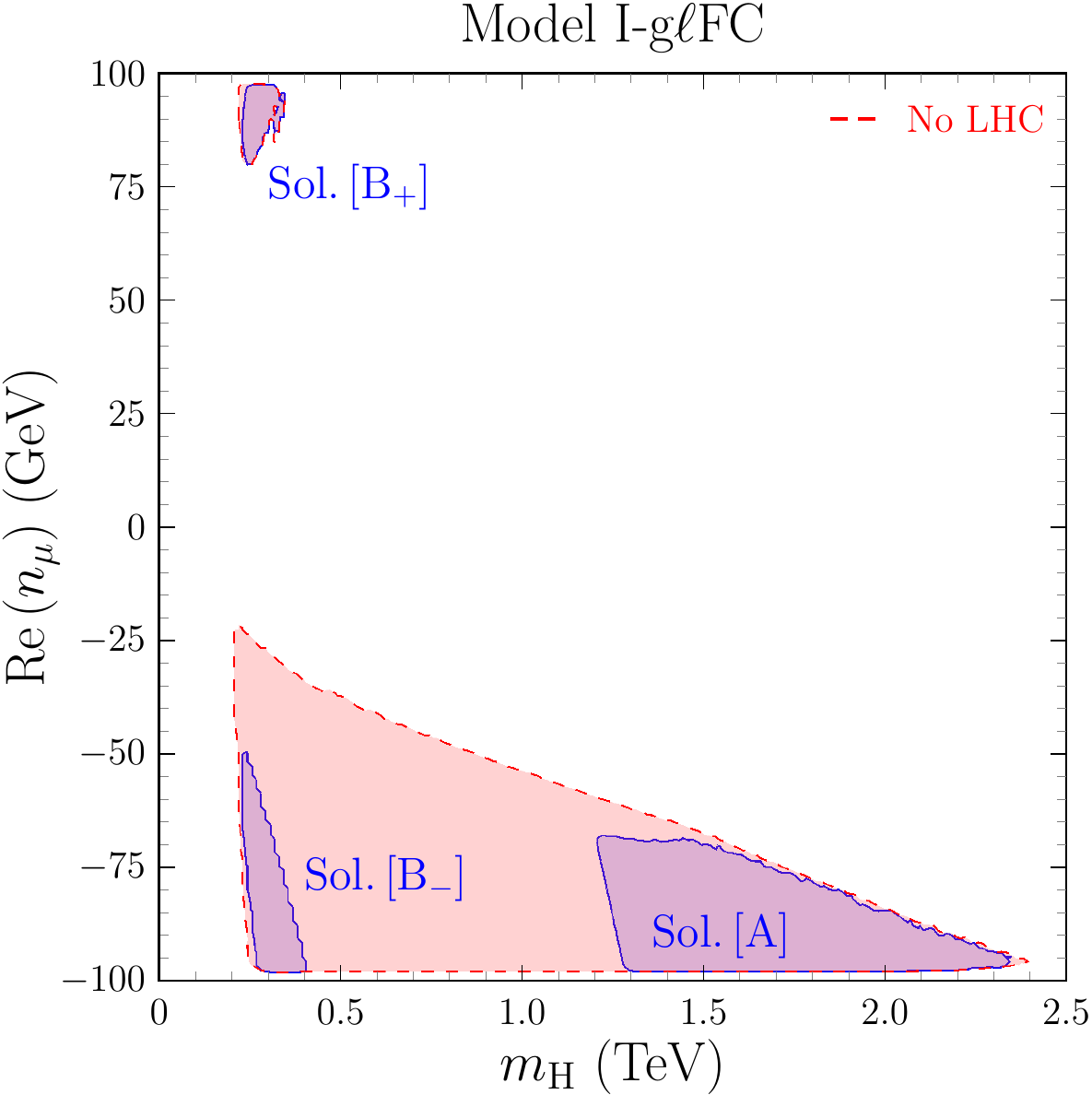}}\quad
\subfloat[\label{fig:couplings:typeI:SBS:nt:mH}]{\includegraphics[width=0.25\textwidth]{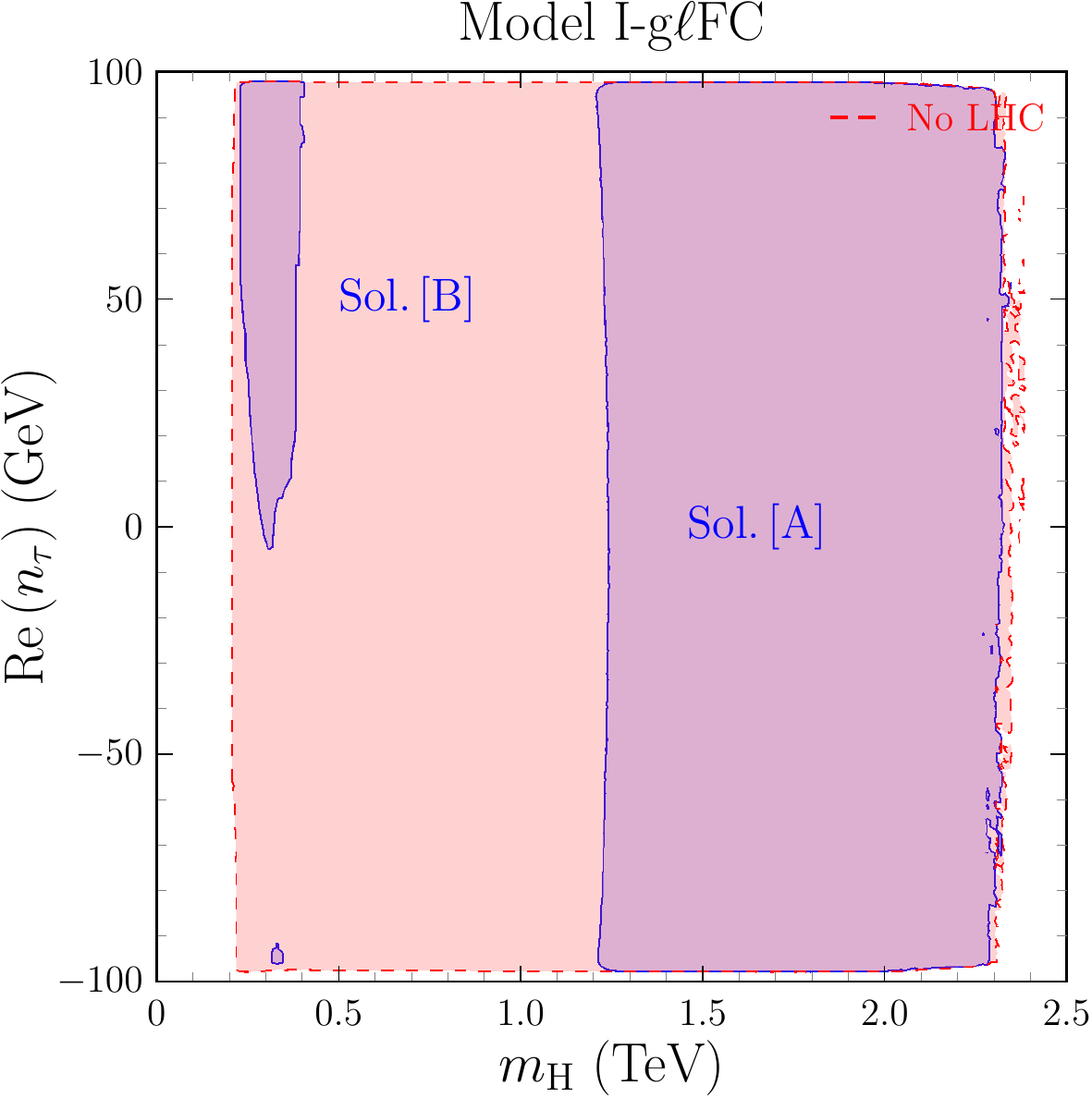}}
\end{center}
\caption{$\nl{\ell}$ couplings versus $\tb$, $\mH$.\label{fig:couplings:typeI:SBS}}
\end{figure}

\begin{figure}[!htb]
\begin{center}
\subfloat[\label{fig:scalar:typeI:SBS:mH:mCh}]{\includegraphics[width=0.22\textwidth]{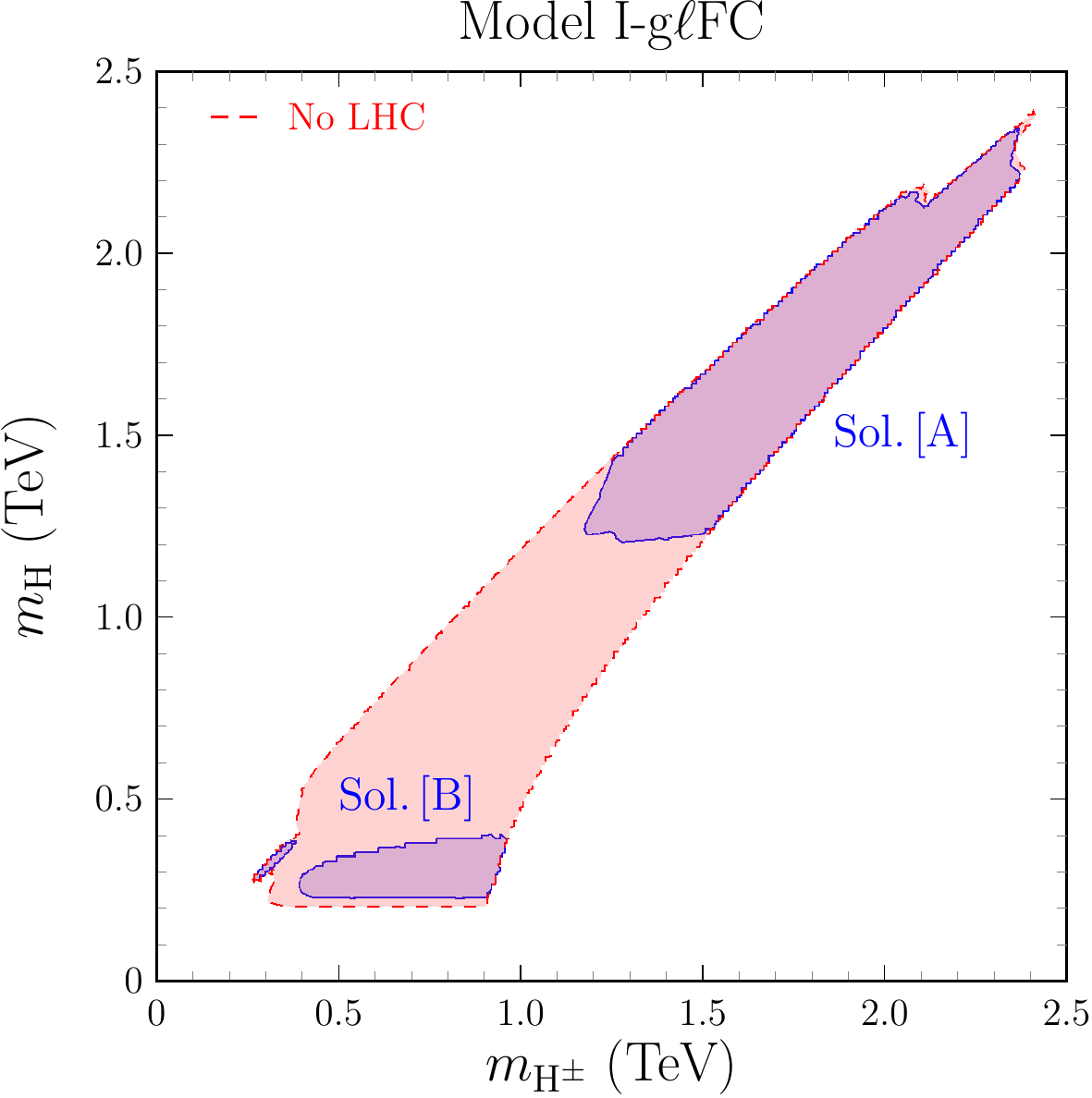}}\quad
\subfloat[\label{fig:scalar:typeI:SBS:mA:mCh}]{\includegraphics[width=0.22\textwidth]{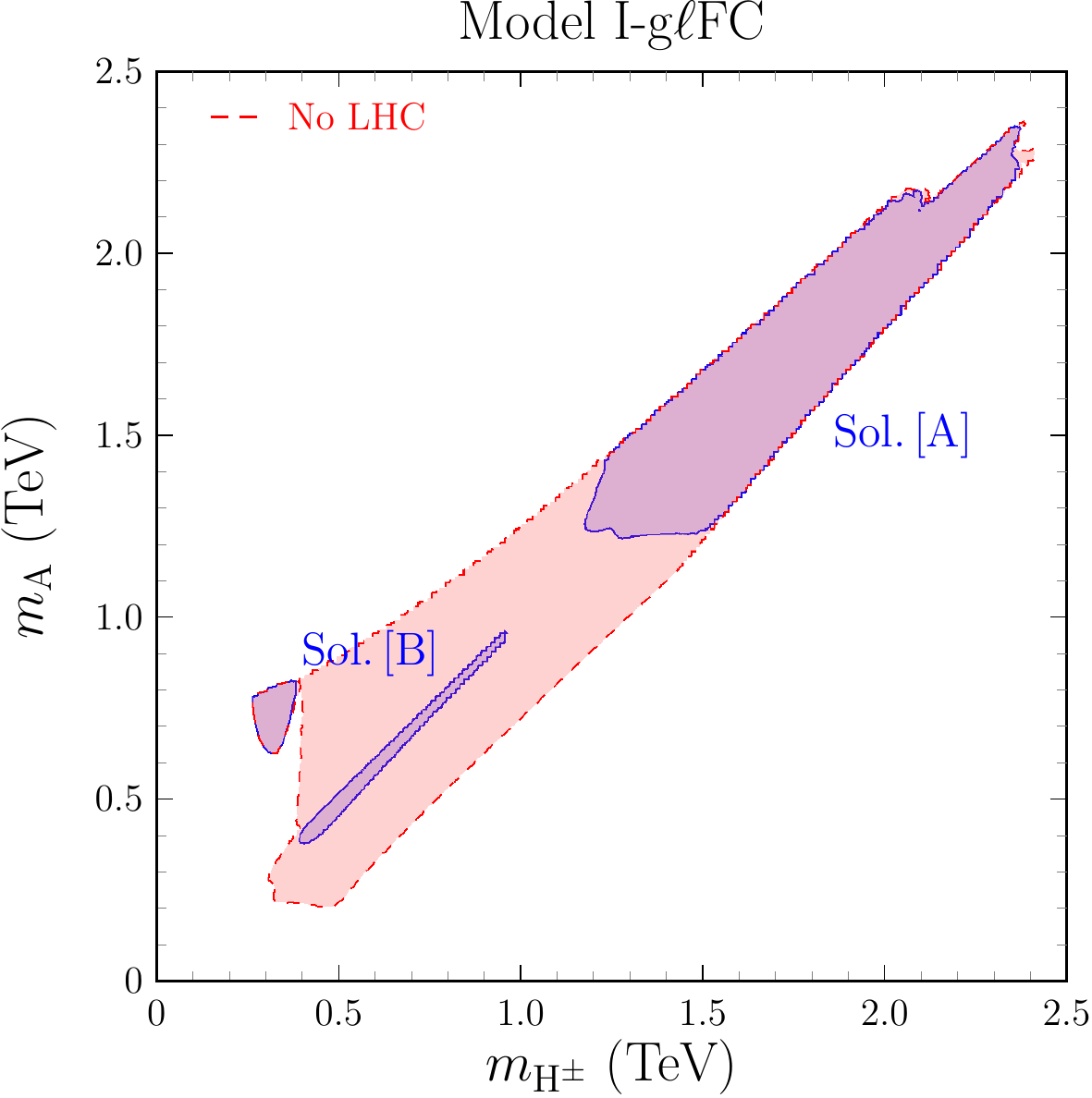}}\quad
\subfloat[\label{fig:scalar:typeI:SBS:mA:mH}]{\includegraphics[width=0.22\textwidth]{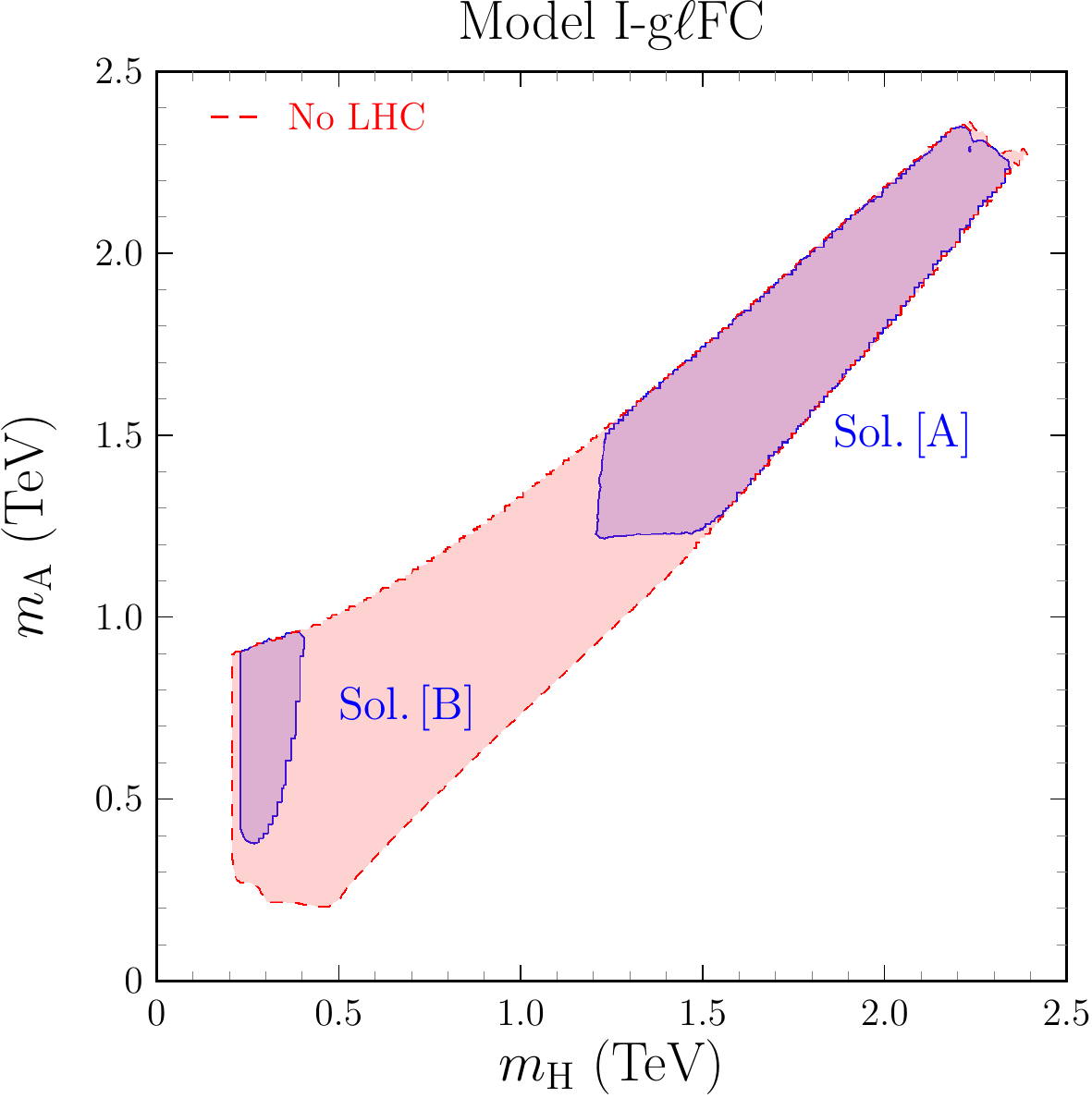}}\\
\subfloat[\label{fig:scalar:typeI:SBS:mCh:tb}]{\includegraphics[width=0.22\textwidth]{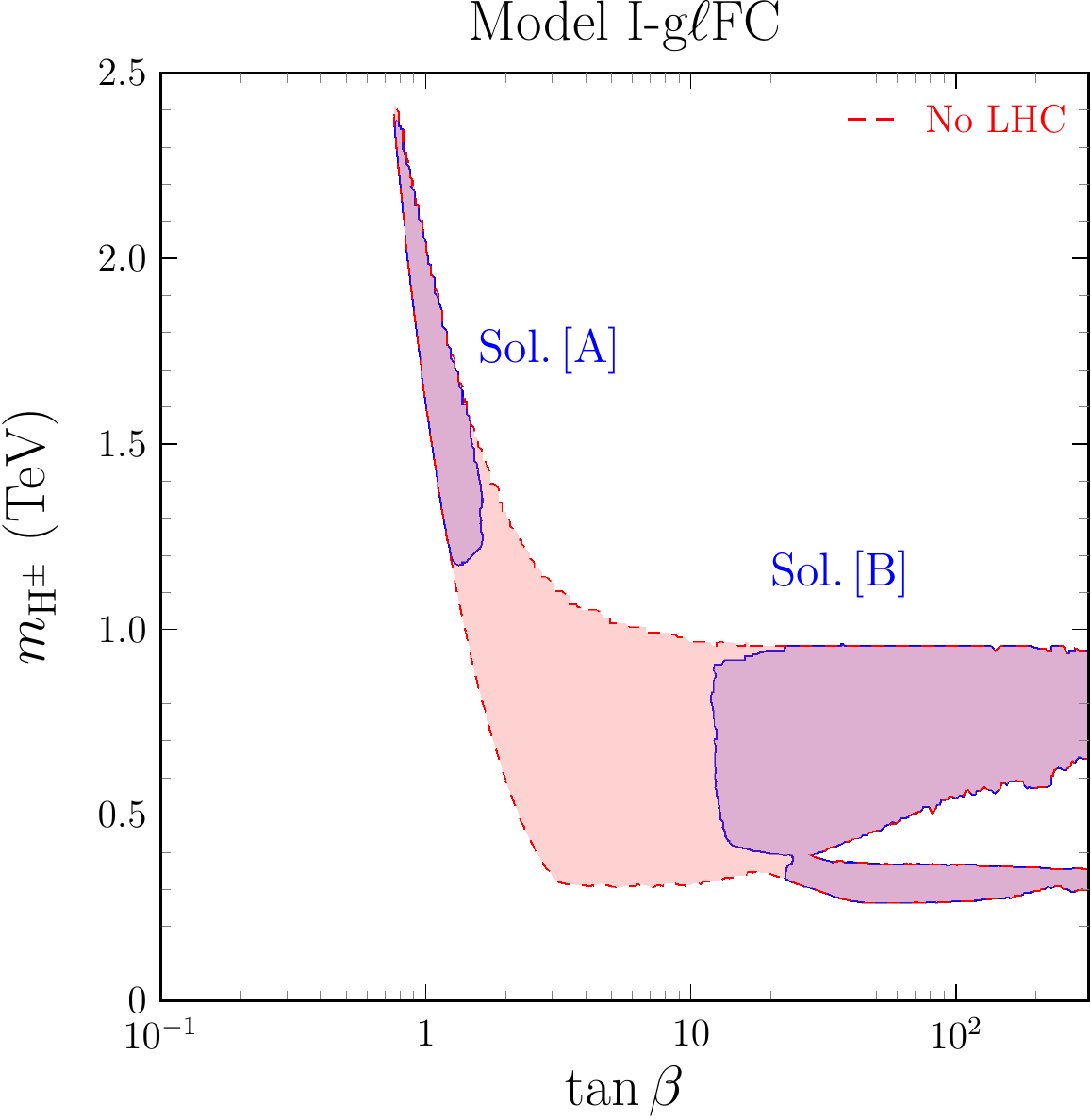}}\quad
\subfloat[\label{fig:scalar:typeI:SBS:mH:tb}]{\includegraphics[width=0.22\textwidth]{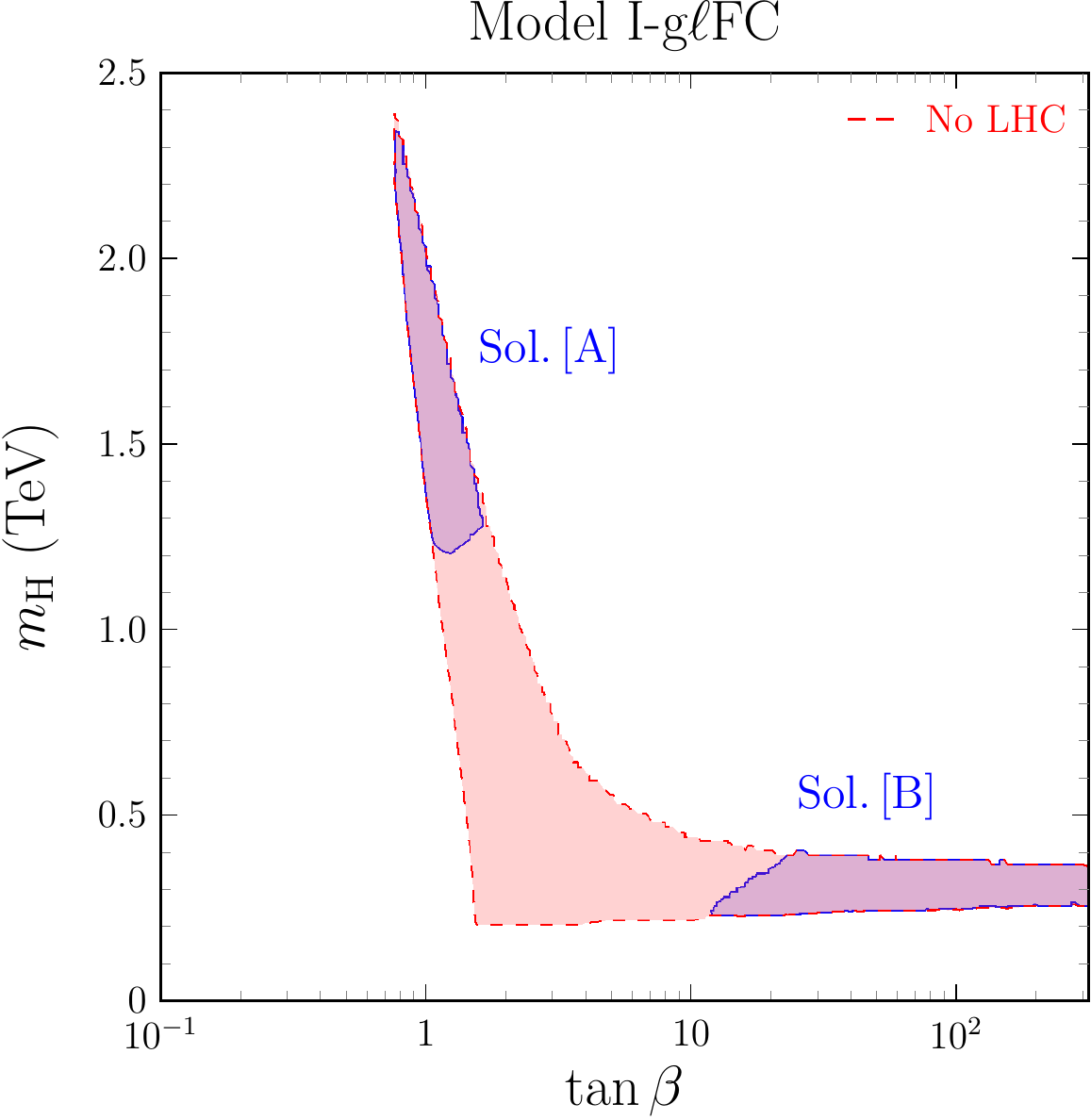}}
\end{center}
\caption{Scalar sector.\label{fig:scalar:typeI:SBS}}
\end{figure}

\begin{figure}[!htb]
\begin{center}
\subfloat[\label{fig:typeI:ppAlmlm-vs-MA}]{\includegraphics[width=0.25\textwidth]{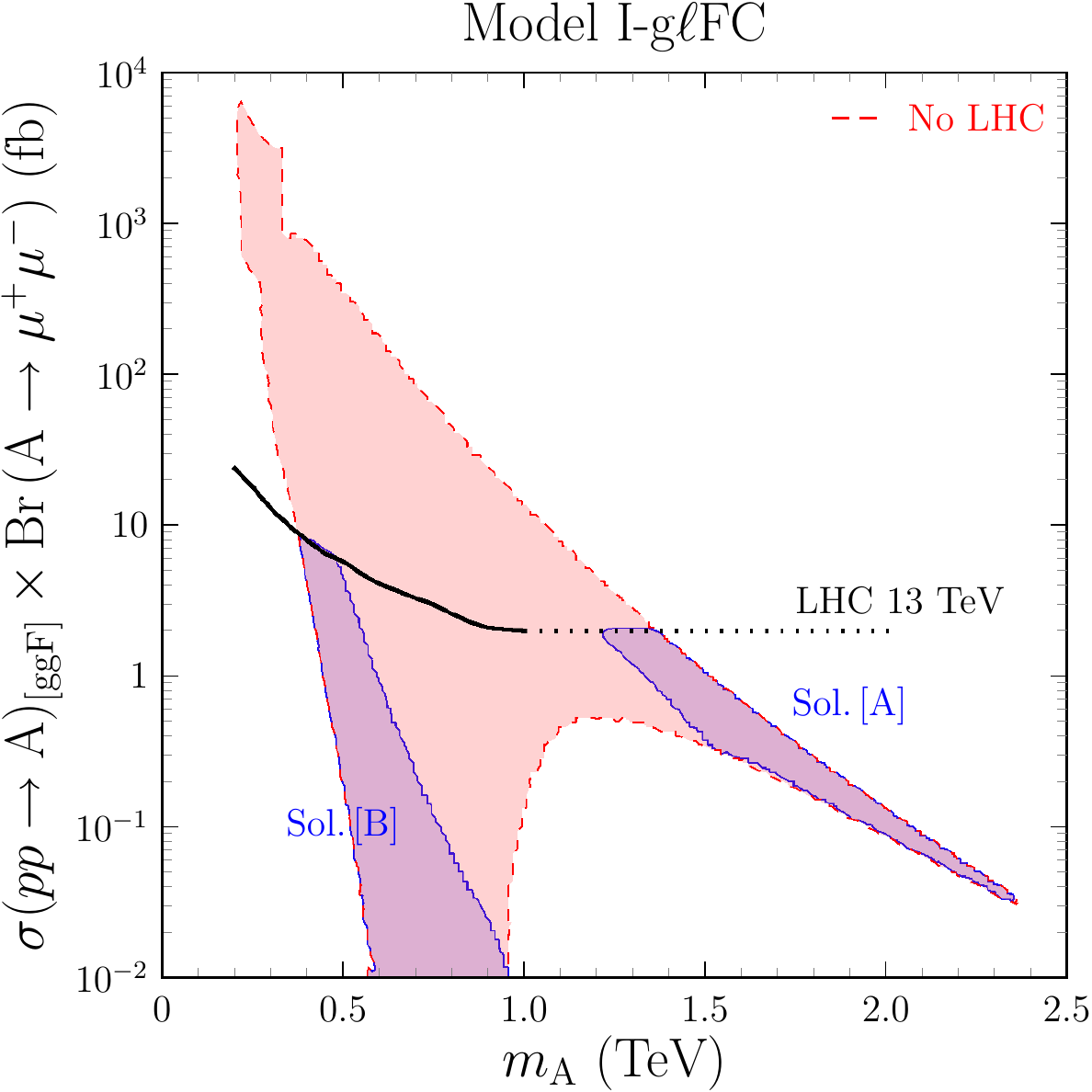}}\quad
\subfloat[\label{fig:typeI:ppHlmlm-vs-MH}]{\includegraphics[width=0.25\textwidth]{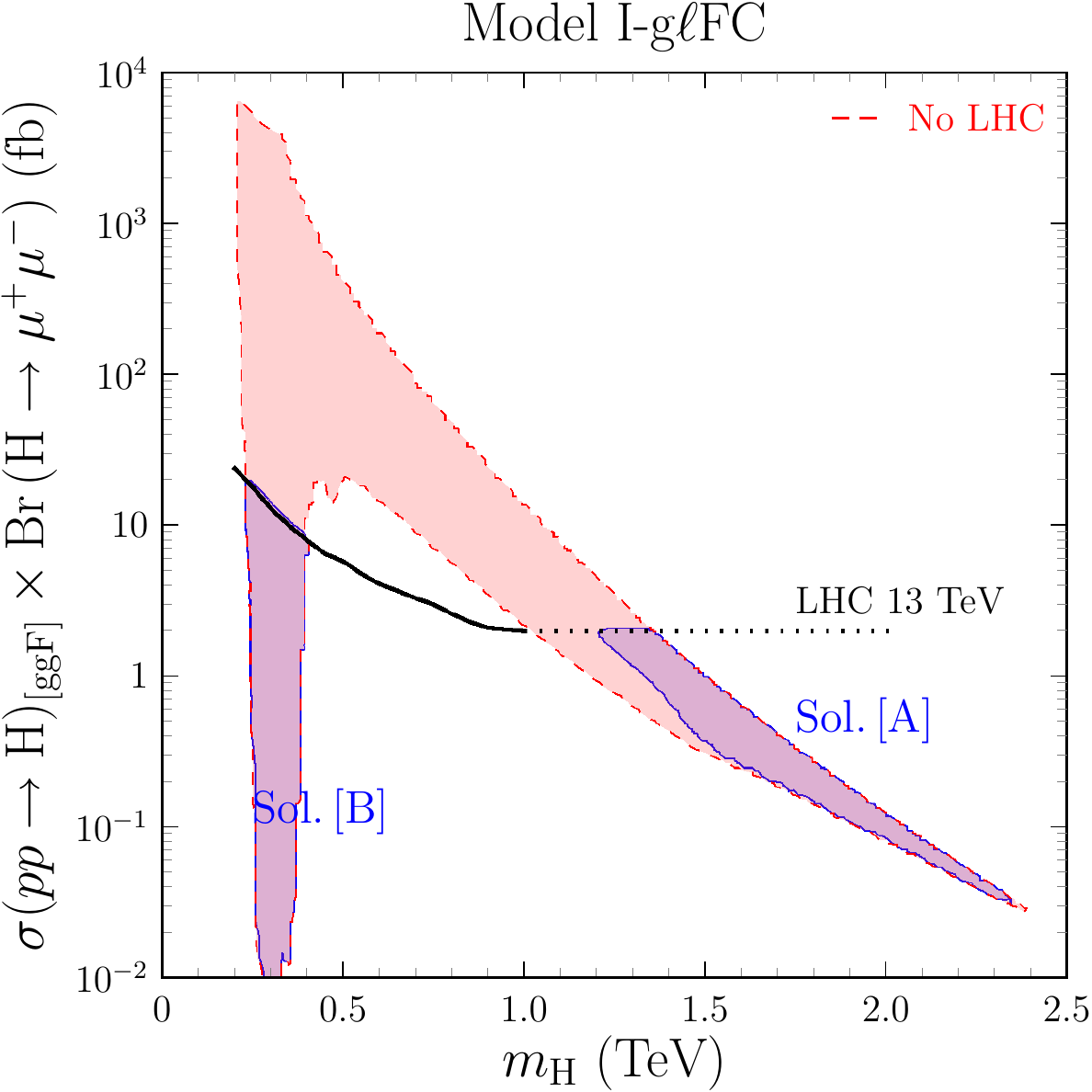}}
\end{center}
\caption{$[pp]_{\rm ggF}\to\mathrm{S}\to\mu^+\mu^-$ versus $\mS$.\label{fig:typeI:ppSlmlm-vs-MS}}
\end{figure}

\begin{figure}[!htb]
\begin{center}
\subfloat[\label{fig:typeI:ppAltlt-vs-MA}]{\includegraphics[width=0.25\textwidth]{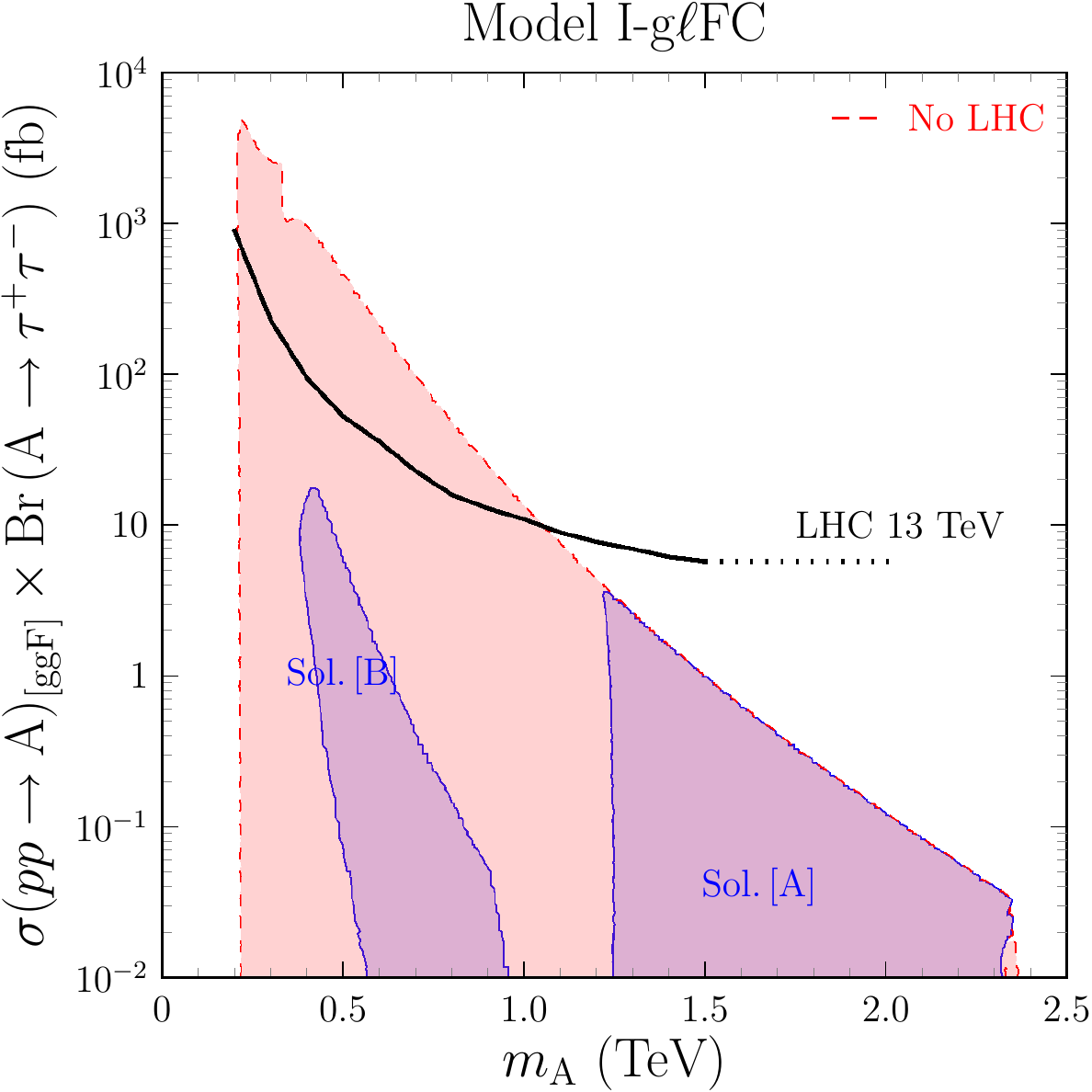}}\quad
\subfloat[\label{fig:typeI:ppHltlt-vs-MH}]{\includegraphics[width=0.25\textwidth]{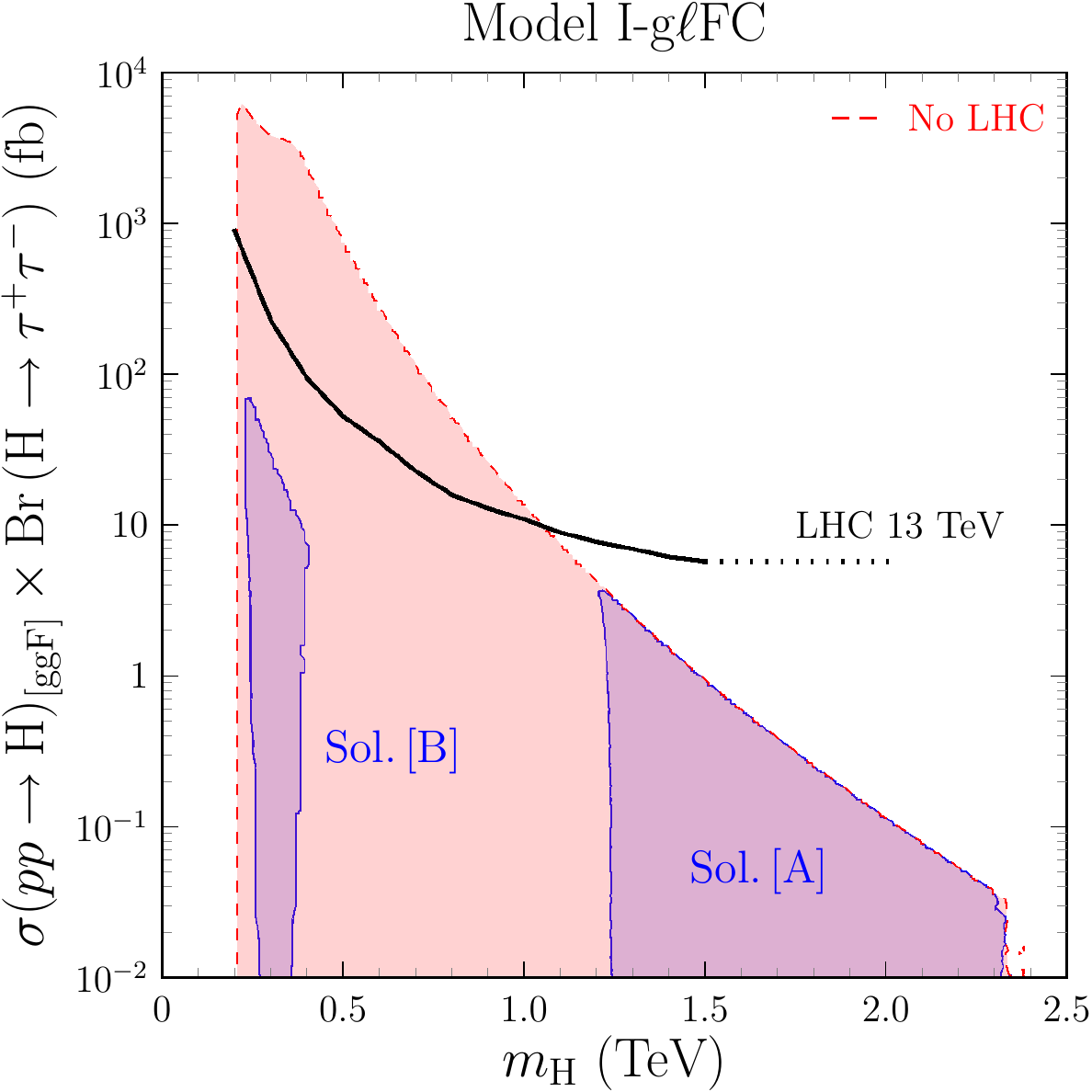}}
\end{center}
\caption{$[pp]_{\rm ggF}\to\mathrm{S}\to\tau^+\tau^-$ versus $\mS$.\label{fig:typeI:ppSltlt-vs-MS}}
\end{figure}

\begin{figure}[!htb]
\begin{center}
\subfloat[\label{fig:decays:typeI:SBS:Hlele}]{\includegraphics[width=0.2\textwidth]{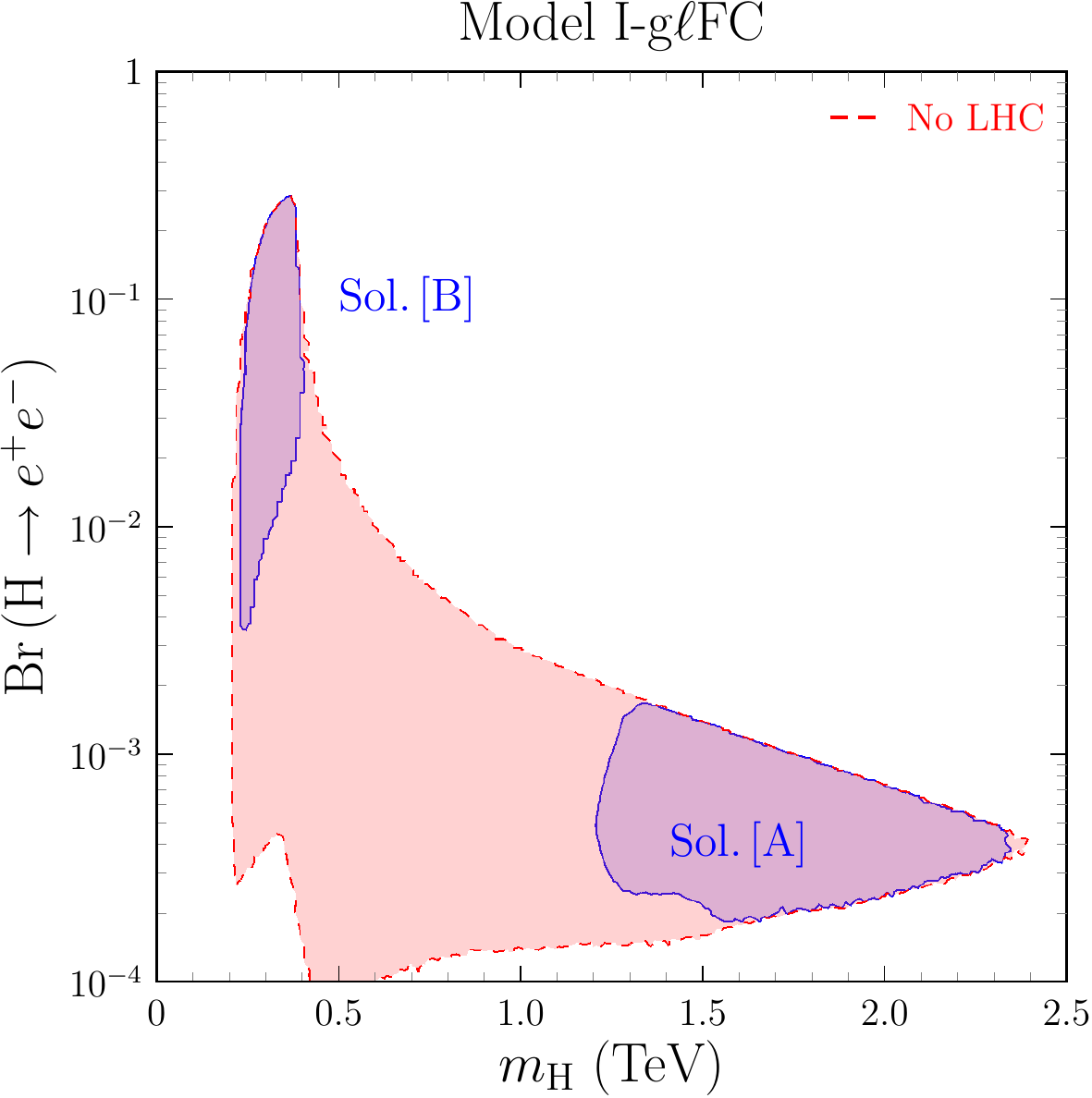}}\quad
\subfloat[\label{fig:decays:typeI:SBS:Hlmlm}]{\includegraphics[width=0.2\textwidth]{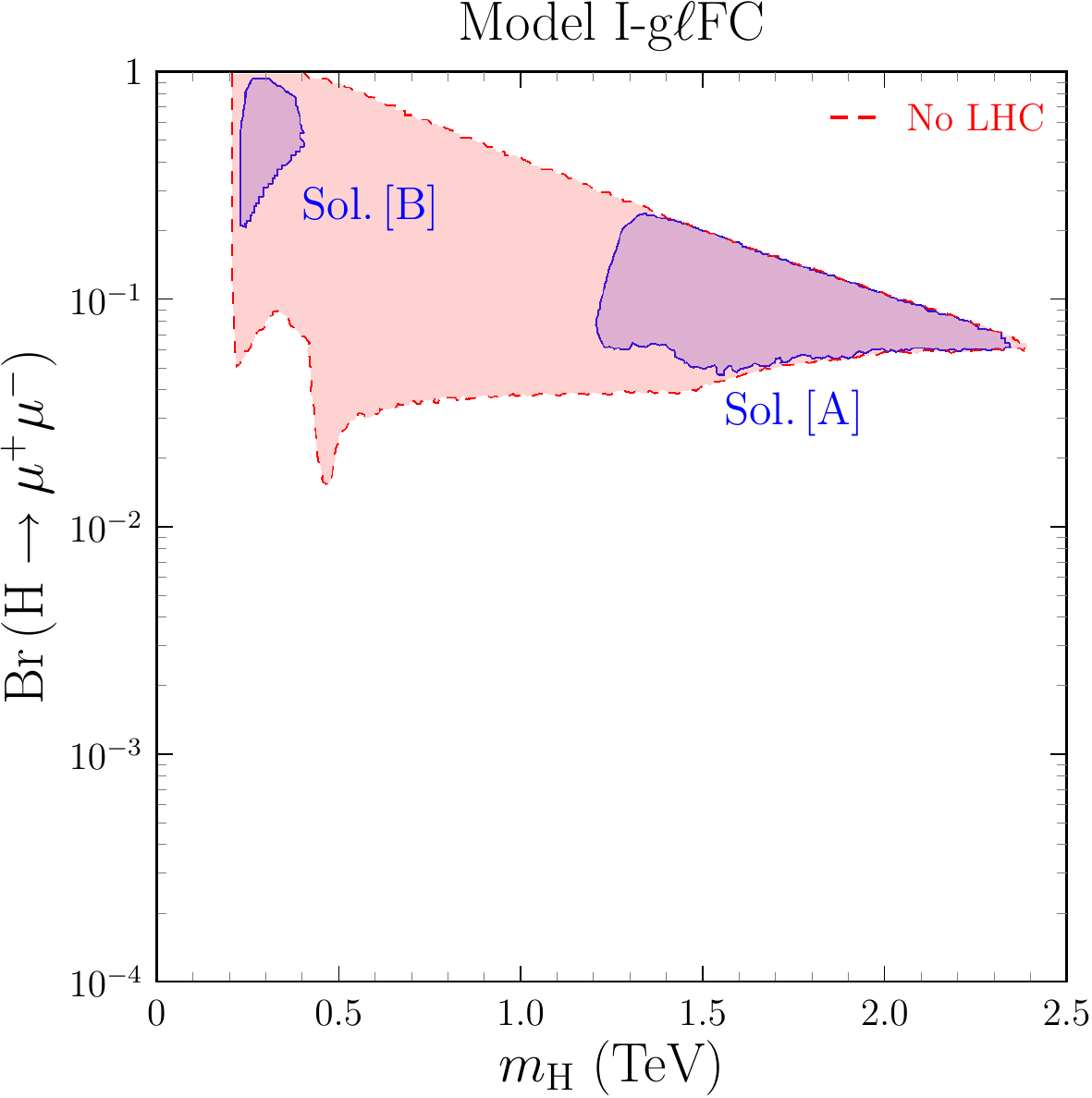}}\quad
\subfloat[\label{fig:decays:typeI:SBS:Hltlt}]{\includegraphics[width=0.2\textwidth]{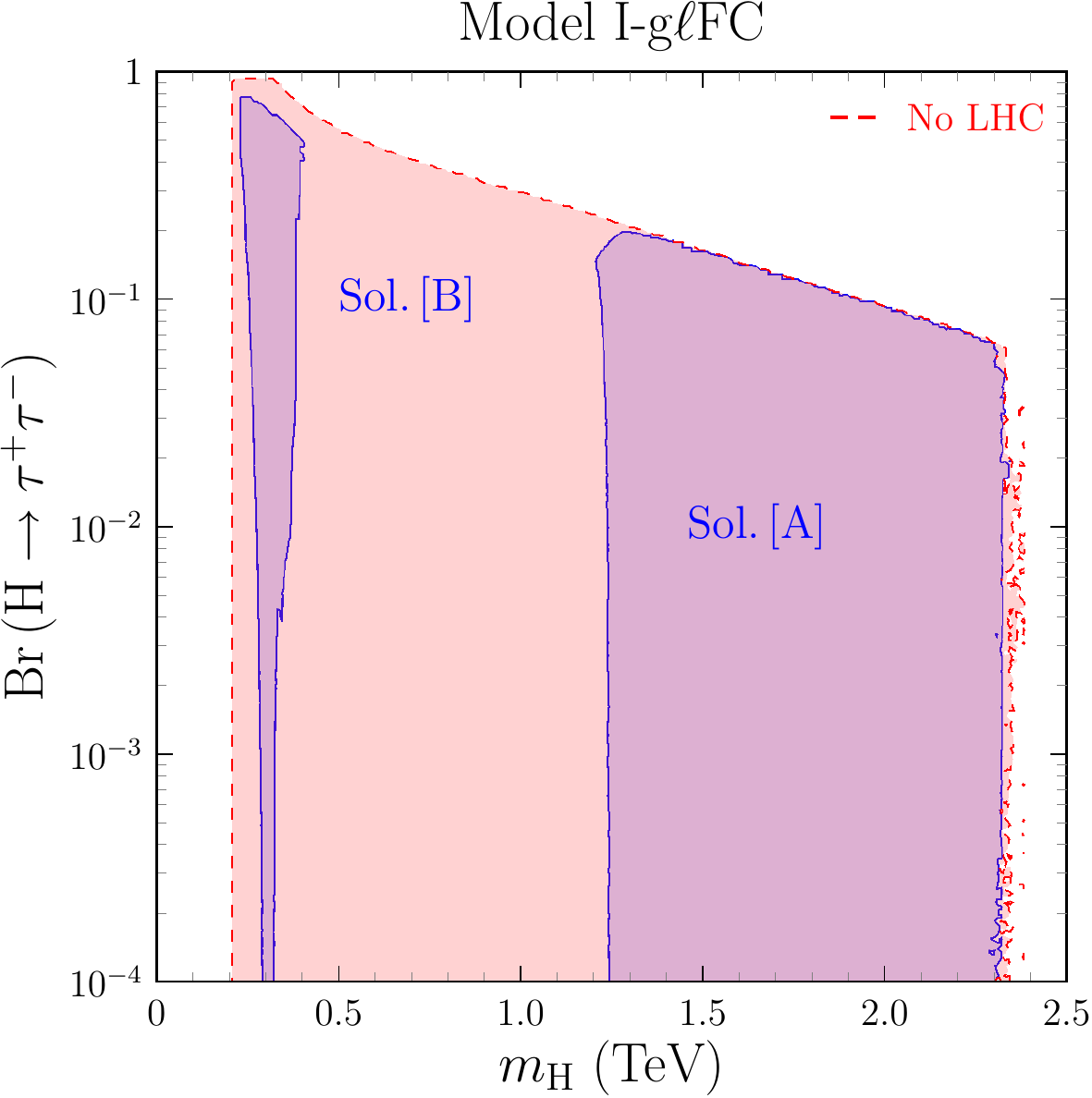}}\quad
\subfloat[\label{fig:decays:typeI:SBS:Hqtqt}]{\includegraphics[width=0.2\textwidth]{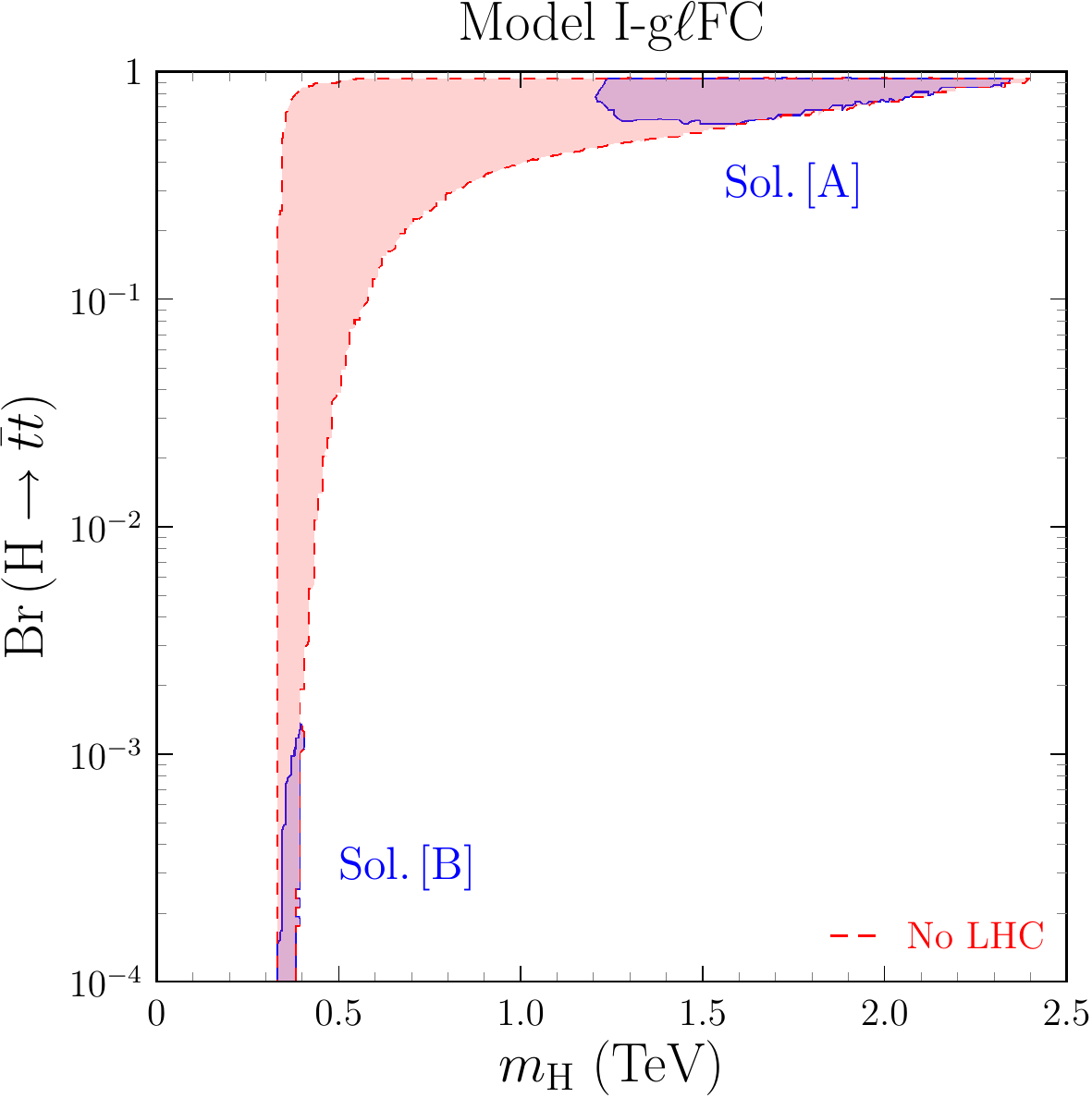}}\\
\subfloat[\label{fig:decays:typeI:SBS:AHZ}]{\includegraphics[width=0.2\textwidth]{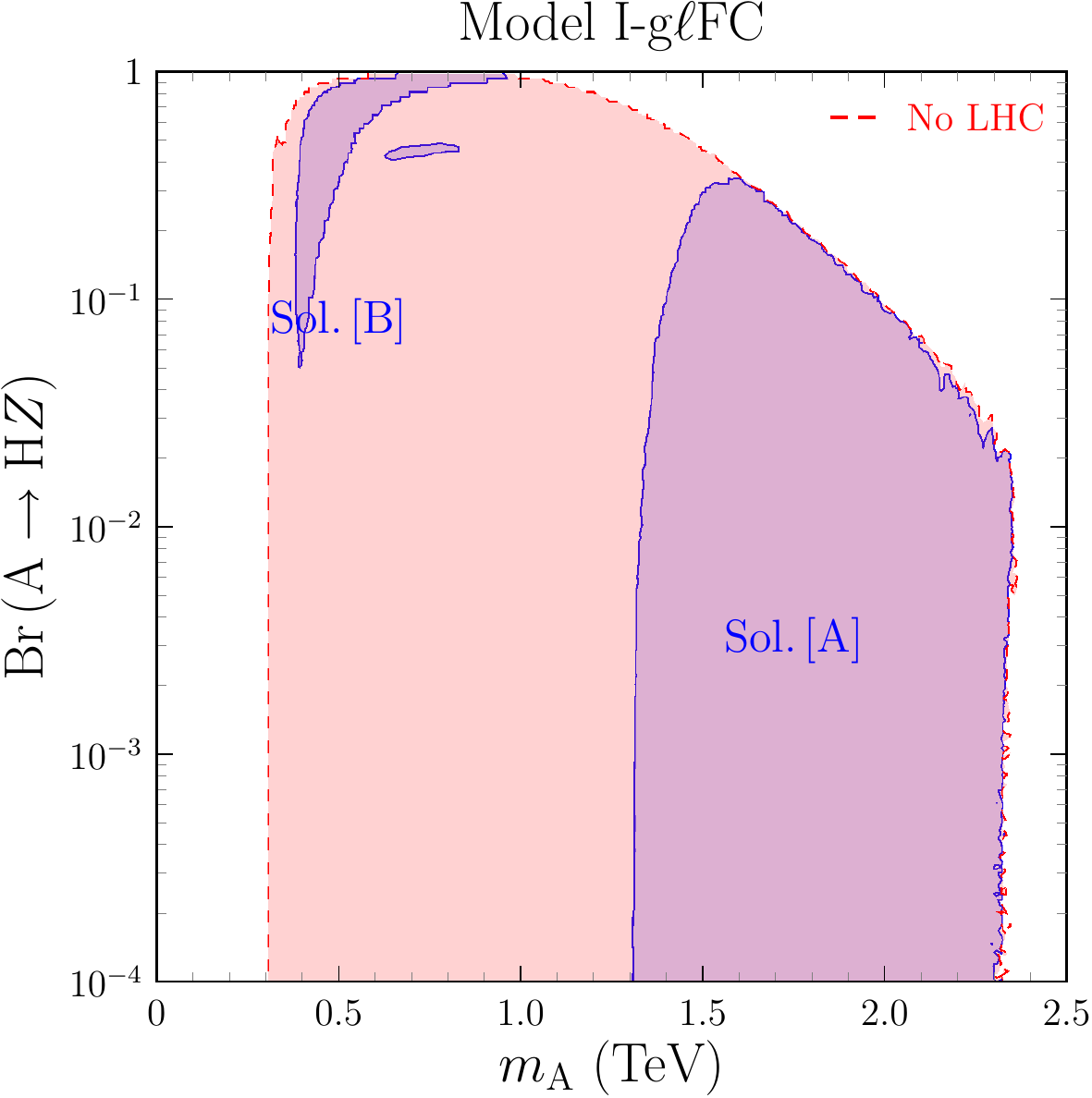}}\quad
\subfloat[\label{fig:decays:typeI:SBS:Almlm}]{\includegraphics[width=0.2\textwidth]{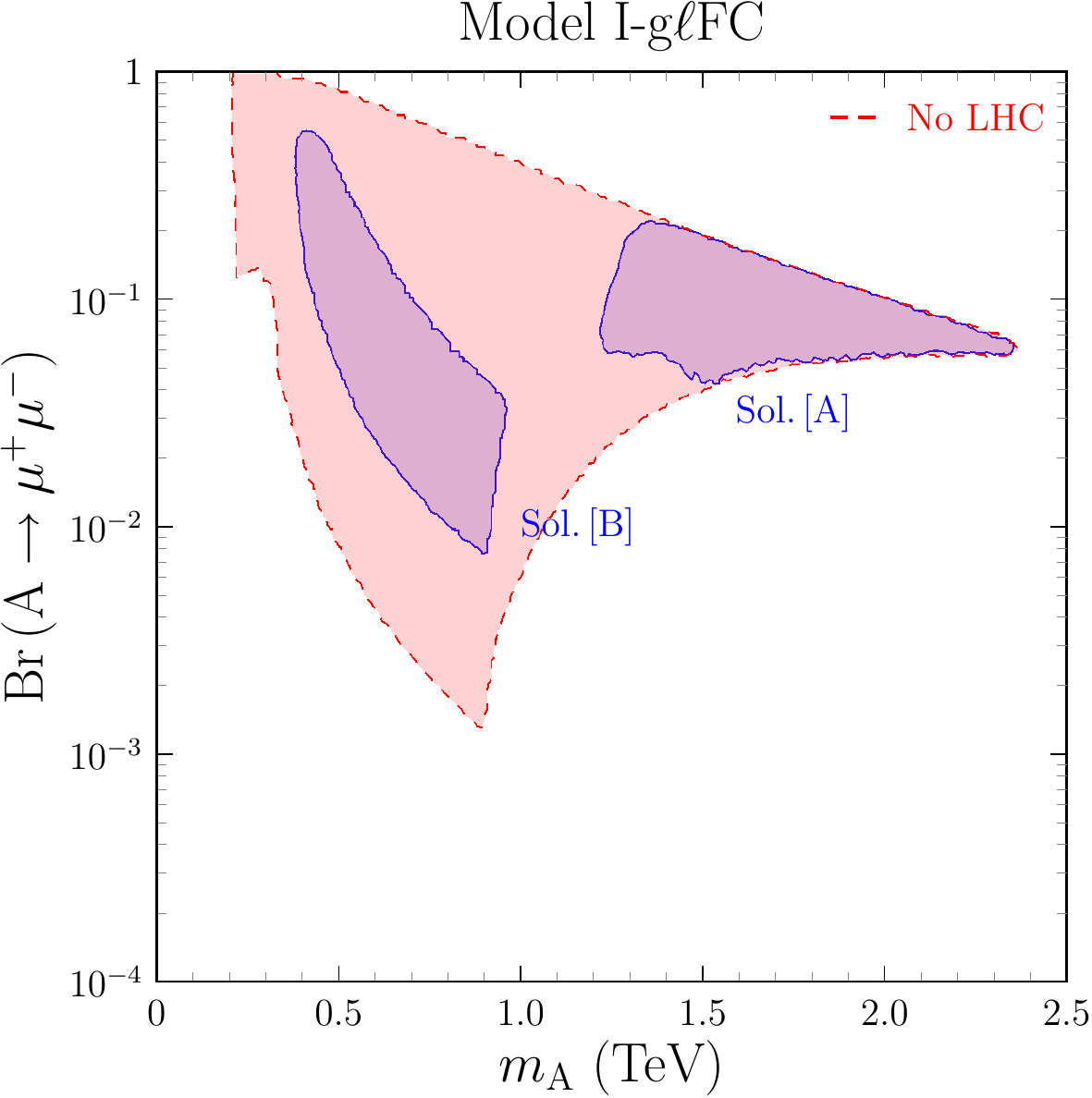}}\quad
\subfloat[\label{fig:decays:typeI:SBS:Altlt}]{\includegraphics[width=0.2\textwidth]{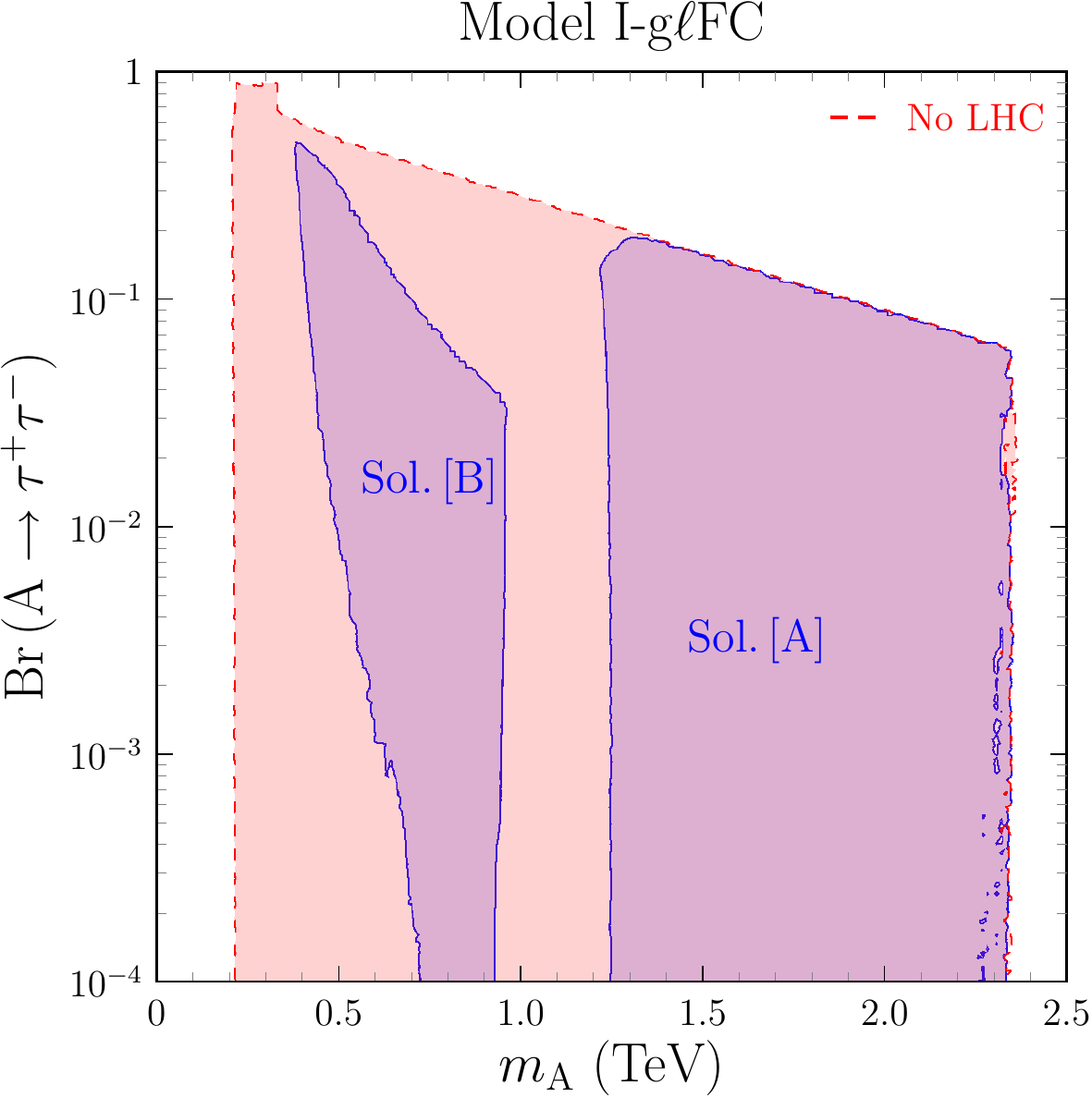}}\quad
\subfloat[\label{fig:decays:typeI:SBS:Aqtqt}]{\includegraphics[width=0.2\textwidth]{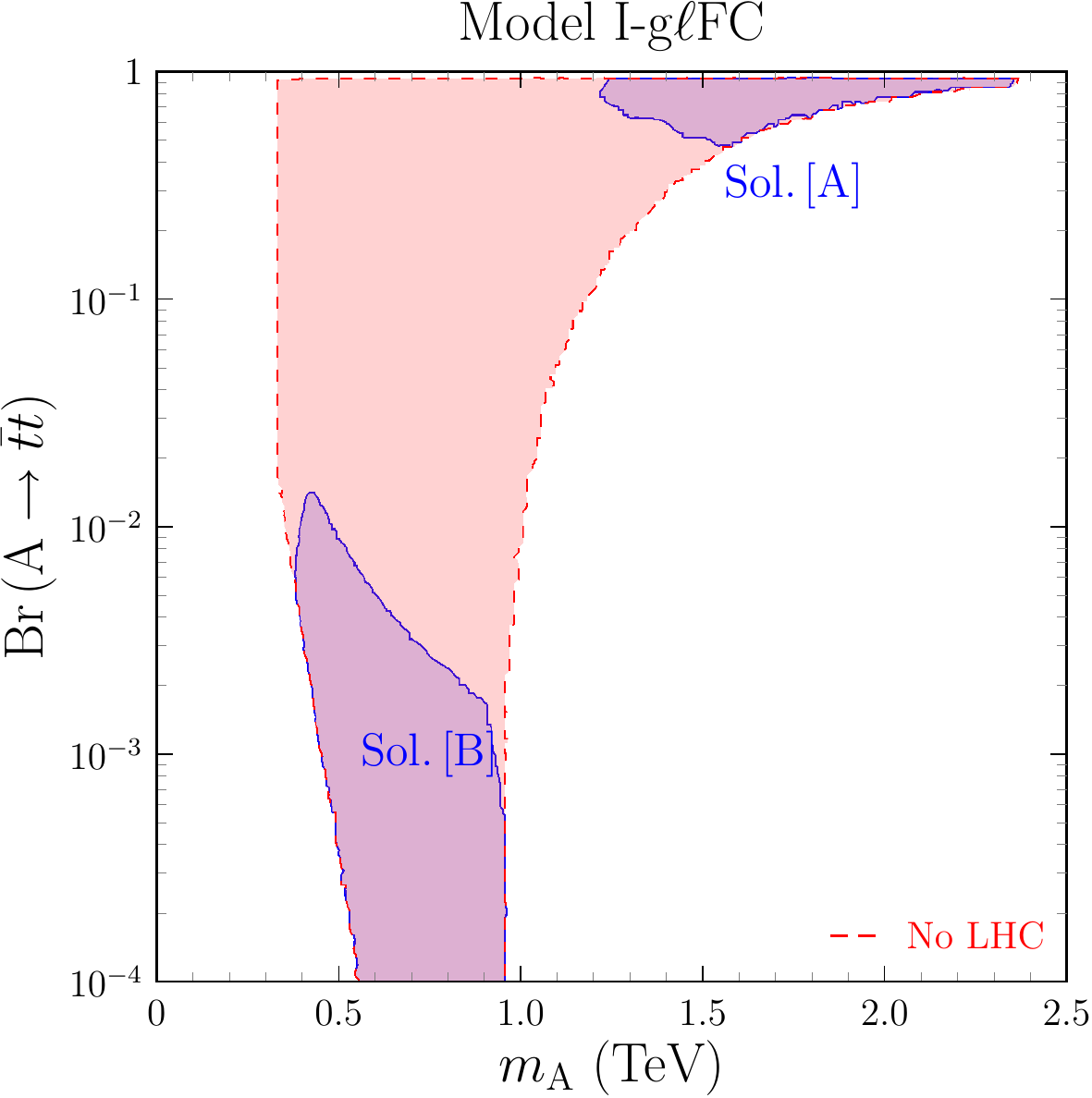}}\\
\subfloat[\label{fig:decays:typeI:SBS:ChHW}]{\includegraphics[width=0.2\textwidth]{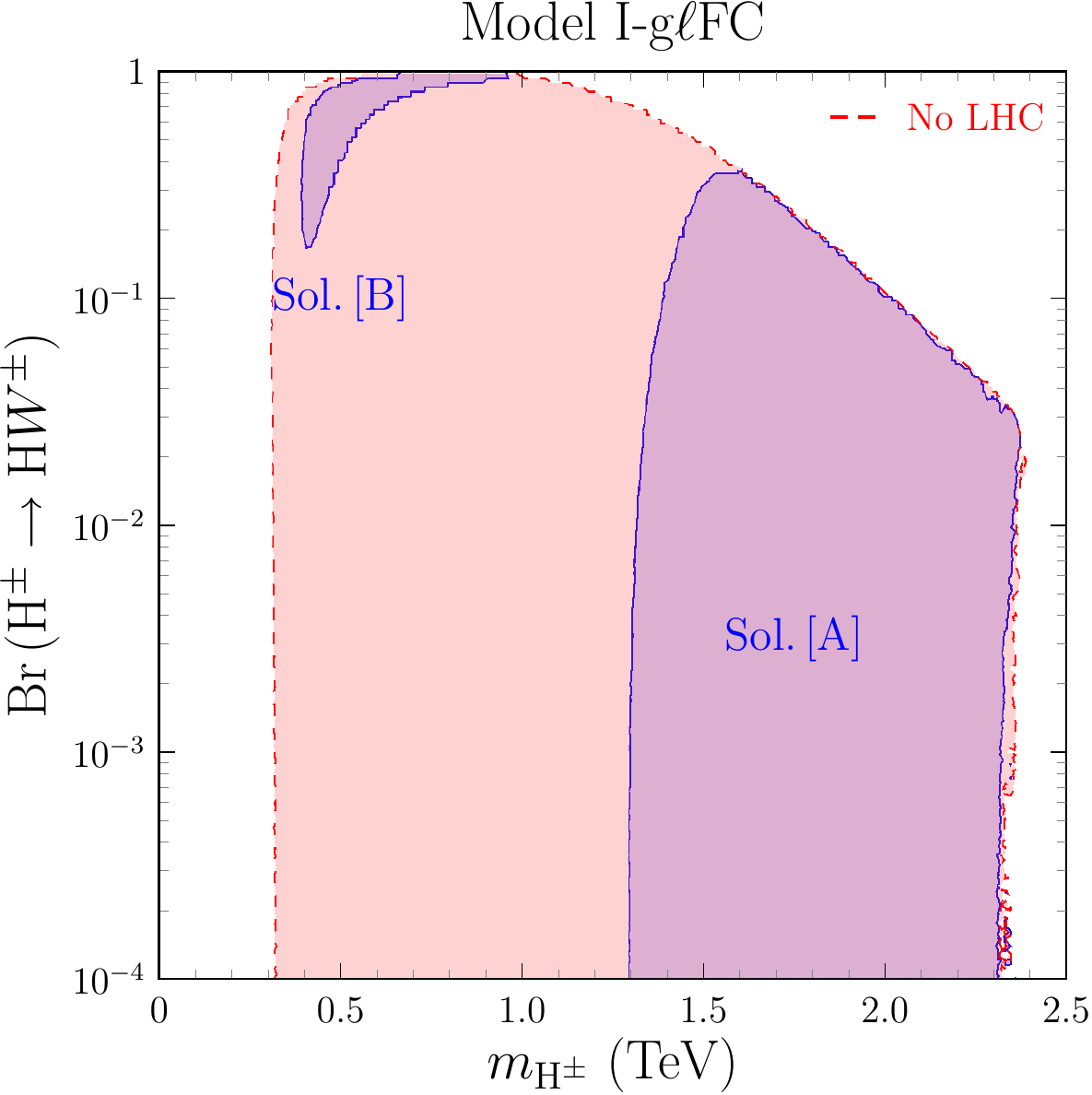}}\quad
\subfloat[\label{fig:decays:typeI:SBS:Chlm}]{\includegraphics[width=0.2\textwidth]{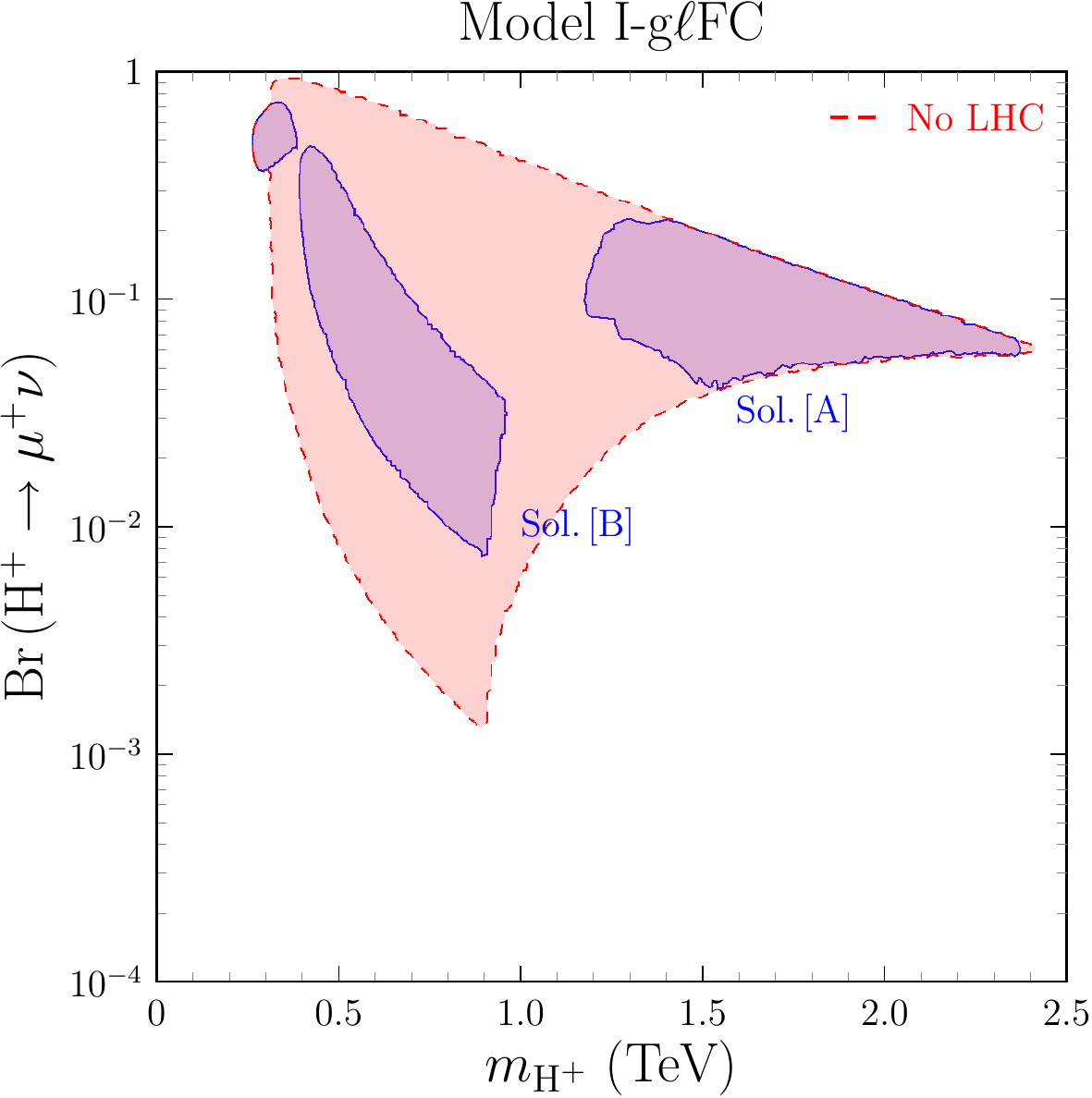}}\quad
\subfloat[\label{fig:decays:typeI:SBS:Chlt}]{\includegraphics[width=0.2\textwidth]{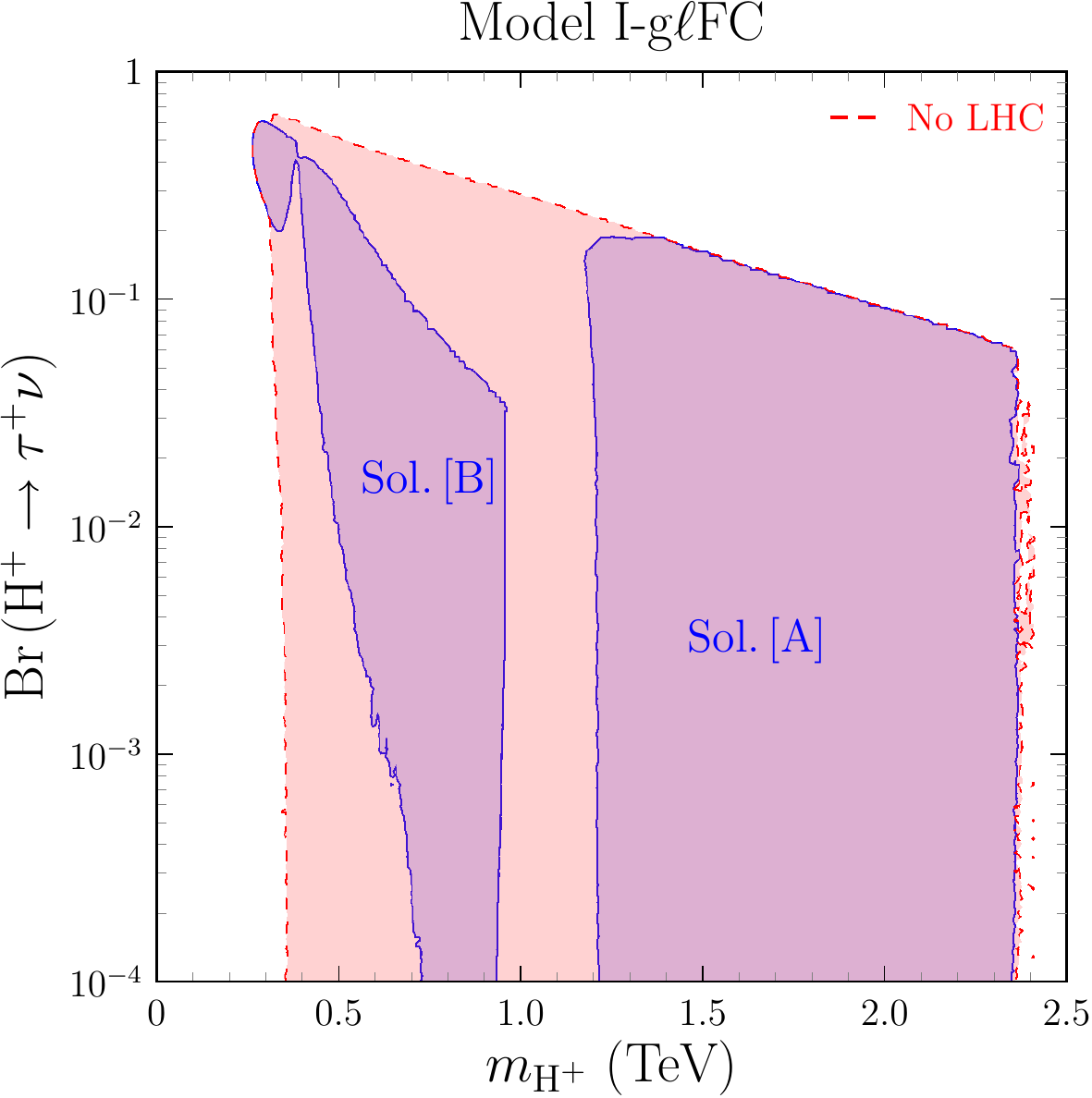}}\quad
\subfloat[\label{fig:decays:typeI:SBS:Chqtqb}]{\includegraphics[width=0.2\textwidth]{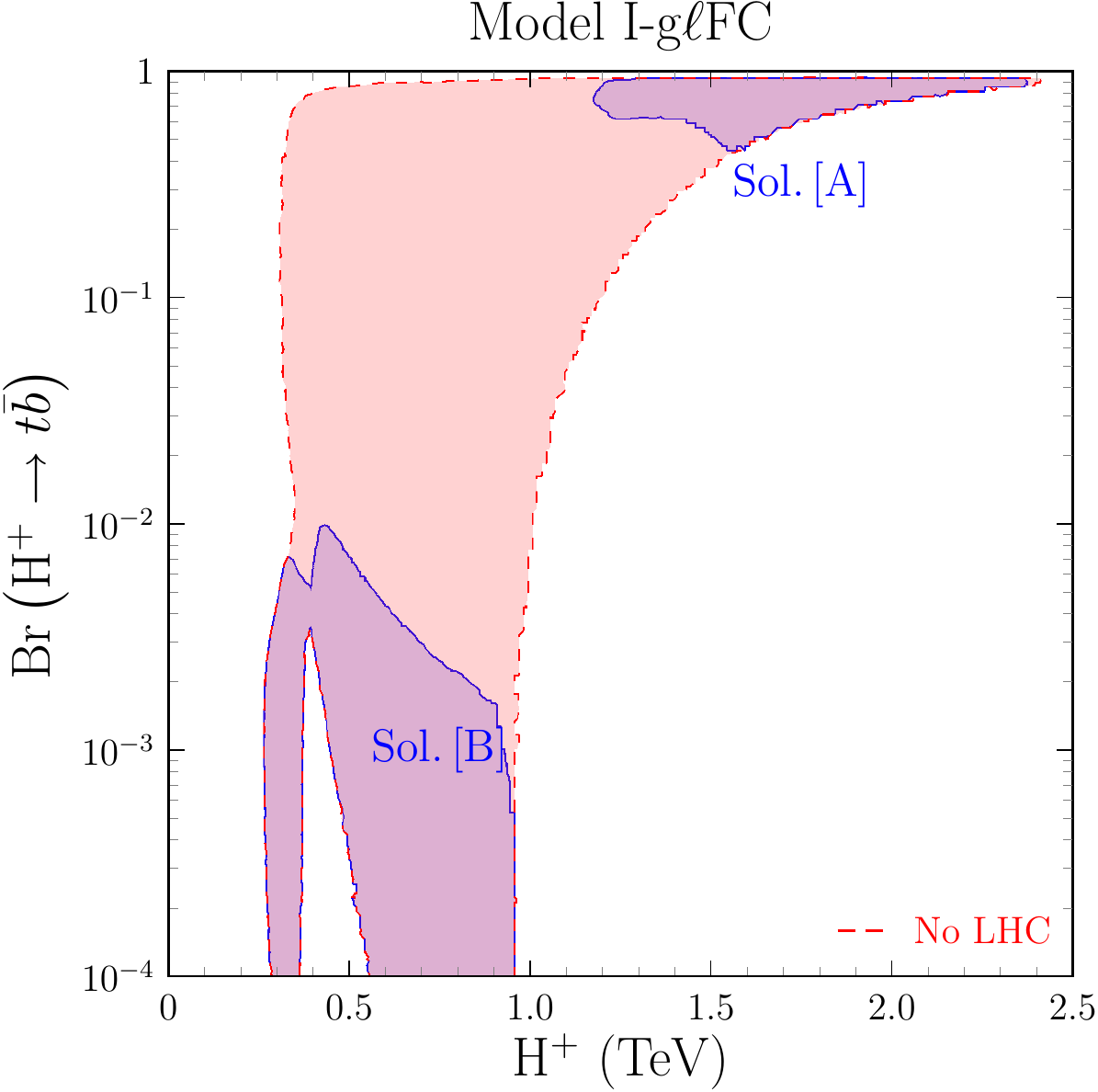}}

\end{center}
\caption{Dominant decay channels of $\nH$, $\nA$, $\cH$.\label{fig:decays:typeI:SBS}}
\end{figure}

\begin{figure}[!htb]
\begin{center}
\subfloat[\label{fig:typeI:ppChHlmln-vs-MChH}]{\includegraphics[width=0.25\textwidth]{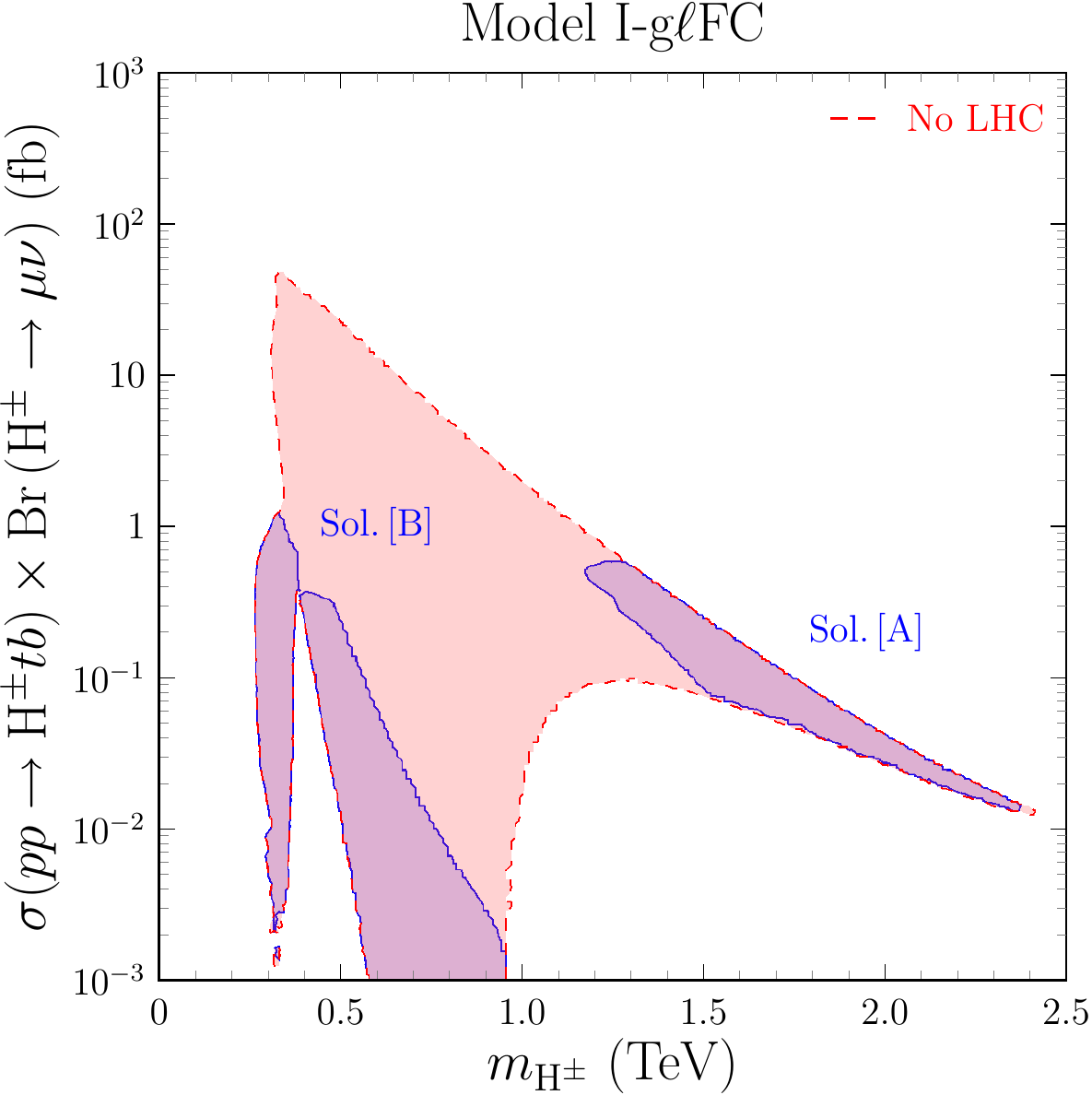}}\quad
\subfloat[\label{fig:typeI:ppChHltln-vs-MChH}]{\includegraphics[width=0.25\textwidth]{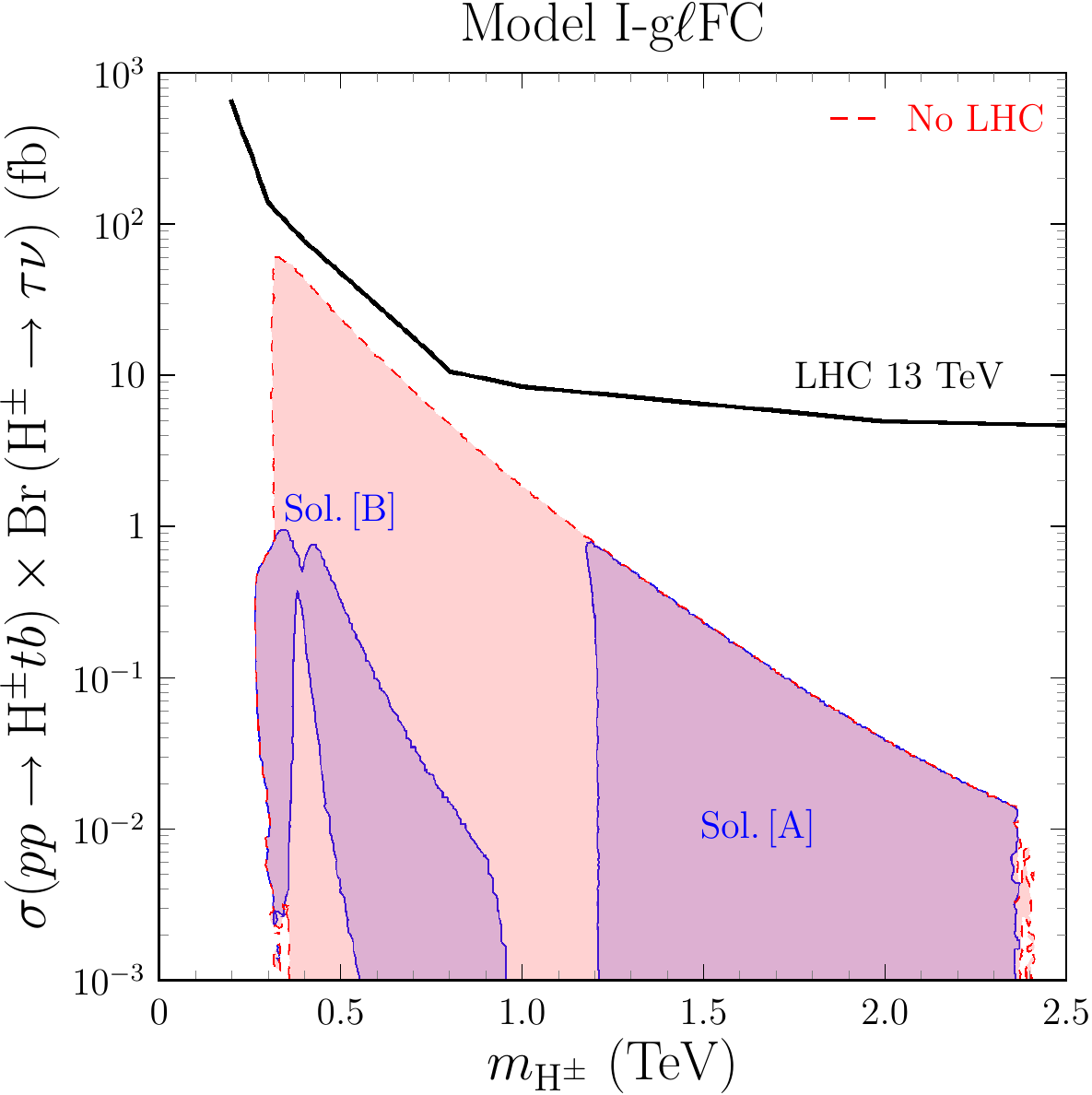}}\quad
\subfloat[\label{fig:typeI:ppChHqtqb-vs-MChH}]{\includegraphics[width=0.25\textwidth]{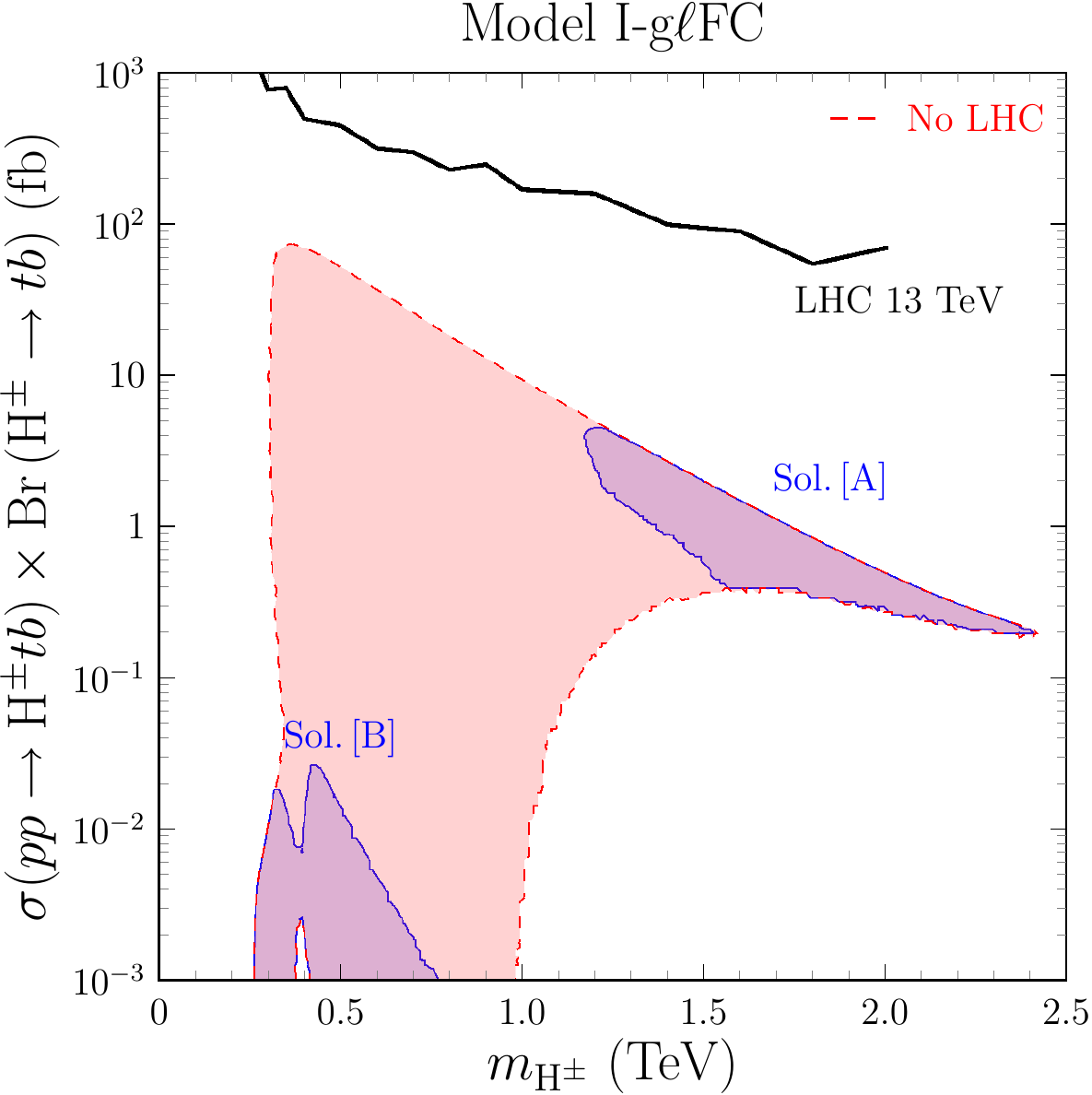}}
\end{center}
\caption{$pp\to\cH(tb)\to\ell\nu,tb$ versus $\mcH$.\label{fig:typeI:ppChH-vs-MChH}}
\end{figure}

\begin{figure}[!htb]
\begin{center}
\includegraphics[width=0.35\textwidth]{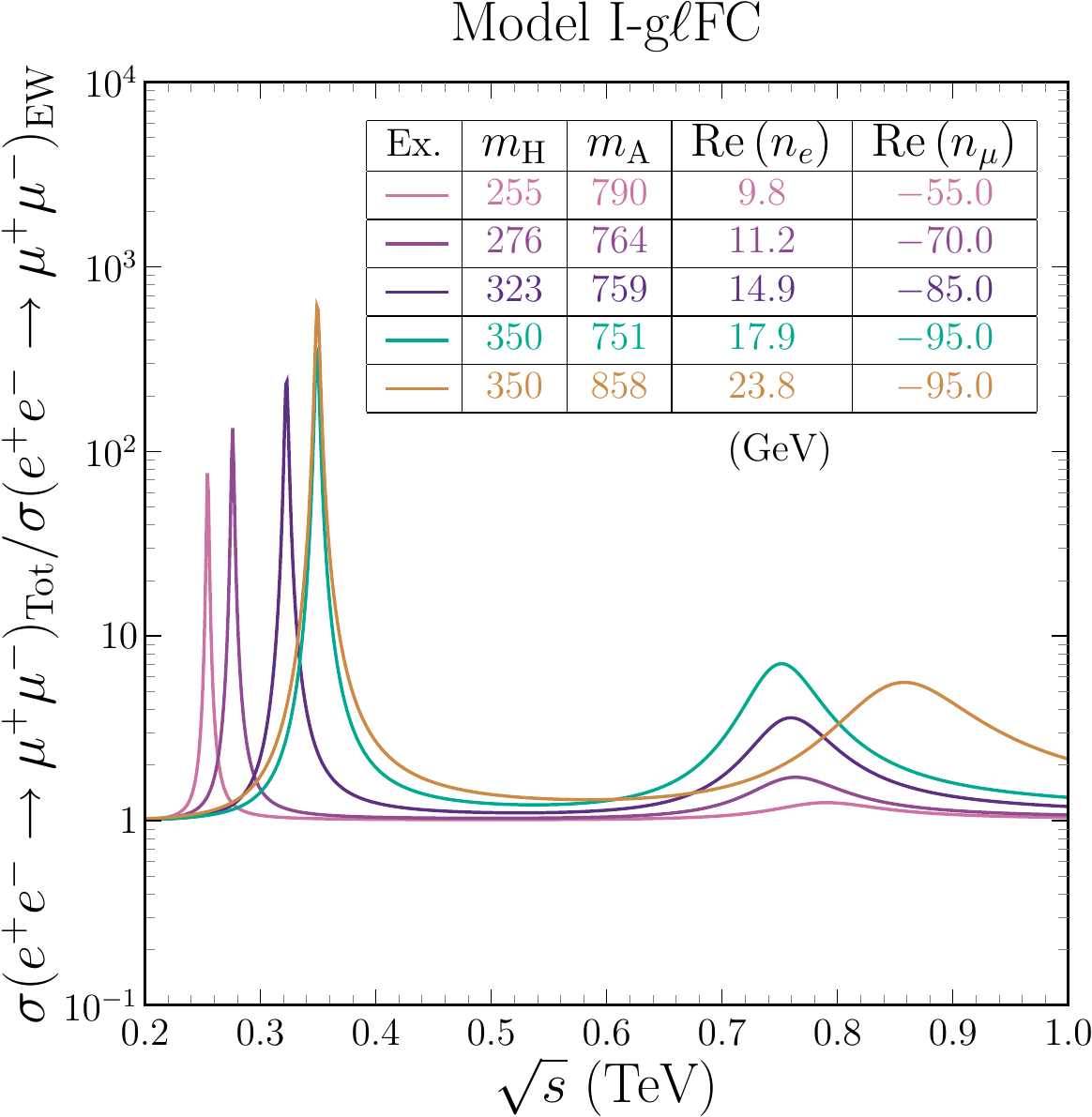}
\end{center}
\caption{$e^+e^-\to\mu^+\mu^-$ for $\sqrt{s}\in [0.2;1.0]$ TeV, examples of solution \solB\ in model \glFC{I}.\label{fig:ILC}}
\end{figure}

%
As discussed in section \ref{SEC:g-2}, we expect, at least, two different types of solution to the $\delta a_\ell$ anomalies. In the following we refer to them, as anticipated, as solutions \solA\ and \solB. Solution \solA\ corresponds to $\tb\sim 1$, heavy neutral new scalars (with masses in the 1--2 TeV range), and both anomalies explained by two loop Barr-Zee contributions. Solution \solB\ corresponds instead to large $\tb$, lighter new scalars, with $\delta a_e$ obtained through two loop Barr-Zee contributions while in $\delta a_\mu$ the most important contributions are one loop and $\nH$-mediated. Note that in general one would expect a set of intermediate solutions between \solA\ and \solB, at least in model \glFC{I}, where we have a priori identified the presence of both solutions. For model \glFC{II} we can only anticipate with some certainty the presence of solution \solA.\\ 
It is therefore very important to find out which constraints, if any, can distinguish among both types of solutions. One should also remember that quite large couplings of the new scalars to leptons are required to explain the anomalies. This fact confers a special role to dilepton resonance and charged scalar searches at the LHC. Consequently the analyses are separated in two stages: (i) one, labelled ``No LHC'', which includes all constraints discussed in section \ref{SEC:Constraints} except for the LHC searches which are not imposed as constraints, and (ii) the complete analysis with all constraints, including these LHC searches.\\
One should also remark, before presenting results, that solutions \solA\ and \solB\ as discussed above, cannot be realized when the scalar potential in \refeq{eq:ScalarPotential:general:01} is exactly $\ZZ$ symmetric, i.e. when $\mu_{12}^2=0$. This was to be expected. The reason to have difficulties obtaining solution \solA\ with the exactly $\ZZ$ symmetric potential is simple: it does not allow a ``decoupling regime'' \cite{Gunion:2002zf,Nebot:2020niz,Faro:2020qyp}, i.e. in that case one cannot have scalars heavier than $\sim 1$ TeV (without violating requirements such as perturbativity). On the other hand, concerning solution \solB, the exact $\ZZ$ symmetry does not allow large $\tb$. Introducing $\mu_{12}^2\neq 0$ removes both obstacles.\\ 
In the plots to follow, the results from the ``No LHC'' analysis correspond to lighter red regions while the results from the full analysis correspond to darker blue regions. The regions represented are allowed at $2\sigma$ (for a $2D-\chi^2$ distribution); the $\chi^2$ or likelihood function used in the numerical analysis implements the constraints of section \ref{SEC:Constraints}.\\ 
In Figure \ref{fig:typeI:nmu_vs_ne} we have $\nrlm$ versus $\nrle$; the full analysis shows, clearly, three disjoint regions. As indicated in the figure, the bottom left small region corresponds to solution \solA, and reproduces the linear relation of \refeq{scaling_of_the_anomaly_1}, arising from the explanation of both anomalies through two loop Barr-Zee contributions. The largest blue region to the bottom right corresponds to solution \solB\ with $\nrlm<0$, where $\delta a_e$ is two loop dominated while $\delta a_\mu$ also receives significant one loop contributions. In this region there is no linear relation among $\nrle$ and $\nrlm$. For $\nrlm>0$, solution \solB\ corresponds to the top blue region (the subindex $\pm$ in \solBpm\ refers to the sign of $\nrlm$). It is clear, from the underlying red region, that excluding LHC searches, there is a smooth transition between solutions \solA\ and \solBm\  where all kinds of contributions must be considered: we recall that the numerical analyses incorporate the complete expressions of appendix \ref{APP:Dipoles}, which consider one and two loop contributions with all possible fermions in the fermion loop of Barr-Zee terms. It is important to stress that, since the lepton couplings to $\nH$ and $\nA$ can be quite large, it is mandatory to include all leptons in the computation of Barr-Zee terms.\\ 
Figure \ref{fig:couplings:typeI:SBS} shows results for $\nrl{\ell}$ versus $\tb$ and $\mH$. From previous discussions, the regions corresponding to solutions \solA\ and \solB\ can be easily identified. For example, in Figure \ref{fig:couplings:typeI:SBS:ne:tb}, the blue region reaching larger values of $\nrle$, with $\tb\geq 13$ and $200\text{ GeV}\leq\mH\leq 370$ GeV is clearly associated to solution \solB. Figures \ref{fig:couplings:typeI:SBS:nm:tb} and \ref{fig:couplings:typeI:SBS:nm:mH} illustrate the same aspects regarding now $\nrlm$. For $\nrlt$ it also follows from Figures \ref{fig:couplings:typeI:SBS:nt:tb} and \ref{fig:couplings:typeI:SBS:nt:mH} that $\nrlt>0$ is required in solution \solB\ (one can indeed check that it gives a subdominant but necessary two loop contribution to obtain the appropriate value of $\delta a_\mu$).\\ 
To characterize more precisely solutions \solA\ and \solB, Figure \ref{fig:scalar:typeI:SBS} shows correlations among scalar masses and with $\tb$. In particular, it is clear that in solution \solA\ all new scalars are heavy, with masses in the $1.2$ -- $2.3$ TeV range, and mass differences not exceeding $\pm 200$ GeV. For solution \solB, some important results can be observed: (i) in addition to the existence of separate regions \solBp\ and \solBm\ for both signs of $\nrlm$, there are two separate manners in which solution \solB\ can arise, one region where $\mcH\in[0.4;0.9]$ TeV and $\mA=\mcH$ to a high degree of accuracy and another smaller region where $\mcH\in[0.25;0.35]$ TeV and $\mH=\mcH$ to a high degree of accuracy; (ii) in all cases, $\mH<\mA$. This last inequality, as analysed later, must allow the decay $\nA\to\nH Z$ (additionally, either $\cH\to\nH W^\pm$ or $\nA\to\cH W^\mp$ would also be allowed); together with the electroweak precision constraints (in particular the oblique parameter $T$), this forces either $\mA=\mcH$ or $\mH=\mcH$. These two results match nicely with the need for $\nH$ to be as light as possible (LEP constraints will force in any case $\mH\geq 210$ GeV) in order to produce the main contribution (at one loop) to $\delta a_\mu$.\\ 
Figure \ref{fig:typeI:ppSlmlm-vs-MS} shows the resonant $[pp]_{\rm ggF}\to\mathrm{S}\to\mu^+\mu^-$ cross sections with respect to $\mS$ for $\mathrm{S}=\nH,\nA$. The black line shows the LHC bounds included in the full analysis. In gluon-gluon fusion production, for the same scalar mass, the gluon-gluon-pseudoscalar amplitude is 2--6 times larger than the corresponding gluon-gluon-scalar amplitude (that is $2^2-6^2$ larger pseudoscalar vs. scalar production cross sections). One could have expected, attending to this fact, that the constraints from LHC searches on $\sigma(pp\to \nA)_{\rm ggF}\times\BR{\nA\to\mu^+\mu^-}$ versus $\mA$ would be responsible for the separation among solutions \solA\ and \solB. Figure \ref{fig:typeI:ppAlmlm-vs-MA} disproves this naive expectation; as Figure \ref{fig:typeI:ppHlmlm-vs-MH} shows it is rather $\sigma(pp\to \nH)_{\rm ggF}\times\BR{\nH\to\mu^+\mu^-}$ which shows how the bounds from LHC searches separate the solutions by excluding $\mH\in[380;1200]$ GeV (i.e. eliminating the red region ``bridge'' connecting the blue regions). Comparing the shape of the allowed regions in Figures \ref{fig:typeI:ppAlmlm-vs-MA} and \ref{fig:typeI:ppHlmlm-vs-MH} it is also clear that, besides the production cross section, the branching ratios $\BR{\nH,\nA\to\mu^+\mu^-}$ may play an important role.\\
On this respect, let us start by observing that, since values of $\abs{\nrlm}$ larger than some minimal $\abs{\nrlm}_{\rm Min}$ are required to explain $\delta a_\mu$, both $\BR{\nH\to\mu^+\mu^-}$ and $\BR{\nA\to\mu^+\mu^-}$ are bounded from below. The dominant decay channels of the new scalars are shown in Figure \ref{fig:decays:typeI:SBS}; Figures \ref{fig:decays:typeI:SBS:Hlmlm} and \ref{fig:decays:typeI:SBS:Almlm} show that $\BR{\nH\to\mu^+\mu^-}$ and $\BR{\nA\to\mu^+\mu^-}$ are indeed bounded from below, but in the case of $\nH\to\mu^+\mu^-$ the lower bound is larger than that of $\nA\to\mu^+\mu^-$ (it can even saturate the decay width of $\nH$). This explains the narrowness of the red and blue regions in Figure \ref{fig:typeI:ppHlmlm-vs-MH} for $\mH>400$ GeV. One should keep in mind that solutions \solA\ and \solB\ also differ quite substantially in the values of $\tb$: in Figure \ref{fig:scalar:typeI:SBS:mH:tb} it is clear that large $\mH>1$ TeV requires $\tb\sim 1$, while $\mH<500$ GeV is compatible with a broad range $\tb\in[1;10^2]$. This is the last ingredient necessary to interpret the shape of Figure \ref{fig:typeI:ppHlmlm-vs-MH}. For $\mH<500$ GeV, without constraints from LHC searches, the broad range of $\tb$ values gives a broad range for $\sigma(pp\to\nH)_{\rm ggF}$: since the gluon-gluon fusion production cross section is proportional to $\tb^{-2}$, and thus for solution \solB\ there is a substantial suppression of $\sigma(pp\to\nH)_{\rm ggF}$ due to $\tb\gg 1$. Due to the larger production cross section of a pseudoscalar, despite the $\tb^{-2}$ suppression, LHC searches might rule out $pp\to\nA\to\mu^+\mu^-$ predictions for solution \solB: as Figure \ref{fig:typeI:ppAlmlm-vs-MA} shows, that is not the case. This is clearly achieved through a reduction of $\BR{\nA\to\mu^+\mu^-}$; Figures \ref{fig:decays:typeI:SBS:Almlm} and \ref{fig:decays:typeI:SBS:AHZ} show that $\nA\to\nH Z$ contributes decisively to reduce $\BR{\nA\to\mu^+\mu^-}$, evade LHC bounds and obtain a viable solution \solB. For this reason, as anticipated, $\mA>\mH+M_Z$. For the charged scalar $\cH$, the behaviour of the most relevant decay channels $\cHp\to\mu^+\nu$, $\tau^+\nu$, $t\bar b$, $\nH W^\pm$ mirrors the corresponding $\nA\to\mu^+\mu^-$, $\tau^+\tau^-$, $t\bar t$, $\nH Z$, as Figures \ref{fig:decays:typeI:SBS:Chlm}--\ref{fig:decays:typeI:SBS:ChHW} show. The only minor difference arises for solution \solB\ in the small region where $\mcH\simeq \mH$: in that region, (i) $\cH\to\nH W^\pm$ is forbidden and (ii) in addition to $\nA\to\nH Z$, also $\nA\to\cH W^\mp$ (not shown) has a large branching ratio. \\
Figure \ref{fig:typeI:ppSltlt-vs-MS} shows that resonant $\tau^+\tau^-$ searches are less constraining than the corresponding $\mu^+\mu^-$ searches in Figure \ref{fig:typeI:ppSlmlm-vs-MS}.
Concerning production of $\cH$, Figure \ref{fig:typeI:ppChH-vs-MChH} shows that current results from searches at the LHC are much less constraining than the results from resonant dilepton searches in Figures \ref{fig:typeI:ppSltlt-vs-MS} and \ref{fig:typeI:ppSlmlm-vs-MS}.
Results in the previous figures concern model \glFC{I}, where, in addition to solution \solA\ in which both $\delta a_e$ and $\delta a_\mu$ arise from 2 loop contributions, a second set of solutions \solB\ exists in which 1 loop contributions are dominant in $\delta a_\mu$. For model \glFC{II} this second possibility is not available, and only solution \solA\ is obtained. Furthermore, since $\tb\sim 1$ in solution \solA, the corresponding allowed regions do not differ much in both models \glFC{I} and \glFC{II}. We do not show figures corresponding to model \glFC{II} since the allowed regions in that case very approximately coincide with ``Sol. \solA'' regions in model \glFC{I} plots.\\ 
Finally, Figure \ref{fig:ILC} illustrates with some examples the kind of clear signal that solution \solB\ in model \glFC{I} gives in $e^+e^-\to\mu^+\mu^-$ scattering at energies beyond the range explored at LEP.

\clearpage

\section*{Conclusions\label{SEC:conc}}
General 2HDMs without SFCNC in the lepton sector are a robust framework, stable under renormalization group evolution, in which the possibility of decoupling the electron, muon and tau interactions is open. In this context, lepton flavour universality is broken beyond the mass proportionality, and a different behaviour among charged leptons can be accommodated in a simple way, without introducing highly constrained SFCNC. We have considered two of these general flavour conserving models in the leptonic sector, to address simultaneously the electron and muon $(g-2)$  anomalies. These two models, I-g$\ell$FC and II-g$\ell$FC, differ in the quark Yukawa couplings, which coincide, respectively, with the ones in type I and in type II 2HDMs.
There are two types of solutions that fully reproduce both the muon and electron $(g-2)$ anomalies, while remaining in agreement with constraints from LEP and LHC, from LFU, from flavour and electroweak physics, and theoretical requirements in the scalar sector. 
In one solution, all the new scalars have masses in the 1--2.5 TeV range, the vevs of both doublets are quite similar and both anomalies are dominated by two loop Barr-Zee contributions. This solution arises in both models, I- and II-g$\ell$FC.
There is a second type of solution, where one loop contributions are dominant in the muon anomaly, the new scalars have masses below 1 TeV, and the vevs quite different, with a ratio in the range 10-100. Among the new scalars, the second neutral one $\mathrm{H}$ is the lighter, with a mass in the range 210-390 GeV, while the pseudoscalar $\mathrm{A}$ is the heavier, with a mass in the range 400-900 GeV. The new charged scalar $\mathrm{H}^\pm$ is almost degenerate either with the scalar or with the pseudoscalar. This solution is only available in the I-g$\ell$FC model.
In both solutions, soft breaking of the $\mathbb{Z}_{2}$ symmetry of the Higgs potential is required, together with lepton Yukawa couplings with values from 1 to 100 GeV. These results imply for LHC searches, in the light scalar solution, that it should be easier to find both charged and neutral Higgses in the muonic channel. The heavy channels, like the top quark channels, are more suited to searches addressing the heavy scalars solution.

\section*{Acknowledgments}
The authors thank Luca Fiorini and Marcel Vos for useful discussions. The authors acknowledge support from Spanish grant FPA2017-85140-C3-3-P (AEI/FEDER, UE) and PROMETEO 2019-113 (Generalitat Valenciana). The work of FCG is funded by \emph{Ministerio de Ciencia, Innovacion y Universidades}, Spain (Grant BES-2017-080070) and partially supported by a Short-Term Scientific Mission Grant from the COST Action CA15108. MN acknowledges support from \emph{Funda\c{c}\~ao para a Ci\^encia e a Tecnologia} (FCT, Portugal) through the following projects: CFTP-FCT Unit 777 (UID/FIS/00777/2013, UID/FIS/00777/2019) and PTDC/FIS-PAR/29436/2017, partially funded through POCTI (FEDER), COMPETE, QREN and EU.

\appendix

\clearpage
\section{Yukawa couplings\label{APP:Yukawa}}
%
For completeness we show in this appendix the form of the Yukawa couplings in the general case with arbitrary scalar mixing $\ROTmat$ and couplings $\matND$, $\matNU$, $\matNL$. For neutral scalars they read\footnote{Flavour indices are omitted for simplicity: e.g. $\matMf{f}$ is the diagonal mass matrix.}
\begin{multline}
\mathscr L_{\nS\bar ff}=
-\frac{\nS}{v}\,\bar f\,\left[\ROT{1s}\matMf{f}+\ROT{2s}\frac{\matNf{f}+\matNfd{f}}{2}+i\epsilon_{(f)}\ROT{3s}\frac{\matNf{f}-\matNfd{f}}{2}\right]\,f\\
-\frac{\nS}{v}\,\bar f\,\gamma_5\,\left(\ROT{2s}\frac{\matNf{f}-\matNfd{f}}{2}+i\epsilon_{(f)}\ROT{3s}\frac{\matNf{f}+\matNfd{f}}{2}\right)\,f\,,
\end{multline}
where $s=1,2,3$ in correspondence with $\nS=\nh,\nH,\nA$; $f=u,d,\ell$, and, in terms proportional to $\ROT{3s}$, $\epsilon_{(d)}=\epsilon_{(\ell)}=-\epsilon_{(u)}=1$.\\
The Yukawa couplings of $\cH$ read
\begin{multline}
\mathscr L_{\cH ud}=
\frac{\cHm}{\sqrt{2} v}\bar\Fd\left[\CKMdag\matNU-\matNDd\CKMdag+\gamma_5\left(\CKMdag\matNU+\matNDd\CKMdag\right)\right]\Fu\\
+\frac{\cHp}{\sqrt{2} v}\bar\Fu\left[\matNUd\CKM-\CKM\matND+\gamma_5\left(\matNUd\CKM+\CKM\matND\right)\right]\Fd\,,
\end{multline}
and
\begin{equation}\label{eq:ChYuk:lep}
\mathscr L_{\cH \ell\nu}=-\frac{\sqrt{2}}{v}\cHp\,\nLb{}\,\PMNSdag\matNL\,\lR{}-\frac{\sqrt{2}}{v}\cHm\,\lRb{}\,\matNLd\PMNS\,\nL{}\,.
\end{equation}
$\CKM$ and $\PMNS$ are, respectively, the CKM and PMNS mixing matrices\footnote{Equation \eqref{eq:ChYuk:lep} assumes massless neutrinos, in which case one can indeed set $\PMNS\to\id$.}.

\section{Contributions to $(g-2)_\ell$\label{APP:Dipoles}}
%
\subsection{One loop contributions}
Yukawa interactions (of neutral scalars $S$) of the form 
\begin{equation}\label{eq:YukLep:generic}
\mathscr L_{S\ell\ell}=-\frac{m_\ell}{\vev{}}S\bar\ell(a_\ell^S+ib_\ell^S\gamma_5)\ell\,,
\end{equation}
give one loop contributions to the anomalous magnetic moment of lepton $\ell$ of the form
\begin{equation}
\Delta a_\ell^{(1)}=\frac{1}{8\pi^2}\frac{m_\ell^2}{\vev{}^2}\sum_{S}\left\{[a_\ell^S]^2\left(2I_2(x_{\ell S})-I_3(x)\right)-[b_\ell^S]^2I_3(x_{\ell S})\right\},
\end{equation}
with $x_{\ell S}\equiv {m_\ell^2}/{m_{S}^2}$ and
\begin{equation}
I_2(x)=1+\frac{1-2x}{2x\sqrt{1-4x}}\ln\left(\frac{1+\sqrt{1-4x}}{1-\sqrt{1-4x}}\right)+\frac{1}{2x}\ln x,
\end{equation}
\begin{equation}
I_3(x)=\frac{1}{2}+\frac{1}{x}+\frac{1-3x}{2x^2\sqrt{1-4x}}\ln\left(\frac{1+\sqrt{1-4x}}{1-\sqrt{1-4x}}\right)+\frac{1-x}{2x^2}\ln x.
\end{equation}
For $x\ll 1$,
\begin{equation}
I_2(x)\simeq x\left(-\frac{3}{2}-\ln x\right)+x^2\left(-\frac{16}{3}-4\ln x\right)+\mathcal O(x^3),
\end{equation}
\begin{equation}
I_3(x)\simeq x\left(-\frac{11}{6}-\ln x\right)+x^2\left(-\frac{89}{12}-5\ln x\right)+\mathcal O(x^3),
\end{equation}
and thus, for $m_\ell\ll m_S$,
\begin{equation}
\Delta a_\ell^{(1)}=\frac{1}{8\pi^2}\frac{m_\ell^2}{m_S^2}\frac{m_\ell^2}{\vev{}^2}\left\{-[a_\ell^S]^2\left(\frac{7}{6}+\ln\left(\frac{m_\ell^2}{m_S^2}\right)\right)+[b_\ell^S]^2\left(\frac{11}{6}+\ln\left(\frac{m_\ell^2}{m_S^2}\right)\right)\right\}\,.
\end{equation}
%
Yukawa interactions (of charged scalars $C^\pm$) of the form 
\begin{equation}
\mathscr L_{C\ell\nu}=-C^-\bar\ell(a_\ell^C+ib_\ell^C\gamma_5)\nu-C^+\bar\nu(a_\ell^{C\ast}+ib_\ell^{C\ast}\gamma_5)\ell\,,
\end{equation}
give one loop contributions to the anomalous magnetic moment of lepton $\ell$ of the form
\begin{equation}
\Delta a_\ell^{(1)}=-\frac{1}{8\pi^2}\sum_{C}\left\{\abs{a_\ell^C}^2+\abs{b_\ell^C}^2\right\}\,H(x_{\ell C})\,,
\end{equation}
where $x_{\ell C}=m_\ell^2/m_{C^\pm}^2$, and
\begin{equation}
H(x)=-\frac{1}{2}+\frac{1}{x}+\frac{1-x}{x^2}\ln(1-x)\,,\quad H(x)\simeq \frac{x}{6}+\frac{x^2}{12}+\mathcal O(x^3)\text{ for } x\ll 1.
\end{equation}
\subsection{Two loop contributions}
\begin{figure}[!htb]
\begin{center}
\includegraphics[width=0.25\textwidth]{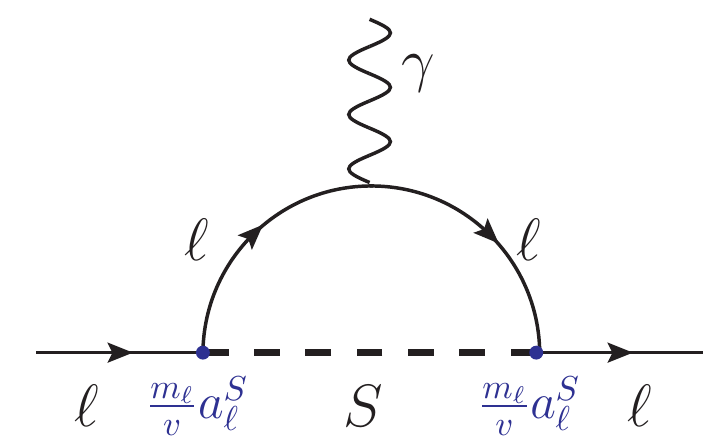}\qquad
\includegraphics[width=0.25\textwidth]{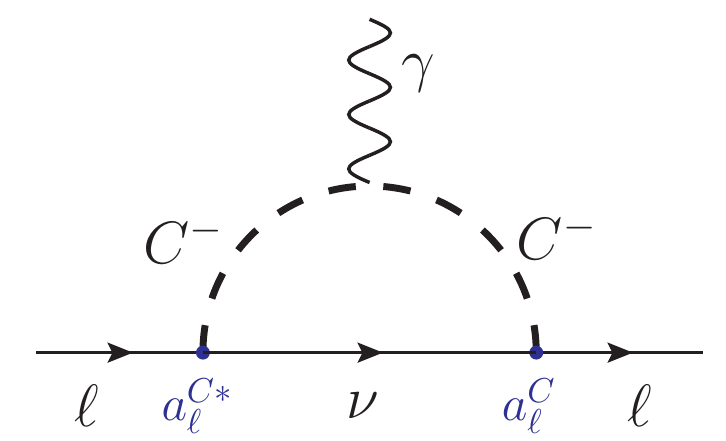}\qquad
\includegraphics[width=0.25\textwidth]{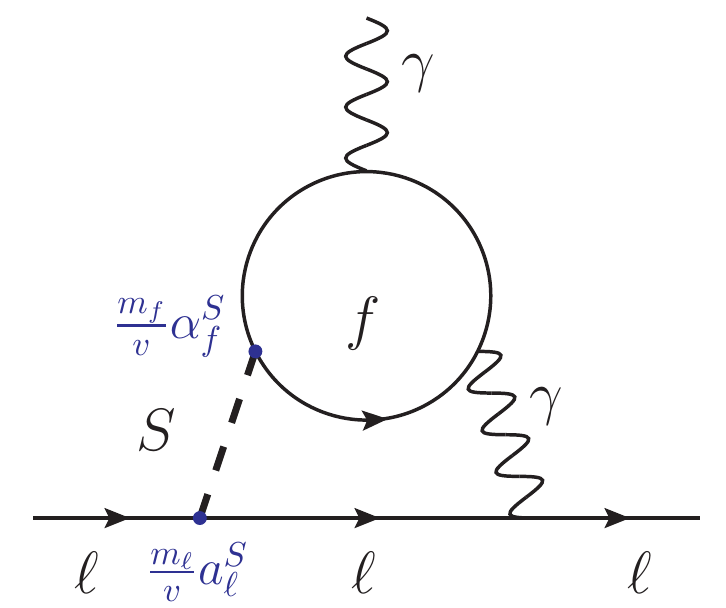}
\end{center}
\caption{Illustrative 1 and 2 loop contributions to $\delta a_\ell$.\label{fig:loops}}
\end{figure}

In addition to \refeq{eq:YukLep:generic}, Yukawa interactions of the form 
\begin{equation}
\mathscr L_{S\bar ff}=-\frac{m_f}{\vev{}}S\bar f(\alpha_f^S+i\beta_f^S\gamma_5)f\,,
\end{equation}
give the following type of two loop Barr-Zee contributions to the anomalous magnetic moment of lepton $\ell$: 
\begin{equation}
\Delta a_\ell^{(2)}=-\frac{\alpha^2}{4\pi^2s_W^2}\frac{m_\ell^2}{M_W^2}\sum_{f}\sum_{S}N_c^fQ_f^2
\left\{
a_\ell^S\alpha_f^Sf(z_{fS})-b_\ell^S\beta_f^Sg(z_{fS})
\right\}\,.
\end{equation}
%
The sum over fermions $f$ corresponds to the different fermions appearing in the closed fermion loop (with $N_c^f$ the number of colours of $f$ and $Q_f$ its electric charge and $z_{fS}=m_f^2/m_S^2$), while the sum over scalars $S$ corresponds to the different neutral scalars connecting the closed fermion loop with the external lepton line, as Figure \ref{fig:loops} illustrates. The functions $f(z)$ and $g(z)$ (see the discussion in section \ref{SEC:g-2}) read:
\begin{equation}
f(z)=\frac{z}{2}\int_0^1\!\!\!\! dx\,\frac{1-2x(1-x)}{x(1-x)-z}\,\ln\left(\frac{x(1-x)}{z}\right)\,,
\end{equation}
\begin{equation}
g(z)=\frac{z}{2}\int_0^1\!\!\!\! dx\,\frac{1}{x(1-x)-z}\,\ln\left(\frac{x(1-x)}{z}\right)\,.
\end{equation}
For other 2 loop contributions see \cite{Cherchiglia:2016eui}.
\clearpage

\providecommand{\href}[2]{#2}\begingroup\raggedright\endgroup


\begin{thebibliography}{114}

\bibitem{Parker:2018vye}
  R.~H.~Parker, C.~Yu, W.~Zhong, B.~Estey and H.~Müller,
  {\it {Measurement of the fine-structure constant as a test of the Standard Model}}
  Science {\bf 360} (2018) 191
  [arXiv:1812.04130 [physics.atom-ph]].


\bibitem{Davoudiasl:2018fbb}
  H.~Davoudiasl and W.~J.~Marciano,
  {\it {Tale of two anomalies}}
  Phys.\ Rev.\ D {\bf 98} (2018) no.7,  075011
  [arXiv:1806.10252 [hep-ph]].


\bibitem{Aoyama:2012wj}
  T.~Aoyama, M.~Hayakawa, T.~Kinoshita and M.~Nio,
  {\it {Tenth-Order QED Contribution to the Electron g-2 and an Improved Value of the Fine Structure Constant}}
  Phys.\ Rev.\ Lett.\  {\bf 109} (2012) 111807
  doi:10.1103/PhysRevLett.109.111807
  [arXiv:1205.5368 [hep-ph]].


\bibitem{Aoyama:2012wk}
  T.~Aoyama, M.~Hayakawa, T.~Kinoshita and M.~Nio,
  {\it {Complete Tenth-Order QED Contribution to the Muon g-2}}
  Phys.\ Rev.\ Lett.\  {\bf 109} (2012) 111808
  [arXiv:1205.5370 [hep-ph]].


\bibitem{Laporta:2017okg}
  S.~Laporta,
  {\it {High-precision calculation of the 4-loop contribution to the electron g-2 in QED}}
  Phys.\ Lett.\ B {\bf 772} (2017) 232
  [arXiv:1704.06996 [hep-ph]].


\bibitem{Aoyama:2017uqe}
  T.~Aoyama, T.~Kinoshita and M.~Nio,
  {\it {Revised and Improved Value of the QED Tenth-Order Electron Anomalous Magnetic Moment}}
  Phys.\ Rev.\ D {\bf 97} (2018) no.3,  036001
  [arXiv:1712.06060 [hep-ph]].


\bibitem{Volkov:2019phy}
  S.~Volkov,
  {\it {Calculating the five-loop QED contribution to the electron anomalous magnetic moment: Graphs without lepton loops}}
  Phys.\ Rev.\ D {\bf 100} (2019) no.9,  096004
  [arXiv:1909.08015 [hep-ph]].


\bibitem{Terazawa:2018pdc}
  H.~Terazawa,
  {\it {Convergence of Perturbative Expansion Series in QED and the Muon g-2: One of the Oldest Problems in Quantum Field Theory and of the Latest Problems in the Standard Model}}
  Nonlin.\ Phenom.\ Complex Syst.\  {\bf 21} (2018) no.3,  268.


\bibitem{Bennett:2006fi}
  G.~W.~Bennett {\it et al.} [Muon g-2 Collaboration],
  {\it {Final Report of the Muon E821 Anomalous Magnetic Moment Measurement at BNL}}
  Phys.\ Rev.\ D {\bf 73} (2006) 072003
  [hep-ex/0602035].


\bibitem{Jegerlehner:2009ry}
  F.~Jegerlehner and A.~Nyffeler,
  {\it {The Muon g-2}}
  Phys.\ Rept.\  {\bf 477} (2009) 1
  [arXiv:0902.3360 [hep-ph]].


\bibitem{Davier:2010nc}
  M.~Davier, A.~Hoecker, B.~Malaescu and Z.~Zhang,
  {\it {Reevaluation of the Hadronic Contributions to the Muon g-2 and to alpha(MZ)}}
  Eur.\ Phys.\ J.\ C {\bf 71} (2011) 1515
   Erratum: [Eur.\ Phys.\ J.\ C {\bf 72} (2012) 1874]
  [arXiv:1010.4180 [hep-ph]].


\bibitem{Davier:2019can}
  M.~Davier, A.~Hoecker, B.~Malaescu and Z.~Zhang,
  {\it {A new evaluation of the hadronic vacuum polarisation contributions to the muon anomalous magnetic moment and to $\mathbf{\boldsymbol\alpha(m_Z^2)}$}}
  Eur.\ Phys.\ J.\ C {\bf 80} (2020) no.3,  241
   Erratum: [Eur.\ Phys.\ J.\ C {\bf 80} (2020) no.5,  410]
  [arXiv:1908.00921 [hep-ph]].


\bibitem{Blum:2018mom}
  T.~Blum {\it et al.} [RBC and UKQCD Collaborations],
  {\it {Calculation of the hadronic vacuum polarization contribution to the muon anomalous magnetic moment}}
  Phys.\ Rev.\ Lett.\  {\bf 121} (2018) no.2,  022003
  [arXiv:1801.07224 [hep-lat]].


\bibitem{Aoyama:2020ynm}
  T.~Aoyama {\it et al.},
  {\it {The anomalous magnetic moment of the muon in the Standard Model}}
  arXiv:2006.04822 [hep-ph].


\bibitem{Roig:2019reh}
  P.~Roig and P.~Sanchez-Puertas,
  {\it {Axial-vector exchange contribution to the hadronic light-by-light piece of the muon anomalous magnetic moment}}
  Phys.\ Rev.\ D {\bf 101} (2020) no.7,  074019
  [arXiv:1910.02881 [hep-ph]].


\bibitem{Giudice:2012ms}
  G.~F.~Giudice, P.~Paradisi and M.~Passera,
  {\it {Testing new physics with the electron g-2}}
  JHEP {\bf 1211} (2012) 113
  [arXiv:1208.6583 [hep-ph]].


\bibitem{Crivellin:2018qmi}
  A.~Crivellin, M.~Hoferichter and P.~Schmidt-Wellenburg,
  {\it {Combined explanations of $(g-2)_{\mu,e}$ and implications for a large muon EDM}}
  Phys.\ Rev.\ D {\bf 98} (2018) no.11,  113002
  [arXiv:1807.11484 [hep-ph]].


\bibitem{Liu:2018xkx}
  J.~Liu, C.~E.~M.~Wagner and X.~P.~Wang,
   {\it {A light complex scalar for the electron and muon anomalous magnetic moments}}
  JHEP {\bf 1903} (2019) 008
  [arXiv:1810.11028 [hep-ph]].


\bibitem{Han:2018znu}
  X.~F.~Han, T.~Li, L.~Wang and Y.~Zhang,
  {\it {Simple interpretations of lepton anomalies in the lepton-specific inert two-Higgs-doublet model}}
  Phys.\ Rev.\ D {\bf 99} (2019) no.9,  095034
  [arXiv:1812.02449 [hep-ph]].


\bibitem{Endo:2019bcj}
  M.~Endo and W.~Yin,
  {\it {Explaining electron and muon $g-2$ anomaly in SUSY without lepton-flavor mixings}}
  JHEP {\bf 1908} (2019) 122
  [arXiv:1906.08768 [hep-ph]].


\bibitem{Bauer:2019gfk}
  M.~Bauer, M.~Neubert, S.~Renner, M.~Schnubel and A.~Thamm,
  {\it {Axionlike Particles, Lepton-Flavor Violation, and a New Explanation of $a_\mu$ and $a_e$}}
  Phys.\ Rev.\ Lett.\  {\bf 124} (2020) no.21,  211803
  [arXiv:1908.00008 [hep-ph]].


\bibitem{Badziak:2019gaf}
  M.~Badziak and K.~Sakurai,
  {\it {Explanation of electron and muon g-2 anomalies in the MSSM}}
  JHEP {\bf 1910} (2019) 024
  [arXiv:1908.03607 [hep-ph]].


\bibitem{Hiller:2019mou}
G.~Hiller, C.~Hormigos-Feliu, D.~F.~Litim and T.~Steudtner,
Phys. Rev. D \textbf{102} (2020) no.7, 071901
doi:10.1103/PhysRevD.102.071901
[arXiv:1910.14062 [hep-ph]].


\bibitem{CarcamoHernandez:2020pxw}
A.~E.~C\'arcamo Hern\'andez, Y.~Hidalgo Vel\'asquez, S.~Kovalenko, H.~N.~Long, N.~A.~P\'erez-Julve and V.~V.~Vien,
Eur. Phys. J. C \textbf{81} (2021) no.2, 191
doi:10.1140/epjc/s10052-021-08974-4
[arXiv:2002.07347 [hep-ph]].


\bibitem{Haba:2020gkr}
N.~Haba, Y.~Shimizu and T.~Yamada,
PTEP \textbf{2020} (2020) no.9, 093B05
doi:10.1093/ptep/ptaa098
[arXiv:2002.10230 [hep-ph]].


\bibitem{Bigaran:2020jil}
I.~Bigaran and R.~R.~Volkas,
Phys. Rev. D \textbf{102} (2020) no.7, 075037
doi:10.1103/PhysRevD.102.075037
[arXiv:2002.12544 [hep-ph]].


\bibitem{Calibbi:2020emz}
  L.~Calibbi, M.~L.~López-Ibáñez, A.~Melis and O.~Vives,
  {\it {Muon and electron $g-2$ and lepton masses in flavor models}}
  JHEP {\bf 2006} (2020) 087
  [arXiv:2003.06633 [hep-ph]].


\bibitem{Chen:2020jvl}
C.~H.~Chen and T.~Nomura,
Nucl. Phys. B \textbf{964} (2021), 115314
doi:10.1016/j.nuclphysb.2021.115314
[arXiv:2003.07638 [hep-ph]].


\bibitem{Jana:2020pxx}
  S.~Jana, V.~P.~K. and S.~Saad,
  {\it {Resolving electron and muon $g-2$ within the 2HDM}}
  Phys.\ Rev.\ D {\bf 101} (2020) no.11,  115037
  [arXiv:2003.03386 [hep-ph]].


\bibitem{Hati:2020fzp}
  C.~Hati, J.~Kriewald, J.~Orloff and A.~M.~Teixeira,
  {\it {Anomalies in $^8$Be nuclear transitions and $(g-2)_{e,\mu}$: towards a minimal combined explanation}}
  JHEP {\bf 2007} (2020) 235
  [arXiv:2005.00028 [hep-ph]].


\bibitem{Dutta:2020scq}
B.~Dutta, S.~Ghosh and T.~Li,
Phys. Rev. D \textbf{102} (2020) no.5, 055017
doi:10.1103/PhysRevD.102.055017
[arXiv:2006.01319 [hep-ph]].


\bibitem{Sabatta:2019nfg}
  D.~Sabatta, A.~S.~Cornell, A.~Goyal, M.~Kumar, B.~Mellado and X.~Ruan,
  {\it {Connecting the Muon Anomalous Magnetic Moment and the Multi-lepton Anomalies at the LHC}}
  Chin.\ Phys.\ C {\bf 44} (2020) no.6,  063103
  [arXiv:1909.03969 [hep-ph]].


\bibitem{Lee:1973iz}
  T.~D.~Lee,
  {\it {A Theory of Spontaneous T Violation}}
  Phys.\ Rev.\ D {\bf 8} (1973) 1226.


\bibitem{Branco:2011iw}
  G.~C.~Branco, P.~M.~Ferreira, L.~Lavoura, M.~N.~Rebelo, M.~Sher and J.~P.~Silva,
  {\it {Theory and phenomenology of two-Higgs-doublet models}}
  Phys.\ Rept.\  {\bf 516} (2012) 1
  [arXiv:1106.0034 [hep-ph]].


\bibitem{Ivanov:2017dad}
  I.~P.~Ivanov,
  {\it {Building and testing models with extended Higgs sectors}}
  Prog.\ Part.\ Nucl.\ Phys.\  {\bf 95} (2017) 160
  [arXiv:1702.03776 [hep-ph]].


\bibitem{Glashow:1976nt}
  S.~L.~Glashow and S.~Weinberg,
  {\it {Natural Conservation Laws for Neutral Currents}}
  Phys.\ Rev.\ D {\bf 15} (1977) 1958.


\bibitem{Haber:1978jt}
  H.~E.~Haber, G.~L.~Kane and T.~Sterling,
  {\it {The Fermion Mass Scale and Possible Effects of Higgs Bosons on Experimental Observables}}
  Nucl.\ Phys.\ B {\bf 161} (1979) 493.


\bibitem{Barger:1989fj}
  V.~D.~Barger, J.~L.~Hewett and R.~J.~N.~Phillips,
  {\it {New Constraints on the Charged Higgs Sector in Two Higgs Doublet Models}}
  Phys.\ Rev.\ D {\bf 41} (1990) 3421.


\bibitem{Pich:2009sp}
  A.~Pich and P.~Tuzon,
  {\it {Yukawa Alignment in the Two-Higgs-Doublet Model}}
  Phys.\ Rev.\ D {\bf 80} (2009) 091702
  [arXiv:0908.1554 [hep-ph]].


\bibitem{Ferreira:2010xe}
  P.~M.~Ferreira, L.~Lavoura and J.~P.~Silva,
  {\it {Renormalization-group constraints on Yukawa alignment in multi-Higgs-doublet models}}
  Phys.\ Lett.\ B {\bf 688} (2010) 341
  [arXiv:1001.2561 [hep-ph]].


\bibitem{Botella:2015yfa}
  F.~J.~Botella, G.~C.~Branco, A.~M.~Coutinho, M.~N.~Rebelo and J.~I.~Silva-Marcos,
  {\it {Natural Quasi-Alignment with two Higgs Doublets and RGE Stability}}
  Eur.\ Phys.\ J.\ C {\bf 75} (2015) 286
  [arXiv:1501.07435 [hep-ph]].


\bibitem{Braeuninger:2010td}
  C.~B.~Braeuninger, A.~Ibarra and C.~Simonetto,
  {\it {Radiatively induced flavour violation in the general two-Higgs doublet model with Yukawa alignment}}
  Phys.\ Lett.\ B {\bf 692} (2010) 189
  [arXiv:1005.5706 [hep-ph]].


\bibitem{Jung:2010ik}
  M.~Jung, A.~Pich and P.~Tuzon,
  {\it {Charged-Higgs phenomenology in the Aligned two-Higgs-doublet model}}
  JHEP {\bf 1011} (2010) 003
  [arXiv:1006.0470 [hep-ph]].


\bibitem{Penuelas:2017ikk}
  A.~Peñuelas and A.~Pich,
  {\it {Flavour alignment in multi-Higgs-doublet models}}
  JHEP {\bf 1712} (2017) 084
  [arXiv:1710.02040 [hep-ph]].


\bibitem{Botella:2018gzy}
  F.~J.~Botella, F.~Cornet-Gomez and M.~Nebot,
  {\it {Flavor conservation in two-Higgs-doublet models}}
  Phys.\ Rev.\ D {\bf 98} (2018) no.3,  035046
  [arXiv:1803.08521 [hep-ph]].


\bibitem{Rodejohann:2019izm}
  W.~Rodejohann and U.~Saldaña-Salazar,
  {\it {Multi-Higgs-Doublet Models and Singular Alignment}}
  JHEP {\bf 1907} (2019) 036
  [arXiv:1903.00983 [hep-ph]].


\bibitem{Georgi:1978ri}
  H.~Georgi and D.~V.~Nanopoulos,
  {\it {Suppression of Flavor Changing Effects From Neutral Spinless Meson Exchange in Gauge Theories}}
  Phys.\ Lett.\  {\bf 82B} (1979) 95.


\bibitem{Donoghue:1978cj}
  J.~F.~Donoghue and L.~F.~Li,
  {\it {Properties of Charged Higgs Bosons}}
  Phys.\ Rev.\ D {\bf 19} (1979) 945.


\bibitem{Botella:1994cs}
  F.~J.~Botella and J.~P.~Silva,
  {\it {Jarlskog - like invariants for theories with scalars and fermions}}
  Phys.\ Rev.\ D {\bf 51} (1995) 3870
  [hep-ph/9411288].


\bibitem{Nebot:2015wsa}
  M.~Nebot and J.~P.~Silva,
  {\it {Self-cancellation of a scalar in neutral meson mixing and implications for the LHC}}
  Phys.\ Rev.\ D {\bf 92} (2015) no.8,  085010
  [arXiv:1507.07941 [hep-ph]].


\bibitem{Andreev:2018ayy}
  V.~Andreev {\it et al.} [ACME Collaboration],
  {\it {Improved limit on the electric dipole moment of the electron}}
  Nature {\bf 562} (2018) no.7727,  355.


\bibitem{Han:2015yys}
  T.~Han, S.~K.~Kang and J.~Sayre,
  {\it {Muon $g-2$ in the aligned two Higgs doublet model}}
  JHEP {\bf 1602} (2016) 097
  [arXiv:1511.05162 [hep-ph]].


\bibitem{Leveille:1977rc}
  J.~P.~Leveille,
  {\it {The Second Order Weak Correction to (G-2) of the Muon in Arbitrary Gauge Models}}
  Nucl.\ Phys.\ B {\bf 137} (1978) 63.


\bibitem{Barr:1990vd}
  S.~M.~Barr and A.~Zee,
  {\it {Electric Dipole Moment of the Electron and of the Neutron}}
  Phys.\ Rev.\ Lett.\  {\bf 65} (1990) 21
   Erratum: [Phys.\ Rev.\ Lett.\  {\bf 65} (1990) 2920].


\bibitem{Chang:1990sf}
  D.~Chang, W.~Y.~Keung and T.~C.~Yuan,
  {\it {Two loop bosonic contribution to the electron electric dipole moment}}
  Phys.\ Rev.\ D {\bf 43} (1991) R14.


\bibitem{Cheung:2001hz}
  K.~m.~Cheung, C.~H.~Chou and O.~C.~W.~Kong,
  {\it {Muon anomalous magnetic moment, two Higgs doublet model, and supersymmetry}}
  Phys.\ Rev.\ D {\bf 64} (2001) 111301
  [hep-ph/0103183].


\bibitem{Cheung:2009fc}
  K.~Cheung, O.~C.~W.~Kong and J.~S.~Lee,
  {\it {Electric and anomalous magnetic dipole moments of the muon in the MSSM}}
  JHEP {\bf 0906} (2009) 020
  [arXiv:0904.4352 [hep-ph]].


\bibitem{Ilisie:2015tra}
  V.~Ilisie,
  {\it {New Barr-Zee contributions to $\mathbf{(g-2)_\mu}$ in two-Higgs-doublet models}}
  JHEP {\bf 1504} (2015) 077
  [arXiv:1502.04199 [hep-ph]].


\bibitem{Cherchiglia:2016eui}
  A.~Cherchiglia, P.~Kneschke, D.~Stöckinger and H.~Stöckinger-Kim,
  {\it {The muon magnetic moment in the 2HDM: complete two-loop result}}
  JHEP {\bf 1701} (2017) 007
  [arXiv:1607.06292 [hep-ph]].


\bibitem{Ivanov:2015nea}
  I.~P.~Ivanov and J.~P.~Silva,
  {\it {Tree-level metastability bounds for the most general two Higgs doublet model}}
  Phys.\ Rev.\ D {\bf 92} (2015) no.5,  055017
  [arXiv:1507.05100 [hep-ph]].


\bibitem{Kanemura:1993hm}
  S.~Kanemura, T.~Kubota and E.~Takasugi,
  {\it {Lee-Quigg-Thacker bounds for Higgs boson masses in a two doublet model}}
  Phys.\ Lett.\ B {\bf 313} (1993) 155
  [hep-ph/9303263].


\bibitem{Akeroyd:2000wc}
  A.~G.~Akeroyd, A.~Arhrib and E.~M.~Naimi,
  {\it {Note on tree level unitarity in the general two Higgs doublet model}}
  Phys.\ Lett.\ B {\bf 490} (2000) 119
  [hep-ph/0006035].


\bibitem{Ginzburg:2005dt}
  I.~F.~Ginzburg and I.~P.~Ivanov,
  {\it {Tree-level unitarity constraints in the most general 2HDM}}
  Phys.\ Rev.\ D {\bf 72} (2005) 115010
  [hep-ph/0508020].


\bibitem{Horejsi:2005da}
  J.~Horejsi and M.~Kladiva,
  {\it {Tree-unitarity bounds for THDM Higgs masses revisited}}
  Eur.\ Phys.\ J.\ C {\bf 46} (2006) 81
  [hep-ph/0510154].


\bibitem{Kanemura:2015ska}
  S.~Kanemura and K.~Yagyu,
  {\it {Unitarity bound in the most general two Higgs doublet model}}
  Phys.\ Lett.\ B {\bf 751} (2015) 289
  [arXiv:1509.06060 [hep-ph]].


\bibitem{Grinstein:2015rtl}
  B.~Grinstein, C.~W.~Murphy and P.~Uttayarat,
  {\it {One-loop corrections to the perturbative unitarity bounds in the CP-conserving two-Higgs doublet model with a softly broken $ {\mathrm{\mathbb{Z}}}_2 $ symmetry}}
  JHEP {\bf 1606} (2016) 070
  [arXiv:1512.04567 [hep-ph]].


\bibitem{Nebot:2020niz}
M.~Nebot,
``Bounded masses in two Higgs doublets models, spontaneous $CP$ violation and $\mathbb{Z_2}$ symmetry,''
Phys. Rev. D \textbf{102} (2020) no.11, 115002
doi:10.1103/PhysRevD.102.115002
[arXiv:1911.02266 [hep-ph]].


\bibitem{Grimus:2008nb}
  W.~Grimus, L.~Lavoura, O.~M.~Ogreid and P.~Osland,
  {\it {The Oblique parameters in multi-Higgs-doublet models}}
  Nucl.\ Phys.\ B {\bf 801} (2008) 81
  [arXiv:0802.4353 [hep-ph]].


\bibitem{Khachatryan:2016vau}
  G.~Aad {\it et al.} [ATLAS and CMS Collaborations],
  {\it {Measurements of the Higgs boson production and decay rates and constraints on its couplings from a combined ATLAS and CMS analysis of the LHC pp collision data at $ \sqrt{s}=7 $ and 8 TeV}}
  JHEP {\bf 1608} (2016) 045
  [arXiv:1606.02266 [hep-ex]].


\bibitem{Aaboud:2017xsd}
  M.~Aaboud {\it et al.} [ATLAS Collaboration],
  {\it {Evidence for the $ H\to b\overline{b} $ decay with the ATLAS detector}}
  JHEP {\bf 1712} (2017) 024
  [arXiv:1708.03299 [hep-ex]].


\bibitem{Sirunyan:2017elk}
  A.~M.~Sirunyan {\it et al.} [CMS Collaboration],
  {\it {Evidence for the Higgs boson decay to a bottom quark–antiquark pair}}
  Phys.\ Lett.\ B {\bf 780} (2018) 501
  [arXiv:1709.07497 [hep-ex]].


\bibitem{Sirunyan:2017khh}
  A.~M.~Sirunyan {\it et al.} [CMS Collaboration],
  {\it {Observation of the Higgs boson decay to a pair of $\tau$ leptons with the CMS detector}}
  Phys.\ Lett.\ B {\bf 779} (2018) 283
  [arXiv:1708.00373 [hep-ex]].


\bibitem{Khachatryan:2014aep}
  V.~Khachatryan {\it et al.} [CMS Collaboration],
  {\it {Search for a standard model-like Higgs boson in the $\mu^+\mu^-$ and $e^+e^-$ decay channels at the LHC}}
  Phys.\ Lett.\ B {\bf 744} (2015) 184
  [arXiv:1410.6679 [hep-ex]].


\bibitem{Aaboud:2017ojs}
  M.~Aaboud {\it et al.} [ATLAS Collaboration],
  {\it {Search for the dimuon decay of the Higgs boson in $pp$ collisions at $\sqrt{s}$ = 13 TeV with the ATLAS detector}}
  Phys.\ Rev.\ Lett.\  {\bf 119} (2017) no.5,  051802
  [arXiv:1705.04582 [hep-ex]].


\bibitem{Kauer:2012hd}
  N.~Kauer and G.~Passarino,
  {\it {Inadequacy of zero-width approximation for a light Higgs boson signal}}
  JHEP {\bf 1208} (2012) 116
  [arXiv:1206.4803 [hep-ph]].


\bibitem{Aad:2015xua}
  G.~Aad {\it et al.} [ATLAS Collaboration],
  {\it {Constraints on the off-shell Higgs boson signal strength in the high-mass $ZZ$ and $WW$ final states with the ATLAS detector}}
  Eur.\ Phys.\ J.\ C {\bf 75} (2015) no.7,  335
  [arXiv:1503.01060 [hep-ex]].


\bibitem{Khachatryan:2016ctc}
  V.~Khachatryan {\it et al.} [CMS Collaboration],
  {\it {Search for Higgs boson off-shell production in proton-proton collisions at 7 and 8 TeV and derivation of constraints on its total decay width}}
  JHEP {\bf 1609} (2016) 051
  [arXiv:1605.02329 [hep-ex]].


\bibitem{Botella:2015hoa}
  F.~J.~Botella, G.~C.~Branco, M.~Nebot and M.~N.~Rebelo,
  {\it {Flavour Changing Higgs Couplings in a Class of Two Higgs Doublet Models}}
  Eur.\ Phys.\ J.\ C {\bf 76} (2016) no.3,  161
  [arXiv:1508.05101 [hep-ph]].


\bibitem{Nebot:2018nqn}
  M.~Nebot, F.~J.~Botella and G.~C.~Branco,
  {\it {Vacuum Induced CP Violation Generating a Complex CKM Matrix with Controlled Scalar FCNC}}
  Eur.\ Phys.\ J.\ C {\bf 79} (2019) no.8,  711
  [arXiv:1808.00493 [hep-ph]].


\bibitem{Kuno:1999jp}
  Y.~Kuno and Y.~Okada,
  {\it {Muon decay and physics beyond the standard model}}
  Rev.\ Mod.\ Phys.\  {\bf 73} (2001) 151
  [hep-ph/9909265].


\bibitem{Tanabashi:2018oca}
M.~Tanabashi \textit{et al.} [Particle Data Group],
``Review of Particle Physics,''
Phys. Rev. D \textbf{98} (2018) no.3, 030001
doi:10.1103/PhysRevD.98.030001

\bibitem{Cirigliano:2007xi}
  V.~Cirigliano and I.~Rosell,
  {\it {Two-loop effective theory analysis of pi (K) ---> e anti-nu/e [gamma] branching ratios}}
  Phys.\ Rev.\ Lett.\  {\bf 99} (2007) 231801
  [arXiv:0707.3439 [hep-ph]].


\bibitem{Pich:2013lsa}
  A.~Pich,
  {\it {Precision Tau Physics}}
  Prog.\ Part.\ Nucl.\ Phys.\  {\bf 75} (2014) 41
  [arXiv:1310.7922 [hep-ph]].


\bibitem{Chun:2016hzs}
  E.~J.~Chun and J.~Kim,
  {\it {Leptonic Precision Test of Leptophilic Two-Higgs-Doublet Model}}
  JHEP {\bf 1607} (2016) 110
  [arXiv:1605.06298 [hep-ph]].


\bibitem{Misiak:2006zs}
  M.~Misiak {\it et al.},
  {\it {Estimate of $\mathcal{B} (\bar B \to X_s \gamma)$ at $O(\alpha_s^2)$}}
  Phys.\ Rev.\ Lett.\  {\bf 98} (2007) 022002
  [hep-ph/0609232].


\bibitem{Crivellin:2013wna}
  A.~Crivellin, A.~Kokulu and C.~Greub,
  {\it {Flavor-phenomenology of two-Higgs-doublet models with generic Yukawa structure}}
  Phys.\ Rev.\ D {\bf 87} (2013) no.9,  094031
  [arXiv:1303.5877 [hep-ph]].


\bibitem{Botella:2014ska}
  F.~J.~Botella, G.~C.~Branco, A.~Carmona, M.~Nebot, L.~Pedro and M.~N.~Rebelo,
  {\it {Physical Constraints on a Class of Two-Higgs Doublet Models with FCNC at tree level}}
  JHEP {\bf 1407} (2014) 078
  [arXiv:1401.6147 [hep-ph]].


\bibitem{Schael:2006wu}
  S.~Schael {\it et al.} [ALEPH Collaboration],
  {\it {Fermion pair production in $e^{+} e^{-}$ collisions at 189-209-GeV and constraints on physics beyond the standard model}}
  Eur.\ Phys.\ J.\ C {\bf 49} (2007) 411
  [hep-ex/0609051].


\bibitem{Aaboud:2017buh}
  M.~Aaboud {\it et al.} [ATLAS Collaboration],
  {\it {Search for new high-mass phenomena in the dilepton final state using 36 fb$^{-1}$ of proton-proton collision data at $ \sqrt{s}=13 $ TeV with the ATLAS detector}}
  JHEP {\bf 1710} (2017) 182
  [arXiv:1707.02424 [hep-ex]].


\bibitem{Aaboud:2019sgt}
  M.~Aaboud {\it et al.} [ATLAS Collaboration],
  {\it {Search for scalar resonances decaying into $\mu^{+}\mu^{-}$ in events with and without $b$-tagged jets produced in proton-proton collisions at $\sqrt{s}=13$ TeV with the ATLAS detector}}
  JHEP {\bf 1907} (2019) 117
  [arXiv:1901.08144 [hep-ex]].


\bibitem{Sirunyan:2019tkw}
  A.~M.~Sirunyan {\it et al.} [CMS Collaboration],
  {\it {Search for MSSM Higgs bosons decaying to $\mu^+\mu^-$ in proton-proton collisions at $\sqrt{s}=$ 13 TeVSearch for MSSM Higgs bosons decaying to $\mu^+ \mu^-$ in proton-proton collisions at s=13TeV}}
  Phys.\ Lett.\ B {\bf 798} (2019) 134992
  [arXiv:1907.03152 [hep-ex]].


\bibitem{Aaboud:2017sjh}
  M.~Aaboud {\it et al.} [ATLAS Collaboration],
  {\it {Search for additional heavy neutral Higgs and gauge bosons in the ditau final state produced in 36 fb$^{-1}$ of pp collisions at $ \sqrt{s}=13 $ TeV with the ATLAS detector}}
  JHEP {\bf 1801} (2018) 055
  [arXiv:1709.07242 [hep-ex]].


\bibitem{Khachatryan:2016qkc}
  V.~Khachatryan {\it et al.} [CMS Collaboration],
  {\it {Search for heavy resonances decaying to tau lepton pairs in proton-proton collisions at $ \sqrt{s}=13 $ TeV}}
  JHEP {\bf 1702} (2017) 048
  [arXiv:1611.06594 [hep-ex]].


\bibitem{Sirunyan:2018zut}
  A.~M.~Sirunyan {\it et al.} [CMS Collaboration],
  {\it {Search for additional neutral MSSM Higgs bosons in the $\tau\tau$ final state in proton-proton collisions at $\sqrt{s}=$ 13 TeV}}
  JHEP {\bf 1809} (2018) 007
  [arXiv:1803.06553 [hep-ex]].


\bibitem{Aaboud:2016dig}
  M.~Aaboud {\it et al.} [ATLAS Collaboration],
  {\it {Search for charged Higgs bosons produced in association with a top quark and decaying via $H^{\pm} \rightarrow \tau\nu$ using $pp$ collision data recorded at $\sqrt{s} = 13$ TeV by the ATLAS detector}}
  Phys.\ Lett.\ B {\bf 759} (2016) 555
  [arXiv:1603.09203 [hep-ex]].


\bibitem{Aaboud:2018cwk}
  M.~Aaboud {\it et al.} [ATLAS Collaboration],
  {\it {Search for charged Higgs bosons decaying into top and bottom quarks at $\sqrt{s}$ = 13 TeV with the ATLAS detector}}
  JHEP {\bf 1811} (2018) 085
  [arXiv:1808.03599 [hep-ex]].


\bibitem{Sirunyan:2019hkq}
  A.~M.~Sirunyan {\it et al.} [CMS Collaboration],
  {\it {Search for charged Higgs bosons in the H$^{\pm}$ $\to$ $\tau^{\pm}\nu_\tau$ decay channel in proton-proton collisions at $\sqrt{s} =$ 13 TeV}}
  JHEP {\bf 1907} (2019) 142
  [arXiv:1903.04560 [hep-ex]].


\bibitem{Sirunyan:2019arl}
  A.~M.~Sirunyan {\it et al.} [CMS Collaboration],
  {\it {Search for a charged Higgs boson decaying into top and bottom quarks in events with electrons or muons in proton-proton collisions at $ \sqrt{\mathrm{s}} $ = 13 TeV}}
  JHEP {\bf 2001} (2020) 096
  [arXiv:1908.09206 [hep-ex]].


\bibitem{Sirunyan:2020hwv}
  A.~M.~Sirunyan {\it et al.} [CMS Collaboration],
  {\it {Search for charged Higgs bosons decaying into a top and a bottom quark in the all-jet final state of pp collisions at $ \sqrt{s} $ = 13 TeV}}
  JHEP {\bf 2007} (2020) 126
  [arXiv:2001.07763 [hep-ex]].


\bibitem{deFlorian:2016spz}
  D.~de Florian {\it et al.} [LHC Higgs Cross Section Working Group],
  {\it {Handbook of LHC Higgs Cross Sections: 4. Deciphering the Nature of the Higgs Sector}}
  arXiv:1610.07922 [hep-ph].


\bibitem{Harlander:2002wh}
  R.~V.~Harlander and W.~B.~Kilgore,
  {\it {Next-to-next-to-leading order Higgs production at hadron colliders}}
  Phys.\ Rev.\ Lett.\  {\bf 88} (2002) 201801
  [hep-ph/0201206].


\bibitem{Ravindran:2003um}
  V.~Ravindran, J.~Smith and W.~L.~van Neerven,
  {\it {NNLO corrections to the total cross-section for Higgs boson production in hadron hadron collisions}}
  Nucl.\ Phys.\ B {\bf 665} (2003) 325
  [hep-ph/0302135].


\bibitem{Pak:2011hs}
  A.~Pak, M.~Rogal and M.~Steinhauser,
  {\it {Production of scalar and pseudo-scalar Higgs bosons to next-to-next-to-leading order at hadron colliders}}
  JHEP {\bf 1109} (2011) 088
  [arXiv:1107.3391 [hep-ph]].


\bibitem{Harlander:2002vv}
  R.~V.~Harlander and W.~B.~Kilgore,
  {\it {Production of a pseudoscalar Higgs boson at hadron colliders at next-to-next-to leading order}}
  JHEP {\bf 0210} (2002) 017
  [hep-ph/0208096].


\bibitem{Anastasiou:2002wq}
  C.~Anastasiou and K.~Melnikov,
  {\it {Pseudoscalar Higgs boson production at hadron colliders in NNLO QCD}}
  Phys.\ Rev.\ D {\bf 67} (2003) 037501
  [hep-ph/0208115].


\bibitem{Ahmed:2016otz}
  T.~Ahmed, M.~Bonvini, M.~C.~Kumar, P.~Mathews, N.~Rana, V.~Ravindran and L.~Rottoli,
  {\it {Pseudo-scalar Higgs boson production at $\text{N}^{\,3}$LO$_{\text {A}}$ +$\text{N}^{\,3}$LL$'$}}
  Eur.\ Phys.\ J.\ C {\bf 76} (2016) no.12,  663
  [arXiv:1606.00837 [hep-ph]].


\bibitem{Flechl:2014wfa}
  M.~Flechl, R.~Klees, M.~Kramer, M.~Spira and M.~Ubiali,
  {\it {Improved cross-section predictions for heavy charged Higgs boson production at the LHC}}
  Phys.\ Rev.\ D {\bf 91} (2015) no.7,  075015
  [arXiv:1409.5615 [hep-ph]].


\bibitem{Degrande:2015vpa}
  C.~Degrande, M.~Ubiali, M.~Wiesemann and M.~Zaro,
  {\it {Heavy charged Higgs boson production at the LHC}}
  JHEP {\bf 1510} (2015) 145
  [arXiv:1507.02549 [hep-ph]].


\bibitem{Alwall:2014hca}
  J.~Alwall {\it et al.},
  {\it {The automated computation of tree-level and next-to-leading order differential cross sections, and their matching to parton shower simulations}}
  JHEP {\bf 1407} (2014) 079
  [arXiv:1405.0301 [hep-ph]].


\bibitem{Alloul:2013bka}
  A.~Alloul, N.~D.~Christensen, C.~Degrande, C.~Duhr and B.~Fuks,
  {\it {FeynRules  2.0 - A complete toolbox for tree-level phenomenology}}
  Comput.\ Phys.\ Commun.\  {\bf 185} (2014) 2250
  [arXiv:1310.1921 [hep-ph]].


\bibitem{Degrande:2014vpa}
  C.~Degrande,
  {\it {Automatic evaluation of UV and R2 terms for beyond the Standard Model Lagrangians: a proof-of-principle}}
  Comput.\ Phys.\ Commun.\  {\bf 197} (2015) 239
  [arXiv:1406.3030 [hep-ph]].


\bibitem{Degrande:2014qga}
  C.~Degrande,
  {\it {Automated Two Higgs Doublet Model at NLO}}
  PoS Charged {\bf 2014} (2015) 024
  [arXiv:1412.6955 [hep-ph]].


\bibitem{Gunion:2002zf}
  J.~F.~Gunion and H.~E.~Haber,
  {\it {The CP conserving two Higgs doublet model: The Approach to the decoupling limit}}
  Phys.\ Rev.\ D {\bf 67} (2003) 075019
  [hep-ph/0207010].


\bibitem{Faro:2020qyp}
  F.~Faro, J.~C.~Romao and J.~P.~Silva,
  {\it {Nondecoupling in Multi-Higgs doublet models}}
  Eur.\ Phys.\ J.\ C {\bf 80} (2020) no.7,  635
  [arXiv:2002.10518 [hep-ph]].
  
  \end{thebibliography}
\end{document}